\documentclass[11pt,a4paper]{revtex4-hack}
\usepackage{wrapfig}
\usepackage{graphicx,setspace,color}
\usepackage{endnotes}
\usepackage{multirow}
\usepackage{pdfpages}

\usepackage[format=plain, font={small, stretch=0.9}, justification=centerlast, labelfont={small, bf}]{caption}
\usepackage{xspace}

\let\footnote=\endnote
\renewcommand{\thefootnote}{\roman{footnote}}

\renewcommand{\thesection}{\arabic{section}}
\renewcommand{\thesubsection}{\thesection.\arabic{subsection}}
\renewcommand{\thesubsubsection}{\thesubsection.\arabic{subsubsection}}
\usepackage[utf8]{inputenc}

\usepackage[paperwidth=210mm,paperheight=297mm,centering,hmargin=2.245cm,vmargin=2.65cm]{geometry}

\usepackage[gen]{eurosym} 
\usepackage{color} 
\usepackage{txfonts}
\usepackage{helvet}

\newcommand{\pot}[2]{#1 \times 10^{#2}}

\newcommand{\Secfgchall}{{3.2}\xspace}
\newcommand{\SecmuFNL}{{2.2}\xspace}
\newcommand{\SecCRR}{{2.6}\xspace}
\newcommand{\SecRM}{{3.3}\xspace}

\newcommand{\Te}{{T_{\rm e}}}
\newcommand{\COBE}{{\it COBE}\xspace}
\newcommand{\COBEF}{{\it COBE}/FIRAS\xspace}
\newcommand{\Planck}{{\it Planck}\xspace}
\newcommand{\PICO}{{\it PICO}\xspace}
\newcommand{\PIXIE}{{\it PIXIE}\xspace}
\newcommand{\SPIXIE}{{\it Super-PIXIE}\xspace}
\newcommand{\PRISM}{{\it PRISM}\xspace}
\newcommand{\PRISTINE}{{\it PRISTINE}\xspace}
\newcommand{\Litebird}{{\it Litebird}\xspace}
\newcommand{\SphereX}{{\it SphereX}\xspace}

\newcommand{\Braket}[1]{{\left<#1\right>}}
\newcommand{\SI}[2]{#1\,{\rm #2}}

\begin{document}
\includepdf[pagecommand={\thispagestyle{empty}},offset= 0mm 0mm]{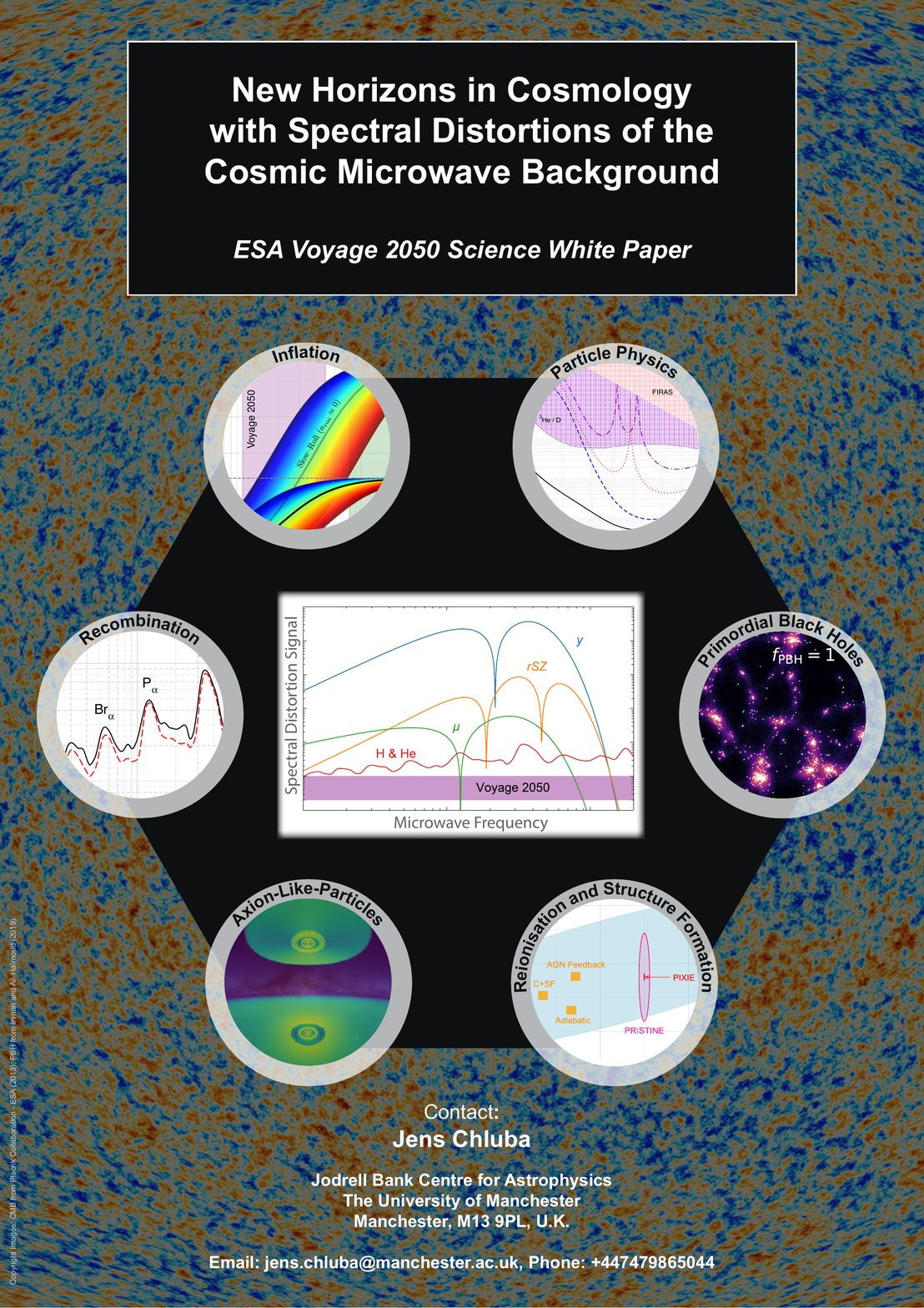}

\begin{center}
{\bf\LARGE Voyage 2050 Science White Paper}

\vspace{5mm}

{\bf\Large New Horizons in Cosmology with Spectral Distortions of the Cosmic Microwave Background}

\end{center}

\thispagestyle{empty}

\begin{center}
{\small
J.~Chluba$^{1}$,
M.~H.~Abitbol$^{2}$,
N.~Aghanim$^{3}$,
Y.~Ali-Ha\"imoud$^{4}$,
M.~Alvarez$^{5,6}$, 
K.~Basu$^{7}$,
B.~Bolliet$^{1}$,
C.~Burigana$^{8}$,
P.~de~Bernardis$^{9, 10}$,
J.~Delabrouille$^{11,12}$, 
E.~Dimastrogiovanni$^{13}$,
F.~Finelli$^{14}$,
D.~Fixsen$^{15}$,
L.~Hart$^{1}$,
C.~Hern\'andez-Monteagudo$^{16}$, 
J.~C.~Hill$^{17, 18}$, 
A.~Kogut$^{19}$,
K.~Kohri$^{20}$,
J.~Lesgourgues$^{21}$,
B.~Maffei$^{3}$, 
J.~Mather$^{19}$, 
S.~Mukherjee$^{22, 23}$, 
S.~P.~Patil$^{24}$,
A.~Ravenni$^{1}$,
M.~Remazeilles$^{1}$, 
A.~Rotti$^{1}$, 
J.~A.~Rubi\~no-Martin$^{25,26}$, 
J.~Silk$^{27, 28}$,
R.~A.~Sunyaev$^{29, 30, 17}$,
E.~R.~Switzer$^{19}$ 
}\\[0mm]

\vspace{3mm}

\vspace{1mm}

\end{center}

\vspace{-5mm}

\noindent
{\scriptsize
$^1$ Jodrell Bank Centre for Astrophysics, Dept. of Physics \& Astronomy, The University of Manchester, Manchester M13 9PL, U.K.
\\[-2mm]
%
$^2$ University of Oxford, Denys Wilkinson Building, Keble Road, Oxford, OX1 3RH, UK
\\[-2mm]
%
$^3$ Institut d'Astrophysique Spatiale (IAS), CNRS (UMR8617), Universit\'e Paris-Sud, Batiment 121, 91405 Orsay, France
\\[-2mm]
%
$^4$ Center for Cosmology and Particle Physics, Department of Physics, New York University, New York, NY 10003, USA
\\[-2mm]
%
$^5$ Berkeley Center for Cosmological Physics, University of California, Berkeley, CA 94720, USA
\\[-2mm]
%
$^6$ Lawrence Berkeley National Laboratory, One Cyclotron Road, Berkeley, CA 94720, USA
\\[-2mm]
%
$^{7}$ Argelander-Institut f\"ur Astronomie, Universit\"at Bonn, Auf dem H\"ugel 71, D-53121 Bonn, Germany
\\[-2mm]
%
$^{8}$ INAF, Istituto di Radioastronomia, Via Piero Gobetti 101, I-40129 Bologna, Italy
\\[-2mm]
%
$^{9}$ Physics department, Sapienza University of Rome, Piazzale Aldo Moro 5, 00185, Rome, Italy
\\[-2mm]
%
$^{10}$ INFN sezione di Roma, P.le A. Moro 2, 00815 Roma, Italy
\\[-2mm]
%
$^{11}$ Laboratoire Astroparticule et Cosmologie, CNRS/IN2P3, 10, rue Alice Domon et L\'eonie Duquet, 75205 Paris Cedex 13, France
\\[-2mm]
%
$^{12}$ D\'epartement d’Astrophysique, CEA Saclay DSM/Irfu, 91191 Gif-sur-Yvette, France 
\\[-2mm]
%
$^{13}$ School of Physics, The University of New South Wales, Sydney NSW 2052, Australia
\\[-2mm]
%
$^{14}$ INAF - Osservatorio di Astrofisica e Scienza dello Spazio, Via Gobetti 101, I-40129 Bologna, Italy
\\[-2mm]
%
$^{15}$ Department of Astronomy, University of Maryland, College Park, MD 20742-2421, USA
\\[-2mm]
%
$^{16}$ Centro de Estudios de F\'isica del Cosmos de Arag\'on (CEFCA), Plaza San Juan, 1, planta 2, E-44001, Teruel, Spain
\\[-2mm]
%
$^{17}$ Institute for Advanced Study, Princeton, NJ 08540, USA
\\[-2mm]
%
$^{18}$ Center for Computational Astrophysics, Flatiron Institute, 162 5th Avenue, New York, NY 10010, USA
\\[-2mm]
%
$^{19}$ NASA/GSFC, Mail Code: 665, Greenbelt, MD 20771, USA
\\[-2mm]
%
$^{20}$ KEK and Sokendai, Tsukuba 305-0801, Japan, and Kavli IPMU, U. of Tokyo, Kashiwa 277-8582, Japan
\\[-2mm]
%
$^{21}$ Institute for Theoretical Particle Physics and Cosmology (TTK), RWTH Aachen University, D-52056 Aachen, Germany
\\[-2mm]
%
$^{22}$ Sorbonne Universit\'e, CNRS, UMR 7095, Institut d'Astrophysique de Paris, 98 bis bd Arago, 75014 Paris, France
\\[-2mm]
%
$^{23}$ Sorbonne Universites, Institut Lagrange de Paris, 98 bis Boulevard Arago, 75014 Paris, France
\\[-2mm]
%
$^{24}$ Niels Bohr International Academy and Discovery Center, Blegdamsvej 17, 2100 Copenhagen, Denmark
\\[-2mm]
%
$^{25}$ Instituto de Astrof\'{\i}sica de Canarias, E-38200 La Laguna, Tenerife, Spain
\\[-2mm]
%
$^{26}$ Departamento de Astrof\'{\i}sica, Universidad de La Laguna, E-38206 La Laguna, Tenerife, Spain
\\[-2mm]
%
$^{27}$ Department of Physics and Astronomy, The Johns Hopkins University, 3701 San Martin Drive, Baltimore MD 21218, USA
\\[-2mm]
%
$^{28}$ Department of Physics and Beecroft Inst. for PAC, University of Oxford, 1 Keble Road, Oxford, OX1 3RH, UK
\\[-2mm]
%
$^{29}$ Max-Planck-Institut f\"ur Astrophysik, Karl-Schwarzschild Str. 1, 85741 Garching, Germany
\\[-2mm]
%
$^{30}$ Space Research Institute (IKI) of the Russian Academy of Sciences, 84/32, Profsoyuznaya str., Moscow, 117810, Russia
\\[-2mm]
}

\vspace{-1mm} 
\hspace{97mm}
\begin{minipage}[t]{0.4\textwidth}
{\scriptsize
\noindent {\it Acronyms}:
\begin{itemize}
\itemsep0em
    \item CMB : Cosmic Microwave Background
    \item SM : Standard Model
    \item CSM : Cosmological Standard Model
    \item SZ : Sunyaev-Zeldovich
    \item FTS  : Fourier Transform Spectrometer
    \item DM  : Dark Matter
    \item CDM  : cold Dark Matter
    \item NG  : non-Gaussian
    \item PBH  : Primordial Black Hole
    \item BBN  : Big Bang Nucleosynthesis
    \item SMBH  : Super-Massive Black Hole
    \item ALP  : Axion-like Particle
    \item IGM  : Intergalactic Medium
    \item CRR  : Cosmic Recombination Radiation
\end{itemize}
}
\end{minipage}

\noindent
\hspace{2mm}\begin{minipage}[t]{0.99\textwidth}
\vspace{-75mm}

\begin{minipage}[t]{0.6\textwidth}
\scriptsize

{\bf Contents}  \\ [1.5mm]
{\it EXECUTIVE SUMMARY}
\\[1mm]
1. {\it COSMOLOGY BEYOND THERMAL EQUILIBRIUM}

\vspace{1mm}
\hspace{3mm}{1.1 Main Types of Spectral Distortions} 

\hspace{3mm}{1.2 CMB Spectral Distortion Signals Across the Sky}
\\[1mm]
2. {\it SPECTRAL DISTORTIONS AS NOVEL TESTS OF $\Lambda$CDM AND BEYOND}

\vspace{1mm}
\hspace{3mm}{2.1 CMB Spectral Distortions as a Probe of Inflation Physics} 

\hspace{3mm}{2.2 Primordial Non-Gaussianity}

\hspace{3mm}{2.3 CMB Spectral Distortions as a Probe of Dark Matter and Particle Physics}

\hspace{3mm}{2.4 Primordial Black Holes}

\hspace{3mm}{2.5 Axion-Like Particles}

\hspace{3mm}{2.6 The Cosmological Recombination Radiation}

\hspace{3mm}{2.7 Reionization and Structure Formation}

\hspace{3mm}{2.8 Line Intensity Mapping}

\hspace{3mm}{2.9 Resonant Scattering Signals}
\\[1mm]
3. {\it THE PATH FORWARD WITH CMB SPECTRAL DISTORTIONS}

\vspace{1mm}
\hspace{3mm}{3.1 Technological challenges} 

\hspace{3mm}{3.2 Foreground Challenge for CMB Spectral Distortion}

\hspace{3mm}{3.3 Possible Mission Concepts and Experimental Roadmap}

\hspace{3mm}{3.4 Synergies}
\\[1mm]
4. {\it CONCLUSIONS}
\end{minipage}

\end{minipage}

\vspace{6mm}
{\scriptsize
\noindent {\it Acknowledgements}: We cordially thank the contributors to the recent Decadal White Paper \citep{Chluba2019WPDEC}, which provided the starting point for this response to the ESA Voyage 2050 call.
This work has received funding from the European Research Council (ERC) under the European Union's Horizon 2020 research and innovation program (grant agreement No 725456, CMBSPEC) as well as the Royal Society (grants UF130435 and RG140523).
}


\newpage
\setcounter{page}{0}
\thispagestyle{empty}

\section*{EXECUTIVE SUMMARY}
\vspace{-3.5mm}
Following the pioneering observations with \COBE in the early 1990s, studies of the cosmic microwave background (CMB) have primarily focused on temperature and polarization anisotropies.
CMB spectral distortions -- tiny departures of the CMB energy spectrum from that of a perfect blackbody -- provide a second, independent probe of fundamental physics, with a reach deep into the primordial Universe. The theoretical foundation of spectral distortions has seen major advances in recent years, highlighting the immense potential of this emerging field. Spectral distortions probe a fundamental property of the Universe -- its {\it thermal history} -- thereby providing additional insight into processes within the cosmological standard model\footnote{When referring to the cosmological standard model (CSM) we assume the $\Lambda$CDM parametrization, supplemented by the Standard Model of particle physics, admitting that the presence of dark matter and dark energy requires physics beyond the latter.} (CSM) as well as new physics beyond.
Spectral distortions are an important tool for understanding inflation and the nature of dark matter. They shed new light on the physics of recombination and reionization, both prominent stages in the evolution of our Universe, and furnish critical information on baryonic feedback processes, in addition to probing primordial correlation functions at scales inaccessible to other tracers. In principle the range of signals is vast: {\it many orders of magnitude of discovery space} can be explored by detailed observations of the CMB energy spectrum. Several CSM signals are predicted and provide clear experimental targets that are observable with present-day technology. 
Confirmation of these signals would extend the reach of the CSM by orders of magnitude in physical scale as the Universe evolves from the initial stages to its present form. {\it Their absence would pose a huge theoretical challenge, immediately pointing to new physics.}

Here, we advocate for a dedicated effort to measure CMB spectral distortions at the largest angular scales ($\gtrsim 1^\circ$) within the ESA Voyage 2050 Program (see Sect.~\SecRM for roadmap). 
We argue that an L-class mission with a pathfinder would allow a precise measurement of {\it all the expected} CSM {\it distortions}. With an M-class mission, the primordial distortions (created at $z\gtrsim 10^3$) would still be detected at modest significance, while the late-time distortions will continue to be measured to high accuracy.
Building on the heritage of \COBEF~\citep{Mather1994, Fixsen1996}, a spectrometer that consists of multiple, cooled ($\simeq 0.1$~K), absolutely-calibrated Fourier Transform Spectrometers (FTS) with wide frequency coverage ($\nu \simeq 10$~GHz to a few$\times$THz) and all-sky spectral sensitivity at the level of $0.1-0.5$~Jy/sr would be the starting point for the M-class option. A scaled and further optimized version of this concept is being envisioned as the L-class option.
Such measurements can only be done from space and would deliver hundreds of absolutely-calibrated maps of the Universe at large scales, opening numerous science opportunities for cosmology and astrophysics (see Sect.~3.4 for synergies). 
This will provide independent probes of inflation, dark matter and particle physics, recombination and the energy output of our Universe from at late times, turning the long-standing spectral distortion limits of \COBEF into clear detections.  

\vspace{-5mm}
\section{Cosmology beyond thermal equilibrium}
\vspace{-3.5mm}
Cosmology is now a precise scientific discipline, with a detailed theoretical model that fits a wealth of very accurate measurements. Of the many cosmological data sets, the CMB temperature and polarization anisotropies provide the most stringent and robust constraints, allowing us to determine the key parameters of our Universe (e.g., the total density, expansion rate and baryon content) with unprecedented precision, while simultaneously addressing fundamental questions about inflation and early-universe physics. By studying the statistics of the CMB anisotropies with different experiments over the past decades we have entered the era of precision cosmology, clearly establishing the highly-successful $\Lambda$CDM concordance model \cite{Smoot1992, WMAP_params, Planck2013params}.

But the quest continues. Despite its many successes, $\Lambda$CDM {\it is known to be incomplete}. It traces the growth of structure in the Universe from primordial density perturbations to the modern era, but the origin of those perturbations remains poorly understood. In addition, in spite of relentless efforts, the nature of dark matter (DM) and dark energy remains a mystery. Together, these enigmatic components comprise 95\% of the energy density of the Universe. Particle and high-energy physics offer candidate solutions for these problems (e.g., inflation and particle dark matter), but these inevitably require new physics beyond the Standard Model of particle physics.

{\it Precision measurements of the} CMB {\it energy spectrum open a new window into the physics of the early Universe, constraining cosmological models in ways not possible using other techniques.} Departures of the CMB energy spectrum from a pure blackbody -- commonly referred to as {\it spectral distortions} -- encode unique information about the thermal history of the Universe, from when it was a few months old until today. Since the measurements with \COBEF in the early '90s, the sky-averaged CMB spectrum is known to be extremely close to a perfect blackbody at a temperature $T_0=(2.7255\pm 0.0006)\,{\rm K}$ \cite{Fixsen1996, Fixsen2009}, with possible distortions limited to one part in $10^5$. This impressive measurement was awarded the 2006 Nobel Prize in Physics and already rules out cosmologies with extended periods of large energy release. 
Here we propose to revisit the measurement of the CMB spectrum with current and upcoming technology.

\begin{figure}
\vspace{-2.5mm}
   \includegraphics[width=0.78\columnwidth]{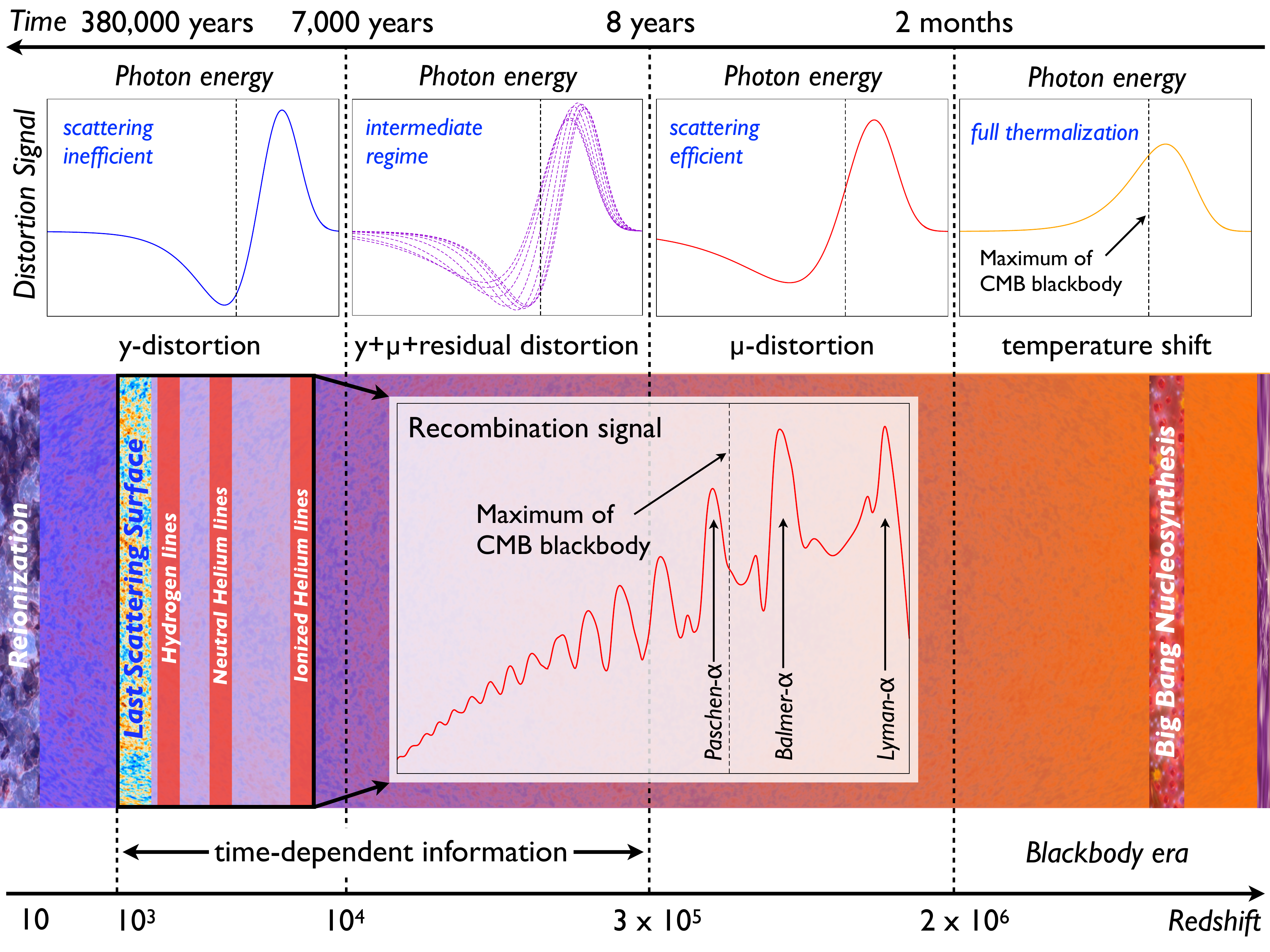}
   \caption{Evolution of spectral distortions across time. Distortions probe the thermal history over long periods deep into the primordial Universe that are inaccessible by other means. The distortion shape contains valuable epoch-dependent information that allows distinguishing different sources of distortions. Line-emission is created during the cosmological recombination eras leaving a detailed 'fingerprint' of the recombination process. The figure is adapted from \citep{Silk2014Science}.}
   \label{fig:stages}
\end{figure}

\vspace{-5mm}
\subsection{Main Types of Spectral Distortions}
\vspace{-3.5mm}
Spectral distortions are created by processes that drive matter and radiation out of thermal equilibrium after thermalization becomes inefficient at redshift $z\lesssim \pot{2}{6}$. Examples are energy-releasing mechanisms that heat the baryonic matter or inject photons or other electromagnetically-interacting particles. The associated signals are usually characterized as $\mu$- and $y$-type distortions, formed by energy exchange between electrons and photons through Compton scattering \citep{Zeldovich1969, Sunyaev1970mu, Illarionov1975b, Burigana1991, Hu1993}. Compton scattering is inefficient at $z \lesssim \pot{5}{4}$, yielding a $y$-type distortion, which probes the thermal history during recombination and reionization (Fig.~\ref{fig:stages}). In contrast, a $\mu$-type (or chemical potential) distortion forms at $z\gtrsim \pot{5}{4}$, when Compton scattering is very efficient. A $\mu$-distortion cannot be generated at recent epochs and thus directly probes events in the pre-recombination era. 

The simple classical picture has been refined in recent years. We now understand that the transition between $\mu$- and $y$-type distortions is gradual (see intermediate regime in Fig.~\ref{fig:stages} at redshifts $10^4\lesssim z \lesssim \pot{3}{5}$) and that the signal contains additional time-dependent information \citep{Chluba2011therm, Khatri2012mix, Chluba2013Green}. This extra information is contained in the residual or $r$-type distortion, which cannot be described by a simple sum of $\mu$ and $y$, and thus can be used to distinguish energy release mechanisms \citep{Chluba2013fore, Chluba2013PCA}. 

It was also shown that distortions created by photon-injection mechanisms can exhibit a rich spectral phenomenology \citep{Chluba2015GreensII}. One prominent example is the distortion created by the cosmological recombination process \citep{Dubrovich1975, Sunyaev2009, Chluba2016CosmoSpec} (see Fig.~\ref{fig:stages}). Additional epoch-dependent information can be imprinted by non-equilibrium processes in the pre-recombination hydrogen and helium plasma \citep{Liubarskii83, Chluba2008c, Chluba2010a} or by non-thermal particles in high-energy particle cascades \citep[e.g.,][]{Ensslin2000, Chluba2010a, Chluba2015GreensII, Slatyer2015, Acharya2018}. 

Spectral distortions thus provide more than just a simple integral constraint for cosmology. They are a unique and powerful probe of a wide range of interactions between particles and CMB photons, reaching back all the way from the present to a few months after the Big Bang and allowing us to access information that cannot be extracted in any other way. Broad overviews of the CMB spectral distortion science case can be found in \citep{Sunyaev1970SPEC, Sunyaev1980ARAA, Sunyaev2009, Chluba2011therm, Sunyaev2013, PRISM2013WPII, Chluba2014Moriond, Tashiro2014, deZotti2015, Chluba2016, Chluba2018}.

\vspace{-3mm}
\subsection{CMB Spectral Distortion Signals Across the Sky}
\vspace{-3.5mm}
CMB distortion signals span a wide range of spectral shapes. Another important way to distinguish distortions is through their distribution across the sky. CMB spectral distortions are usually {\it isotropic} signals, directly imprinted in the energy distribution of the sky-averaged CMB (i.e., the monopole). To extract spectral distortions one therefore has to measure the absolute photon flux at different frequencies, while the direction on the sky is secondary. 
This requires accurate absolute calibration \citep{Mather1994,Kogut2011} or accurate channel inter-calibration \citep{Sunyaev2009,Mukherjee2018FSD}, which can be achieved with experimental concepts like \PIXIE \citep{Kogut2011PIXIE, Kogut2016SPIE}.
To minimize foreground contaminations, prior knowledge (e.g., from \Planck) can be used to optimize the scanning strategy and beam size. This implies that spectral distortion measurements at large angular scales ($\gtrsim 1^\circ$) are optimal.

However, CMB spectral distortions can also have {\it anisotropic} components.
One prominent example is due to the Sunyaev-Zeldovich (SZ) effect \citep{Sunyaev1970mu, Sunyaev1972CoASP} caused by the scattering of photons by energetic electrons inside clusters of galaxies, which has become an important tool for  cosmology \citep[e.g.,][]{Carlstrom2002}. 
In contrast, anisotropic $\mu$- or $y$-distortions from the pre-recombination era ($z>10^3$) are expected to be negligible, can, however, be boosted to visible levels due to super-horizon mode correlations, e.g., caused by primordial non-Gaussianity \citep{Pajer2012, Ganc2012, Emami:2015xqa, Ota2016muE} (see Sect.~\SecmuFNL).  The CMB dipole spectrum is furthermore distorted due to our motion with respect to the CMB restframe \citep{Danese1981, Balashev2015, Burigana2018CORE}.
Line and resonance scattering effects also leave anisotropic imprints (Sect.~2.8 and 2.9).
All these signals can be correlated against tracers of both primordial density perturbations and large-scale structure to further probe cosmic evolution \citep{Refregier2000, Zhang2004, Pitrou2010, Pajer2012, Ganc2012, Hill2013, Alvarez2016}. Measurements of distortion anisotropies may furthermore shed new light on the origin the large-scale CMB anomalies~\citep{Dai2013}.

The focus of this white paper is to optimize towards measurements of distortions signals at large angular scales ($\gtrsim 1^\circ$), primarily targeting the monopole signals. A more detailed discussion of possible experimental concepts is left to Sect.~\ref{sec:observ}; however, in the following, \PIXIE \citep{Kogut2011PIXIE, Kogut2016SPIE} and an enhanced version, \SPIXIE \citep{Kogut2019WP}, will be used as benchmarks. 
In the presence of foregrounds, \PIXIE could reach one standard deviation errors of $\sigma(y)\simeq \pot{3.4}{-9}$ and $\sigma(\mu)\simeq \pot{3}{-8}$, while \SPIXIE could reach $\sigma(y)\simeq \pot{1.6}{-9}$ and $\sigma(\mu)\simeq \pot{7.7}{-9}$ (see Sect.~\Secfgchall, Fig.~\ref{fig:future_errors}). Both concepts would improve the long-standing distortion limits of \COBEF $|y|<\pot{1.5}{-5}$ and $|\mu|<\pot{9}{-5}$ (95\% c.l.) \citep{Mather1994, Fixsen1996} by several orders of magnitude.
A dedicated high-resolution CMB imager approach is discussed in \citep{Basu2019WP} to enable precise spectral measurements at small angular scales, targeting SZ clusters, high-$\ell$ line-scattering signals and other secondary CMB anisotropies.

Finally, we mention that polarized CMB spectral distortions are usually negligible, such that one can in principle focus on intensity measurements only. However, polarization sensitivity could be useful for component separation. It was furthermore demonstrated that polarization-sensitive spectrometers like {\it PIXIE} \citep{Kogut2011PIXIE, Kogut2016SPIE} could place tight constraints on the tensor-to-scalar ratio, $r$, reaching $\sigma(r) \simeq 10^{-3}$ \citep{Kogut2011PIXIE,Kogut2016SPIE}, and could deliver a cosmic-variance-limited measurement of the Thomson optical depth, $\tau$ \citep{Calabrese2017}, to complement future ground-based experiments in their efforts to measure neutrino masses (e.g., CMB-S4). Polarization capabilities should thus still be considered when designing future CMB spectrometers.

\vspace{3mm}
\section{Spectral Distortions as Novel Tests of $\Lambda$CDM and Beyond}

\vspace{-2mm}
\subsection{CMB Spectral Distortions as a Probe of Inflation Physics}
\vspace{-3.5mm}
A central question in modern cosmology is the origin of the observed primordial density perturbations. Measurements from CMB anisotropies and large-scale structure find a nearly scale-invariant power spectrum $\mathcal{P}(k) \simeq k^{n_{\rm S} - 1}$ with spectral index $n_{\rm S}= 0.965 \pm 0.004$, sampled over a range of spatial scales $k\simeq 10^{-4}$ to $0.1$ Mpc$^{-1}$ \citep{Planck2018params}. Their phase coherence is evidence for their super-Hubble nature, and their near scale-invariance is evidence of a weakly broken shift symmetry in the underlying theory. However, their precise origin is as of yet unknown. 

Inflation provides a widely accepted framework for generating these initial fluctuations \citep{Starobinsky:1980te, Guth:1980zm, Linde:1981mu, Albrecht:1982wi}, with the simplest models generically predicting a small departure from scale-invariance (with $n_{\rm S} < 1$) as the inflaton rolls down its potential \citep{mukhanov, Hawking:1982cz, Starobinsky:1979ty, Guth:1982ec}.
However, various alternatives to inflation have been proposed \citep{Gasperini:1992em,Wands:1998yp, Khoury:2001bz, Hollands:2002yb, Brandenberger:2006vv, Craps:2007ch, Creminelli:2010ba, Ijjas:2016tpn, Dobre2018} and no clear theoretical consensus has yet emerged. Searches for primordial $B$-mode patterns in CMB polarization could yield additional evidence for the simplest inflationary models. CMB polarization measurements so far only provide upper limits \cite{Planck2018params, Ade:2018gkx} with no firm target from theory for a guaranteed detection. However, detection of a tensor to scalar ratio of $r \simeq 10^{-3}$ is a distinguishing benchmark for large-field models, which in certain realizations further manifest the specific relation $r \simeq (1 - n_{\rm S})^2$ \citep[e.g.,][]{Starobinsky:1980te, Bezrukov2008}.

\begin{figure}
\includegraphics[width=0.95\textwidth]{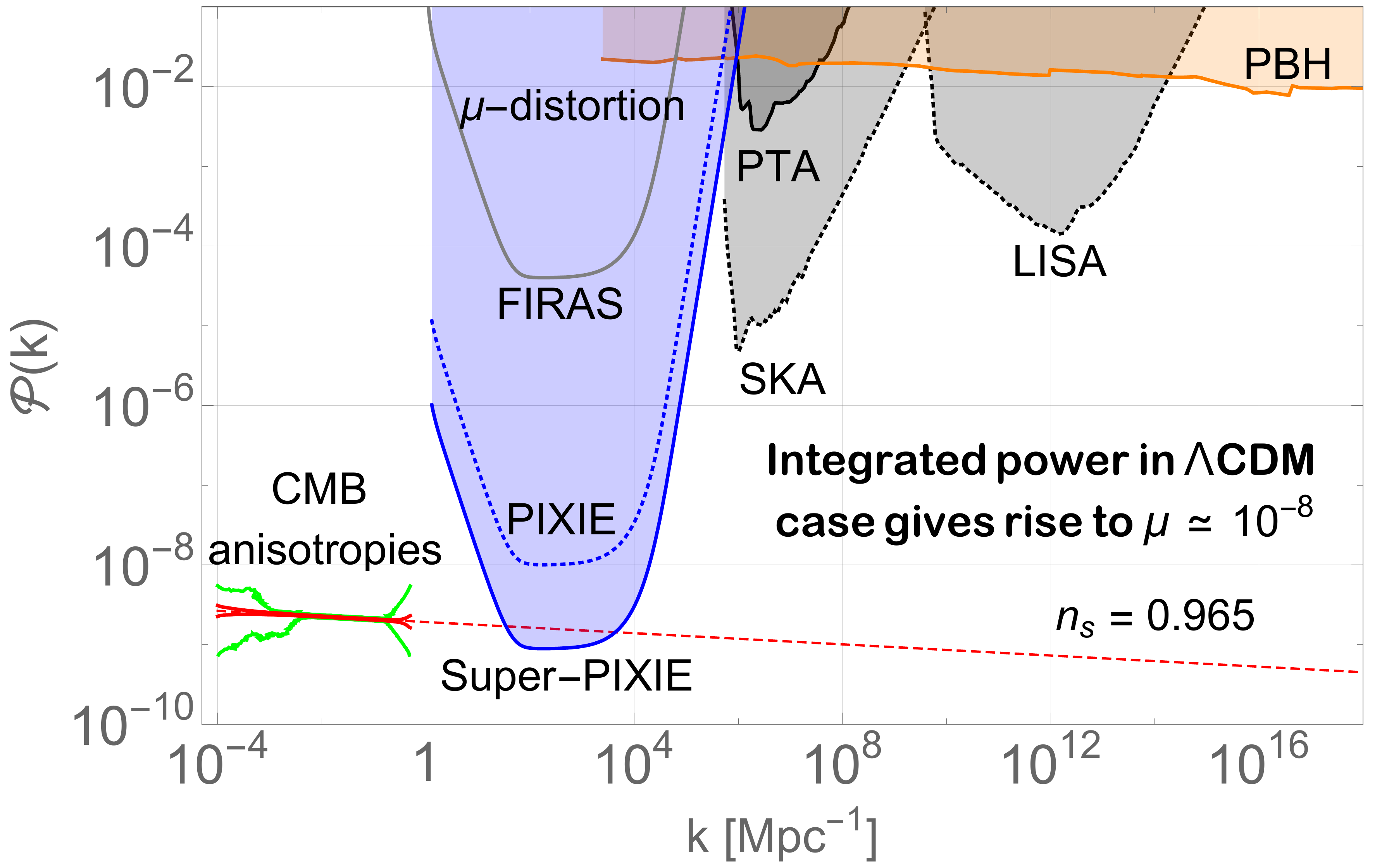}
\caption{Forecast constraints (95 \% c.l.) on the primordial power spectrum for features with a $k^4$ profile that cuts off sharply at some larger wavenumber\protect\footnote{To avoid the unrealistic GW spectrum generated by a $\delta$-function scalar power spectrum, we plot all integrated constraints using a $k^4$ spectrum -- see \citep{Byrnes:2018txb} for the reason for this choice. The peak sensitivity for $\mu$-distortions is effectively unchanged were we to instead plot constraints for $\delta$-function features in the power spectrum with the same integrated power (see Fig.~9 therein), and also \cite{Inomata2018}.} $k_p$ \citep[see][for more details]{Byrnes:2018txb}. $\mu$-distortions constrain perturbations at scales and levels inaccessible to other probes. Early-universe models with enhanced small-scale power at $k\simeq 10-10^4\,{\rm Mpc^{-1}}$ will be immediately ruled out if no distortion with $\mu> \pot{2}{-8}$ is detected. The figure is adapted from \citep{Chluba2019WPDEC}.}
\vspace{2mm}
\label{fig:forecast}
\end{figure}

{\it Spectral distortions provide a unique new probe of primordial density perturbations.}
Inflation may or may not be a valid description of the early Universe, but density perturbations are known to exist; regardless of their origin, dissipation of these perturbations through photon diffusion ($\leftrightarrow$ Silk damping) in the early Universe 
will distort the CMB spectrum at observable levels \citep{Sunyaev1970diss, Daly1991, Hu1994, Chluba2012, Khatri2012short2x2}.
The signal ($\mu+y+r$-distortion) can be accurately calculated using simple linear physics and depends on the amplitude of primordial perturbations at scales with wavenumbers $k\simeq 1-10^4\,{\rm Mpc^{-1}}$, some ten e-folds further than what can be probed by CMB anisotropies (Fig.~\ref{fig:forecast}). 
Given an initial curvature power spectrum, $P(k)=2\pi^2 k^{-3}\,\mathcal{P}(k)$, the average $\mu$-distortion can be estimated  with \citep[e.g.,][]{Chluba2012inflaton, Chluba2015IJMPD}:
\begin{equation}
	\langle\mu\rangle 
	\approx
	\int \frac{k^2\,{\rm d} k}{2\pi^2} \; P(k) \,W_\mu(k),
\label{eq:mu-averaged}
\end{equation}
using an appropriate $k$-space window function, $W_\mu(k)$, which receives most of its contributions from $k\simeq 10^2\,{\rm Mpc^{-1}}-10^4\,{\rm Mpc^{-1}}$. If the near scale-invariance of the power spectrum observed on large scales persists to these much smaller scales, then the predicted distortion, $\mu\simeq \pot{(2.3\pm0.14)}{-8}$ \citep{Chluba2012, Cabass2016, Chluba2016}, could be observed using current technology (Sect.~\ref{sec:observ}). Detecting this signal extends our grasp on primordial density perturbations by over three orders of magnitude in scale, covering epochs that cannot be probed directly in any other way. {\it A non-detection at this level would be a serious challenge for} $\Lambda$CDM, {\it immediately requiring new physics.} 

Measurements of $\mu$-distortions are directly sensitive to the power spectrum amplitude and its scale-dependence around $k\simeq 10^3\,{\rm Mpc^{-1}}$ \citep{Chluba2012inflaton, Pajer2012b, Khatri2013forecast, Chluba2013PCA}. Within the slow-roll paradigm, this provides a handle on higher-order slow-roll parameters (often parametrized as running of the tilt or running of the running), benefiting from a vastly extended lever arm \citep{Powell2012, Chluba2013PCA, Clesse2014, Cabass2016SRparams}.
Outside of standard slow-roll inflation, large departures from scale-invariance are well-motivated and often produce excess small-scale power (e.g., features \citep{Starobinsky:1992ts, Adams:2001vc, Hazra:2014jka} or inflection points \citep{Polnarev:2006aa, Kohri:2007qn, Ido2010, Choudhury:2013woa, Clesse:2015wea, Germani:2017bcs} in the potential, particle production \citep{Barnaby:2009dd, Cook:2011hg, Battefeld:2013bfl, Dimastrogiovanni2017, Domcke:2018eki}, waterfall transitions \citep{Linde:1993cn, Lyth:1996kt,Juan1996PhRvD, Abolhasani:2010kn, Clesse:2014pna}, axion inflation~\citep{Barnaby2011, Barnaby2012, Meerburg:2012id}, etc.~\citep{Chluba2015IJMPD}), implying the presence of new physical scales that can be probed with spectral distortions (Fig.~\ref{fig:forecast}). 
In this respect, a non-detection of the predicted $\mu$-distortions could establish a link to a possible primordial origin of the small-scale structure crisis \citep{Nakama2017, Cho2017}. 
Spectral distortions are also created by the dissipation of small-scale {\it tensor} perturbations \citep{Ota2014, Chluba2015} and depend on the perturbation-type (i.e., adiabatic vs. iso-curvature) \citep{Hu1994isocurv, Dent2012, Chluba2013iso, Haga2018}, providing additional ways to test inflation scenarios in uncharted territory. 

\begin{figure}
\vspace{-6mm}
\includegraphics[width=0.8\textwidth]{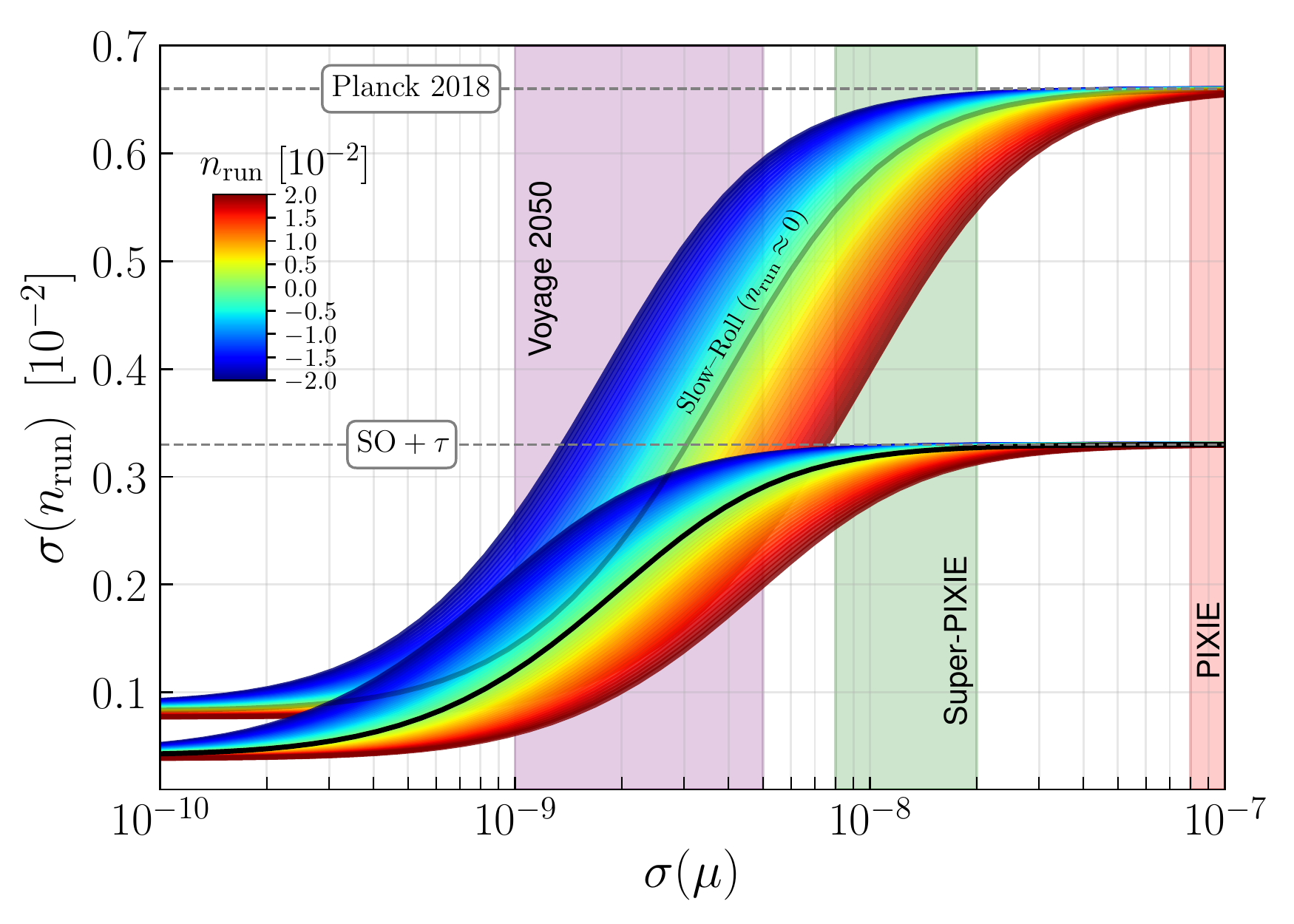}
\vspace{-2mm}
\caption{Expected sensitivity to running of the spectral index, $n_{\rm run}$, when combining CMB anisotropy measurements with a spectrometer of varying sensitivity to $\mu$. Across the colored band the fiducial value for $n_{\rm run}$ is varied. In all cases, a spectrometer leads to improvements of the constraint once $\sigma(\mu)\simeq 2\times 10^{-8}$ can be reached. The improvement depends on the fiducial value of $n_{\rm run}$. For reference, the case $n_{\rm run}=0$ as expected from generic slow-roll scenarios is shown as solid black line. The shaded regions give plausible ranges for $\sigma(\mu)$ expected for the annotated concepts~(Sect.~\Secfgchall).}
\vspace{3mm}
\label{fig:forecast_nrun}
\end{figure}
Working within the slow-roll inflation paradigm, in Fig.~\ref{fig:forecast_nrun}, we further illustrate the gains in estimating the running of the spectral index when combining CMB anisotropy measurements with a CMB spectrometer. Due to the extended lever arm, small changes in $n_{\rm run}$ significantly affect the $\mu$-distortion amplitude \citep[see][for illustrations and approximations]{Chluba2012}. Thus, CMB spectral distortion measurements allow improving constraints on $n_{\rm run}$ \citep{Powell2012, Khatri2013forecast, Chluba2013PCA, Clesse2014, Cabass2016SRparams}. The expected improvement is larger for $n_{\rm run}>0$, while it is lower for $n_{\rm run}<0$, due to the change in the value of $\mu$ \citep[e.g.,][]{Chluba2013PCA}. 

For standard slow-roll inflation models, $n_{\rm run}$ is very close to zero, consistent with current best constraints from \Planck: $n_{\rm run}=-0.0041\pm 0.0067$ \citep{Planck2018params}.
Assuming a fiducial value of $n_{\rm run}=0$ and combining \Planck\ with a future spectrometer could tighten the error on $n_{\rm run}$ by a factor of $\simeq 1.7$ to $\sigma(n_{\rm run})\simeq 0.004$ if a distortion sensitivity $\sigma(\mu)\simeq 5\times 10^{-9}$ is achieved (see Fig.~\ref{fig:forecast_nrun}). 
To reach $\sigma(n_{\rm run})\simeq 0.0033$, as plausible for the funded Simons Observatory (SO) together with a cosmic-variance-limited measurement of $\tau$ \citep{SOWP2018}, in combination with \Planck one requires $\sigma(\mu)\simeq 3\times 10^{-9}$. A cosmic-variance-limited measurement of $\tau$ itself is possible with a polarizing spectrometer, yielding $\sigma(\tau)\simeq 0.002$ \citep{Calabrese2017}, but is also expected to become available with \Litebird \citep{Suzuki2018}.
By combining a spectrometer with\footnote{In practice we rescaled the \Planck covariance matrix assuming factors of 2 error improvements for $A_{\rm S}$, $n_{\rm S}$ and $n_{\rm run}$.} SO$+\tau$, at $\sigma(\mu)\simeq 2\times 10^{-9}$ we can further improve the error on $n_{\rm run}$ to $\sigma(n_{\rm run})\simeq 0.002$, another factor of $\simeq 1.7$ better than SO$+\tau$ alone. 
This highlights some of the potential for spectral distortions as a probe of standard slow-roll inflation physics.

\vspace{-5mm}
\subsection{Primordial Non-Gaussianity}
\label{sec:mu-aniso-FNL}
\vspace{-3.5mm}
Spectral distortion anisotropies also can be used to probe local-type primordial non-Gaussianity at small scales \citep{Pajer2012, Ganc2012, Biagetti2013, Ota2015aniso_iso, Emami:2015xqa, Khatri2015aniso, Chluba2017, Ota:2016mqd, Ravenni2017, Cabass:2018jgj}, an exciting direction that complements other probes and could shed light on multi-field inflation scenarios \citep{Dimastrogiovanni2016}.
As discussed above, the dissipation of primordial acoustic modes on small scales generates a guaranteed contribution to the isotropic $\mu$- and $y$-distortions. Non-Gaussian (NG) couplings between short- and long-wavelength modes create inhomogeneities in the amplitude of the small-scale power, which in turn lead to an anisotropic spectral distortion that correlate with tracers of the long-wavelength modes \cite{Pajer2012, Ganc2012}.

Broadly speaking, most of the information about the non-Gaussianity generated by different early-universe models can be captured by the Fourier transform of the 3- and 4-point correlation functions, respectively, the primordial bispectrum and trispectrum. At large scales, these have been tightly constrained by the \Planck collaboration analysis of the CMB temperature and polarization anisotropies (respectively $T$ and $E$) bispectrum and trispectrum \cite{Akrami:2019izv}.
In contrast, cross correlations of $T$ and $E$ with $\mu$-distortions anisotropies probe an interesting class of bispectra \citep[e.g.][]{Ganc2012, Biagetti2013, Ota2015aniso_iso} and poly-spectra \citep[e.g.][]{Bartolo:2015fqz} that peak in squeezed configurations, with one of the momenta much smaller than the others. In this category falls the local model bispectrum, whose amplitude $f_{\rm NL}^{\rm loc}$ can discriminate between single and multi-field inflation. The measurement of $\mu$-$T$ cross correlation will set the first upper bound on $f_{\rm NL}^{\rm loc}$ on small scales ($k \approx \SI{740}{Mpc^{-1}}$), shedding light on possible scale-dependence of the NG parameters \cite{Biagetti2013, Emami:2015xqa}, thus complementing parametric searches performed on the vastly different CMB anisotropy scales \cite{Becker:2012je, Oppizzi:2017nfy}.

\begin{figure}
\begin{center}
\hspace{-3mm} 
\includegraphics[width=0.502\textwidth]{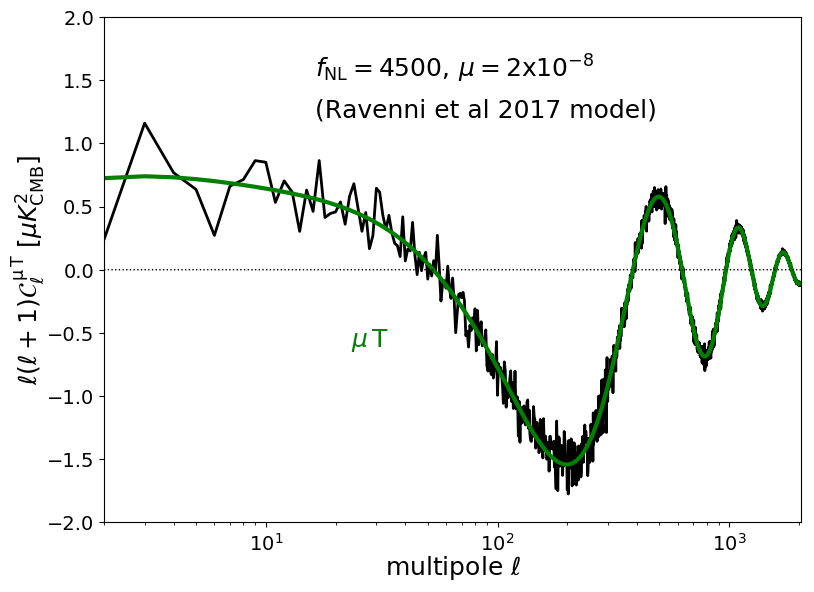}~
\includegraphics[width=0.488\textwidth]{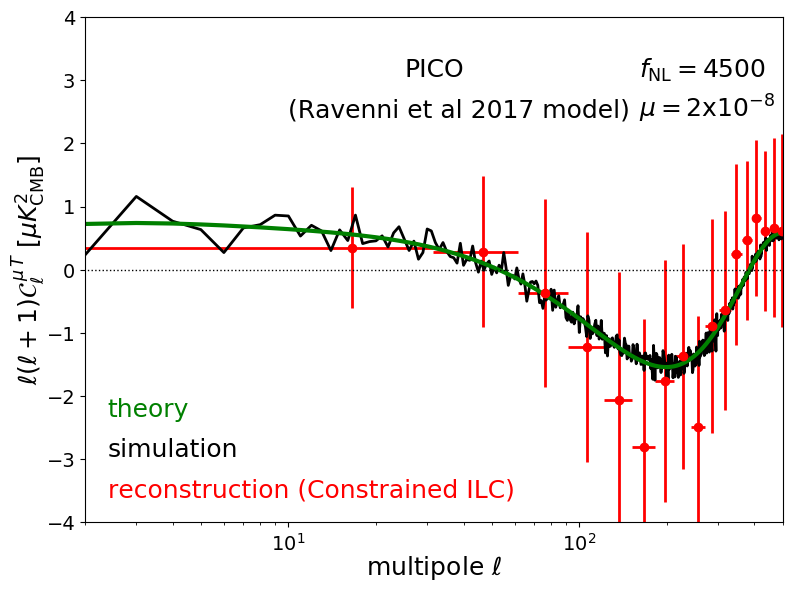}~
  \end{center}
\caption{\small \textit{Left}: Expected $\mu$-$T$ cross-power spectrum Eq.~(\ref{eq:ClXi-X_squeezed}) between CMB temperature and $\mu$-distortion anisotropies for $\langle \mu \rangle=2\times 10^{-8}$ and ${f_{\rm NL}^{\rm loc}(k\simeq 740\,{\rm Mpc}^{-1})=4500}$. \textit{Right}: Reconstructed $\mu$-$T$ correlation signal after foreground removal with the Constrained-ILC method for \PICO. Figures from \cite{Remazeilles2018:mu}.}
\label{fig:mu-T-forecasts}
\end{figure}

Spectral distortion anisotropies can be targeted by both CMB spectrometers or differential CMB imagers. 
The angular cross-correlation $\mu$-$T$ can be expressed as \cite{Ganc2012, Chluba2017}
\vspace{-1mm}
\begin{equation}
	C_\ell^{\mu T} 
	\approx
	- 12 \, f_{\rm NL}^{\rm loc} \, \langle\mu\rangle 
	\int \! {\rm d} k \,\frac{2}{\pi}\, k^2 \,
	\mathcal{T}_\ell^T \! (k)
	\,\frac{j_{\ell}(k \, r_{\rm ls})}{5}\,
	P(k) \, {\rm e}^{-\frac{15k^2}{8k_{{\rm rec}}^2}}
\label{eq:ClXi-X_squeezed}
\end{equation}
\vspace{-1mm}
where $\Braket{\mu}$ is the average dissipation $\mu$-distortion (e.g., Eq.~\ref{eq:mu-averaged}), $\mathcal{T}_\ell^T$ is the temperature transfer function, $r_{\rm ls}$ is the comoving distance to last scattering and $k_{{\rm rec}}$ is the diffusion-damping scale at the epoch of recombination (see Fig.~\ref{fig:mu-T-forecasts}; \textit{left}).
The exact degeneracy between $f_{\rm NL}^{\rm loc}$ and the spectral distortion monopole means that to interpret the data -- in principle measurable with a differential CMB imaging instrument -- also requires an absolute measurement \citep{Chluba2017}. A larger monopole would enhance the signal, and thus render NG signals more observable. While we limit our discussion to the correlation with temperature anisotropies, further improvements in sensitivity to $f_{\rm NL}^{\rm loc}$ can be achieved by considering $y$-$T$, $y$-$E$ and $\mu$-$E$ correlations \cite{Ota:2016mqd, Chluba2017, Ravenni2017}.

Figure~\ref{fig:mu-T-forecasts} (\textit{right}) shows the reconstruction of the $\mu$-$T$ correlation signal between CMB temperature and $\mu$-distortion anisotropies for the \PICO experiment \cite{PICO2019} after foreground mitigation and "deprojection" of residual CMB temperature anisotropies in the reconstructed $\mu$-map with the Constrained-ILC method \cite{Remazeilles2011a} (to eliminate spurious residual $TT$ correlations in the $\mu$-$T$ cross-power spectrum). For a \PICO-type space mission, the $\mu$-$T$ cross-power spectrum for $\langle \mu \rangle=2\times 10^{-8}$ and ${f_{\rm NL}^{\rm loc}(k\simeq 740\,{\rm Mpc}^{-1})=4500}$ is recovered without bias at large angular scales and detected at $2\sigma$ significance when including the recovered modes at $2\leq \ell \leq 500$.  This result is not biased by secondary and line-of-sight effects \cite{Cabass:2018jgj}. 
For $f^{\rm loc}_{\rm NL}(k_0) \simeq 5$ at CMB pivot scale, $k_0 = 0.05\,{\rm Mpc}^{-1}$, this would impose a limit of $n_{\rm NL}\lesssim 1.6$ on the spectral index of $f^{\rm loc}_{\rm NL}(k) \simeq f_{\rm NL}(k_0)(k/k_0)^{n_{\rm NL}-1}$ for scale-dependent non-Gaussianity, providing a new way to constrain non-standard early-universe models (e.g. multifield inflation). These limits complement those from \Planck and future experiments like the SKA and \SphereX, which could reach $f^{\rm loc}_{\rm NL}\simeq \mathcal{O}(1)$ \citep{Camera2015, Dore2014} at much larger scales.

As demonstrated in \cite{Remazeilles2018:mu}, coverage at frequencies below $40$\,GHz is more important for a detection of the enhanced $\mu$-$T$ correlation than high frequencies. This is because the $\mu$-distortion energy spectrum is more degenerate with the CMB temperature blackbody spectrum at high frequencies. Nevertheless, high frequencies are needed to clean dust foregrounds at large angular scales, necessitating broad spectral coverage ($\nu \simeq 20$-$800$\,GHz) for this science objective. In addition, it was shown in \cite{Remazeilles2018:mu} that extended spectral coverage at frequencies $\nu \lesssim 40$\,GHz and $\nu \gtrsim 400$\,GHz provides more leverage for constraining the $\mu$-$T$ cross-power spectrum than increased channel sensitivity over a narrower spectral range.  Finally, since most of the $\mu$-$T$ correlation is contained at large angular scales $2\leq \ell \leq 500$ (Fig.~\ref{fig:mu-T-forecasts}; \textit{left}), a space mission scanning the full sky with broad spectral coverage and moderate angular resolution ($\delta \theta \simeq 0.5^\circ$) is highly motivated. 

The magnitude of this constraint prima-facie is poor compared to the benchmark set by \Planck, and, e.g., to the prospect of measuring NG signatures in the galaxy bias \cite{Dalal:2007cu, Matarrese:2008nc}. However, such a comparison implicitly assumes the lack of any scale-dependence of the NG parameters. Albeit plausible, the absence of any running of $f^{\rm loc}_{\rm NL}$ over more than 4 orders of magnitude is \textit{per se} a valuable hint that can lead to further understanding of the underlying physics. 
Any measurement of this kind would also be a first step toward the invaluable goal of reaching cosmic-variance-limited determination of the $\mu$-$T$ cross correlation which is, to this day, the only proposed way to reach the lower bound set by the Maldacena consistency relation \cite{Maldacena:2002vr}. 

\begin{figure}
\vspace{-9.0mm}
\includegraphics[width=0.56\textwidth]{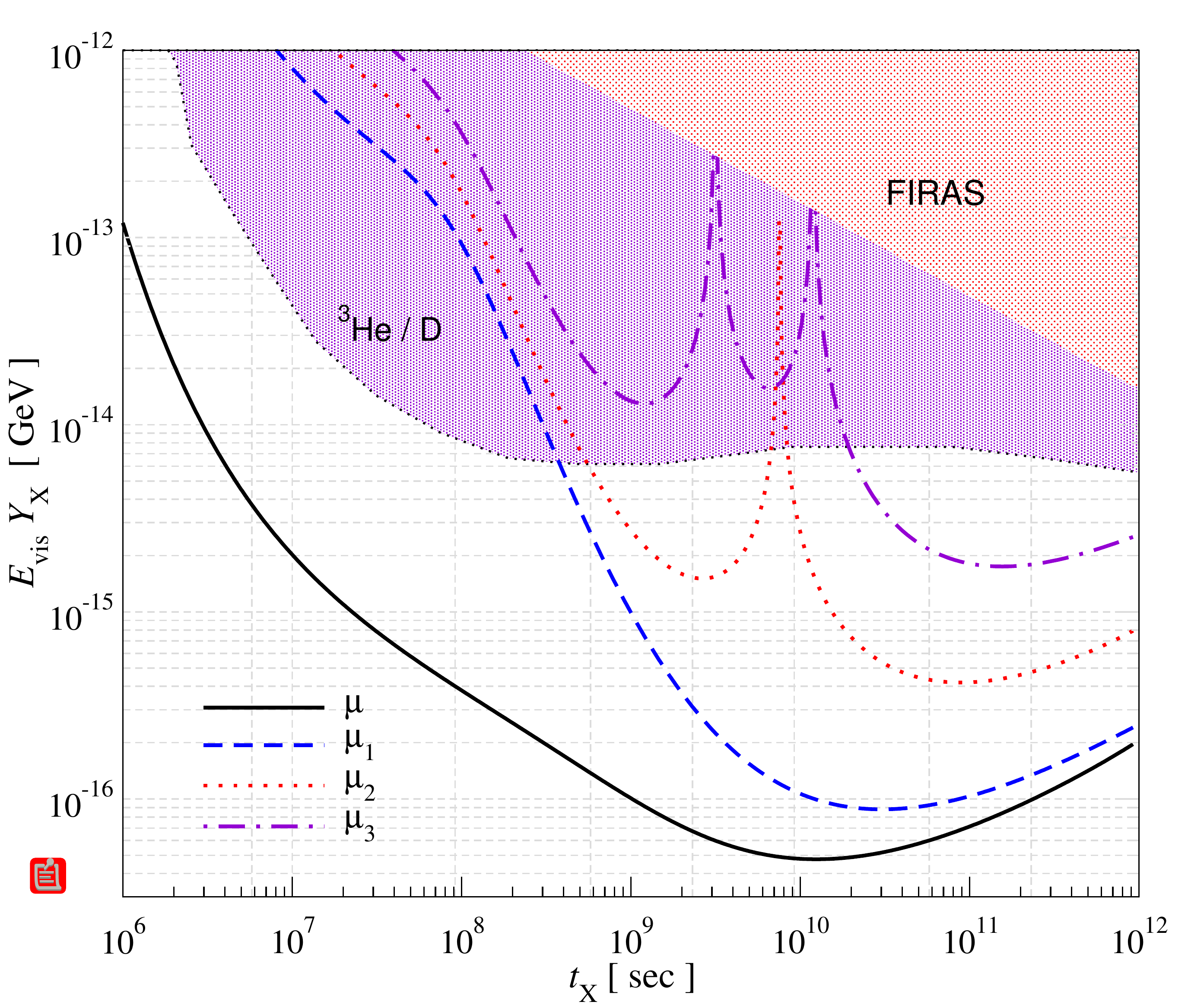}
\vspace{-2.0mm}
\caption{Constraints on the yield variable, $E_{\rm vis} Y_X$ \citep[e.g., see][for details]{Kawasaki2005}, from electromagnetic particle decay for varying lifetime, $t_X$. For the distortion forecast a spectral sensitivity of $\sigma(\mu)\simeq 10^{-8}$ (aka \SPIXIE) was assumed. The parameters $\mu_i$ describe extra time-dependent information available from the $r$-type distortion \citep[see][for details]{Chluba2013PCA}. For comparison we quote the constraints from \citep{Kawasaki2005, Kawasaki2018} for decays into $e^+ e^-$ derived from the $^3$He/D abundance ratio. Future spectral distortion measurements could improve the constraint on decaying particles with lifetimes $t_X\simeq 10^7-10^{12}$ by orders of magnitude. Using the $r$-type distortion we could furthermore break the degeneracy between particle yield and lifetime, should a significant distortion signal be detected \citep{Chluba2011therm, Chluba2013PCA}.  Figure adapted from \citep{Chluba2013PCA}}
\label{fig:decay_limits}
\end{figure}

\vspace{-0mm}
\subsection{CMB Spectral Distortions as a Probe of Dark Matter and Particle Physics}
\vspace{-3.5mm}
The search for dark matter is another example of how spectral distortions probe new physics. Non-baryonic matter constitutes $\simeq25$\% of the energy density of the Universe, but its nature remains unknown. The long-favored WIMP-scenario is under increasing pressure \citep{Ahmed:2009zw, Aprile:2012nq, Angloher:2015ewa, Agnese:2015nto, Tan:2016zwf, Akerib:2016vxi}, and emphasis is gradually shifting focus towards alternatives, prominent examples being axions, sterile neutrinos, sub-GeV DM or primordial black holes \citep{Jungman1996, Feng2003PhRvL, Feng2003, Kusenko2009, Feng2010, Carr2010, Marsh2016Rev}.

To solve this puzzle, a coordinated multi-tracer approach that combines different particle physics and cosmological probes is needed. Measurements of the CMB anisotropies themselves have clearly helped to establish the presence of DM on cosmological scales and provided tight constraints on DM annihilation and decay \citep{Ellis1992,Adams1998nr, Chen2004, Padmanabhan2005, Galli2009, Slatyer2009, Slatyer2017, Poulin2017} and DM-SM-interactions \citep{Wilkinson2014, Dvorkin2014, Wilkinson2014b, Gluscevic2018, Boddy2018}. However, for DM annihilation and decay CMB anisotropies quickly lose constraining power before recombination ($z\gtrsim 10^3$), being impeded by cosmic variance. 
Similarly, measurements of light-element abundances \citep{Ellis1992, Kawasaki2005, Jedamzik2008, Kawasaki2018}, which are only sensitive to non-thermal energy release above nuclear dissociation thresholds in the pre-recombination era \citep{Chluba2013PCA, Poulin2015Loop}, saturated their limits due to astrophysical uncertainties. {\it This is where CMB spectral distortions offer a valuable complementary probe.} 
For decaying particle scenarios, distortions are sensitive to particles with lifetimes $t_X\simeq 10^6-10^{12}\,{\rm s}$ 
\citep{Sarkar1984, Ellis1985, Kawasaki1986, Hu1993b, Chluba2011therm, Chluba2013PCA, Aalberts2018}, providing direct measurement of particle lifetimes via $r$-type distortions \citep{Chluba2013fore, Chluba2013PCA}. 
Existing limits from light-element abundances on the particle yield variable, which provides a measure of the relic abundance and mass of the particle \citep[e.g., see][for details]{Kawasaki2005}, could be improved by orders of magnitude (see Fig.~\ref{fig:decay_limits}).
Similarly, annihilating particles can be constrained using distortions: the $\mu$-distortion is sensitive to light particles ($m \lesssim 100$ keV) and complements $\gamma$-ray searches for heavier particles, being sensitive to s- and p-wave annihilation \citep{McDonald2001, Chluba2013fore}. The rich spectral information added by various non-thermal processes \citep{Liubarskii83, Chluba2008c, Chluba2010a, Chluba2015GreensII, Slatyer2015, Acharya2018} will allow us to glean even more information about the nature of dark matter, placing limits on the importance of different decay or annihilation channels. 

More work is required, although it is already clear that in addition to the aforementioned examples distortions can meaningfully probe scenarios involving axions \citep{Tashiro2013, Ejlli2013, Mukherjee2018}, gravitino decays \citep{Ellis1985, Dimastrogiovanni2015}, cosmic strings \citep{Ostriker1987, Tashiro2012b}, DM-SM-interactions \citep{Yacine2015DM, Diacoumis2017, Slatyer2017}, macros \citep{Kumar2018} and primordial magnetic fields \citep{Jedamzik2000, Sethi2005, Kunze2014, Wagstaff2015}. This opens a path for studying a wide range of new physics.

\vspace{-3mm}
\subsection{Primordial Black Holes}
\vspace{-3.5mm}
CMB spectral distortions can also place stringent limits on the abundance of primordial black holes (PBHs) \citep[e.g.,][]{Carr2010, Pani2013, Clesse2015PBH, Nakama2017xvq}. 
There is good motivation to study these scenarios, because PBHs with masses of $m_{\rm PBH} \simeq {\cal O}(10) M_{\odot}$ may explain the gravitational wave signals \cite{Bird2016,Sasaki:2016jop,Sasaki:2018dmp}
emitted in the merger events of (primordial) binary black holes reported by LIGO / Virgo~\cite{Abbott:2016blz}. 
PBHs with masses in the range $m_{\rm PBH} \simeq 10^{-17} M_{\odot} - 10^{-11} M_{\odot}$ \cite{Carr2010, Niikura:2017zjd} can furthermore still constitute $\simeq 100 \%$ of cold dark matter (see also \cite{Carr:2016drx,Juan1996PhRvD,Clesse2015PBH}). 
Lastly, PBHs with masses $m_{\rm PBH} \simeq 3\times 10^{3} M_{\odot} - 10^{5} M_{\odot}$ may
form the seeds for super-massive black holes (SMBHs) that grow to their current sizes merely by continuous (sub-)Eddington accretion, solving a long-standing problem in cosmology \cite{Kawasaki:2012kn,Kohri2014SMBH,Kawasaki:2019iis}. 

To produce PBHs, we expect a large primordial curvature perturbation $P_{\zeta} \gtrsim 0.03$ with a critical density perturbation $\delta_c \simeq  0.3 - 0.4$ \cite{Carr:1975qj,Harada:2013epa}) at small scales to have collapsed directly into a black hole during the radiation-dominated Universe (see also~\cite{Harada:2017fjm} for the
case of an early matter-dominated Universe). It is known that large curvature perturbations can indeed be produced by some classes of inflation \cite{Kohri:2007qn, Clesse2015PBH, Inomata:2017vxo, Lyth:2011kj}, curvaton \cite{Kawasaki:2012wr,Kohri:2012yw} and preheating scenarios \cite{Frampton:2010sw}, providing further motivation.

With this picture in mind, various ways to limit the abundance of PBHs have been proposed in cosmology and astrophysics \citep[e.g., see][and references therein]{Josan:2009qn,Carr2010}. Among these methods, the limits from observations of CMB in terms of spectral distortions and modifications to the recombination history of atoms are most robust. Here, we focus on the former effects (See also references about the latter effects from evaporations of PBHs~\cite{Carr2010, Poulin2017,Stocker:2018avm} and accretions onto PBHs~\cite{Ali-Haimoud:2016mbv,Poulin:2017bwe}).
There are two types of constraints on PBHs from the spectral distortions of CMB, each of which is individually induced by
1) dissipation of large density perturbation which is expected to collapse into PBHs and 2) electromagnetic particles emitted by evaporating PBHs. 
For case 1), it is notable that this effect occurs whenever PBHs are
formed by a Gaussian density perturbation. The current bound on the spectral 
distortions has already excluded PBH masses of $m_{\rm PBH} \gtrsim 3\times 10^{4}
M_{\odot}$~\cite{Kohri2014SMBH}. In addition, from limits set by Big Bang Nucleosynthesis (BBN) in order
not to dilute baryon due to the dissipation~\cite{Jeong2014, Nakama:2014vla}, we can exclude masses in the range
$3\times 10^{3} M_{\odot} \lesssim m_{\rm PBH} \lesssim 3\times 10^{4} M_{\odot}$~\cite{Inomata:2016uip}.
Overall, PBHs with masses $m_{\rm PBH} \gtrsim 3\times 10^{3} M_{\odot} $ can thus not make more than a faction $f_{\rm dm}\simeq 10^{-8}$ of the DM when considering the dissipation of Gaussian density perturbation. This already puts PBHs as seeds of SMBHs under strong pressure, and future spectral distortion experiments could further tighten these limits.
However, if large non-Gaussian curvature perturbation were created at the relevant small scales (not probed by CMB anisotropies), the above bound could become much milder \cite{Nakama:2017xvq}. In this case, {\it PBH clusters} would be expected \citep[e.g.,][]{Inman2019}, which could furthermore lead to anisotropic distortion signals. Future measurements of CMB spectral distortions could shed further light on these scenarios.

On the other hand, for case 2), existing limits from CMB spectral distortion~\cite{Tashiro:2008sf} are currently $\simeq 10^3$ times weaker than those from BBN~\cite[see Fig. 6 of][]{Carr2010}, which tightly constrain masses in the range $10^{9} {\rm g} \lesssim m_{\rm PBH} \lesssim 3\times 10^{13} {\rm g}$. 
However, in the future, the $\mu$-distortion constraints could be improved beyond the BBN limits, probing masses in the range $10^{10} {\rm g} \lesssim m_{\rm PBH} \lesssim 3\times 10^{12} {\rm g}$.

We close mentioning that CMB spectral distortions are also {\it indirectly} sensitive to PBHs with masses $m_{\rm PBH} \simeq {\cal O}(10) M_{\odot}$ \citep{Clesse2015PBH, Nakama:2017xvq}. Perturbations forming PBHs with $m_{\rm PBH} \simeq {\cal O}(10) M_{\odot}$ correspond to wavenumbers $k\simeq 10^6\,{\rm Mpc^{-1}}$, which is well outside the range $1\,{\rm Mpc^{-1}}\lesssim k \lesssim 10^4\,{\rm Mpc^{-1}}$ to which dissipation spectral distortions are directly sensitive. However, assuming a large enhancement of the primordial power spectrum at $k\simeq 10^6\,{\rm Mpc^{-1}}$ also means that the perturbations at $1\,{\rm Mpc^{-1}} \lesssim k\lesssim 10^6\,{\rm Mpc^{-1}}$ ought to be modified. Depending on the mechanism creating the large perturbation at $k\simeq 10^6\,{\rm Mpc^{-1}}$, the transition from the low power at large angular scales is more or less rapid and thus can be probed using future CMB spectral distortion measurements \citep{Chluba2012inflaton}.

\vspace{-5mm}
\subsection{Axion-Like Particles}
\vspace{-3.5mm}
Axions or Axion-Like Particles (ALPs) are predicted in multiple particle physics scenarios \cite{PhysRevLett.38.1440, PhysRevLett.40.223, PhysRevLett.40.279, Svrcek:2006yi, Arvanitaki:2009fg, Acharya:2010zx}, and their discovery would mark a paradigm shift in the framework of the standard models of cosmology and particle physics. Several particle physics experiments \cite{2015ARNPS..65..485G} such as CAST \cite{2015PhPro..61..153R}, ALPS-II \cite{Bastidon:2015efa}, MADMAX \cite{Majorovits:2016yvk}, ADMX \cite{2010PhRvL.104d1301A}, CASPER \cite{Budker:2013hfa} are looking for the signatures of axions or ALPs over a wide range of masses. Along with the particle physics experiments, cosmological probes such as CMB anisotropies and large-scale structure are exploring the gravitational effects of ALPs on the matter density with the potential to discover ALPs if it constitutes dark matter \cite{Hu:2000ke, Marsh2016Rev, PhysRevD.86.083535, Hlozek:2014lca, Hlozek:2016lzm, Hlozek:2017zzf}. The other possibility to probe ALPs (even if they are fraction of DM) is by studying their coupling with photons in the presence of an external magnetic field \cite{PhysRevLett.51.1415, PhysRevD.37.1237, PhysRevD.37.2001}. 

The coupling between ALPs and photons $g_{\gamma\gamma a}$ leads to oscillations between photons and ALPs and vice versa in the presence of an external magnetic field. This effect is one of the cleanest windows for detecting ALPs. The signatures of this non-gravitational interaction of ALPs with photons distort their energy spectrum and thus can be detected robustly if the energy spectrum of the source is well-known. The radiation field of CMB provides us with an excellent source which can be used to detect the distortions due to ALPs \cite{ Mukherjee2018}. The  ALPs distortion ($\alpha$-distortion) is imprinted on the CMB while it is passing through the external magnetic field of the intergalactic medium (IGM), inter-cluster medium, voids and Milky Way. The conversion from photons to ALPs can be classified into two types, namely the resonant conversion and the non-resonant conversion. 

\begin{figure}
\vspace{-15.0mm}
\includegraphics[width=0.69\textwidth]{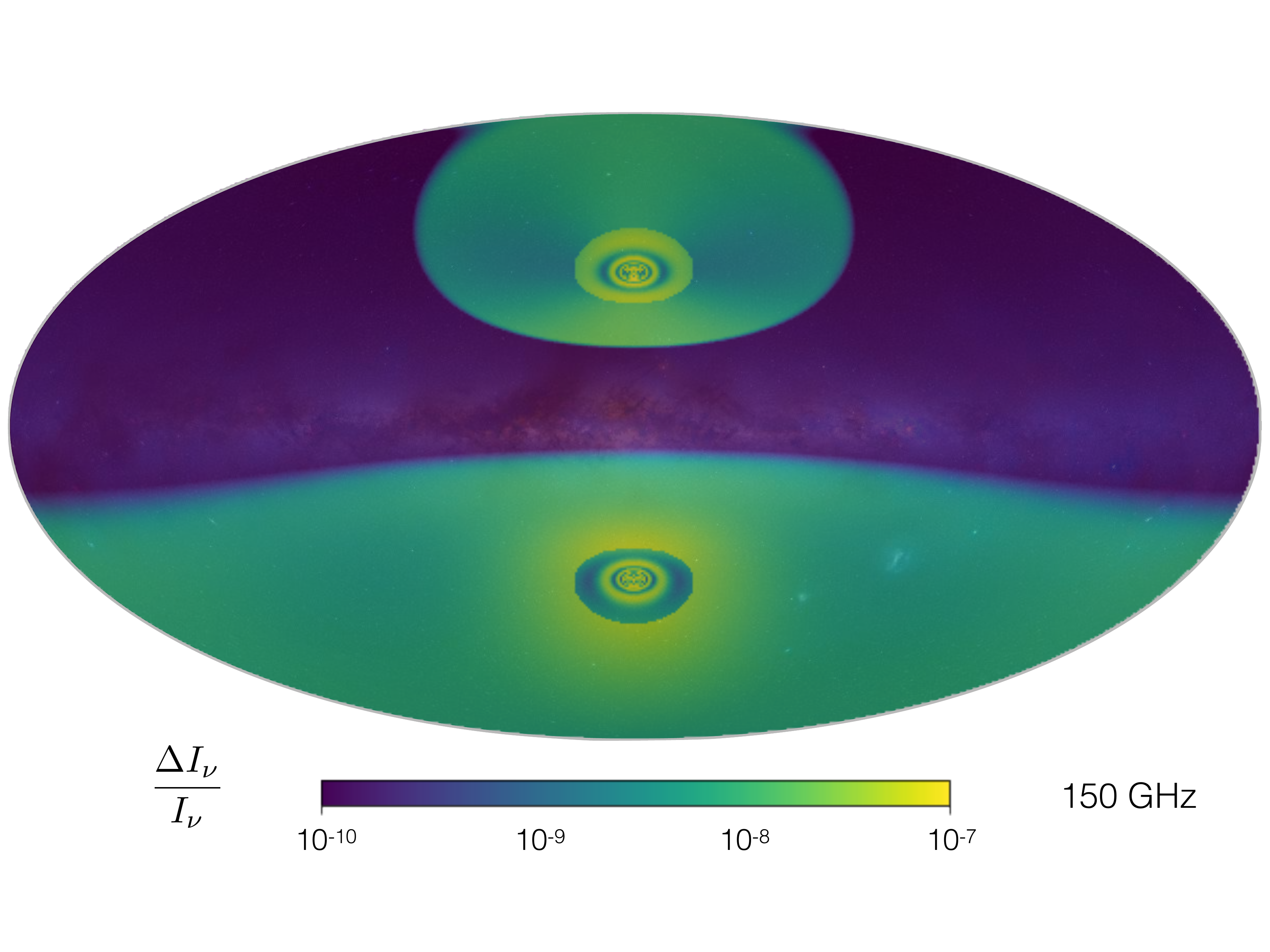}
\vspace{-7.0mm}
\caption{The resonant conversion between CMB photons into ALPs in the presence of galactic magnetic field produces spatially varying spectral distortion. We plot the expected spectral distortion signal $\Delta I_\nu/I_\nu$ at $150$ GHz for ALP of mass $m_a= 5 \times 10^{-13}$ eV and photon-axion coupling $g_{\gamma \gamma a}= 10^{-11}$ GeV$^{-1}$ using the best-fit model of galactic electron density \cite{Cordes:2002wz} and magnetic field \cite{2012ApJ...757...14J, 2012ApJ...761L..11J}. The intensity and the shape of the distortion varies with the mass of ALPs $m_a$ and the coupling $g_{\gamma \gamma a}$ and is most pronounced at large angular scales. An optical image of our galaxy was overlaid for reference. The north-south asymmetry of the signal stems from the structure of the galactic magnetic field and electron density.}
\vspace{-2mm}
\label{fig:Axion_illustration}
\end{figure}

The resonant conversion of CMB photons into ALPs takes place when the photon mass in the plasma equals the mass of ALP. The polarization state of the CMB photon which is parallel to the external magnetic field gets converted into ALPs depending upon the strength of the magnetic field. As a result, it leads to a polarized spectral distortion of the CMB blackbody with a unique spectral shape. Also due to inhomogeneities in the magnetic field of the astrophysical systems, the observed polarized distortion varies spatially which leads to a unique spatial structure that differs from any other known spectral distortions and foreground contaminations. Though the resonant conversion of CMB photons can take place in different kinds of astrophysical systems it can be best measured in Milky Way and galaxy clusters. 

The Milky way's galactic magnetic field induces a large angular scale spectral distortion as shown in Fig.~\ref{fig:Axion_illustration}. This signal can be targeted with a low-resolution spectrometer like \PIXIE or \SPIXIE.
While polarization information increases the sensitivity, even intensity distortion measurements can be used to derive stringent constraints.
The shape of the ALPs distortion depends upon the mass of the axions and the density of electrons in the Milky Way. For the best-fit model of electron density \cite{Cordes:2002wz} and magnetic field \cite{2012ApJ...757...14J, 2012ApJ...761L..11J} of the Milky Way, ALPs in the mass range from a $m_{\rm ALP}\simeq$ few $\times 10^{-13}$ eV to a few $\times 10^{-12}$ eV can be probed by the process of resonant conversion. The measurement of this large angular scale spectral distortion signal requires both wide frequency- and sky-coverage, which is possible only with space-based CMB missions.  
The same physical effect also arises in galaxy clusters \citep{Mukherjee:2019dsu} and produces polarized spectral distortions that can be measured using high-resolution CMB experiments with an imaging telescope \citep{Basu2019WP, Delabrouille2019WP}.
 
Along with the resonant conversion of CMB photons into ALPs, there will also be a non-resonant conversion of CMB photons into ALPs, as the CMB photons propagate through the turbulent magnetic field of our galaxy, IGM and voids \cite{Mukherjee2018}. This leads to an unpolarized spectral distortion of the CMB blackbody. This avenue will provide stringent constraints on the coupling strength $g_{\gamma\gamma a}$ for all the masses of ALPs below $\simeq 10^{-11}$ eV. The first constraint of this kind of distortion is obtained from the data of \Planck satellite \cite{Mukherjee:2018zzg}.  

This new probe of ALP physics will be accessible with a CMB spectrometers like \PIXIE or \SPIXIE. In this way, we can explore a new parameter space of the coupling strength $g_{\gamma \gamma a}$ and ALP masses, which are currently beyond the reach of particle-physics experiments. Spectral distortions are capable of discovering ALPs even if they are a fraction of DM and hence will open a completely new complementary window for studying ALPs in nature. The discovery space is enormous and provides a direct cosmological probe into the string axiverse \cite{Arvanitaki:2009fg}.

\begin{figure}
\vspace{-8.0mm}
\includegraphics[width=0.69\textwidth]{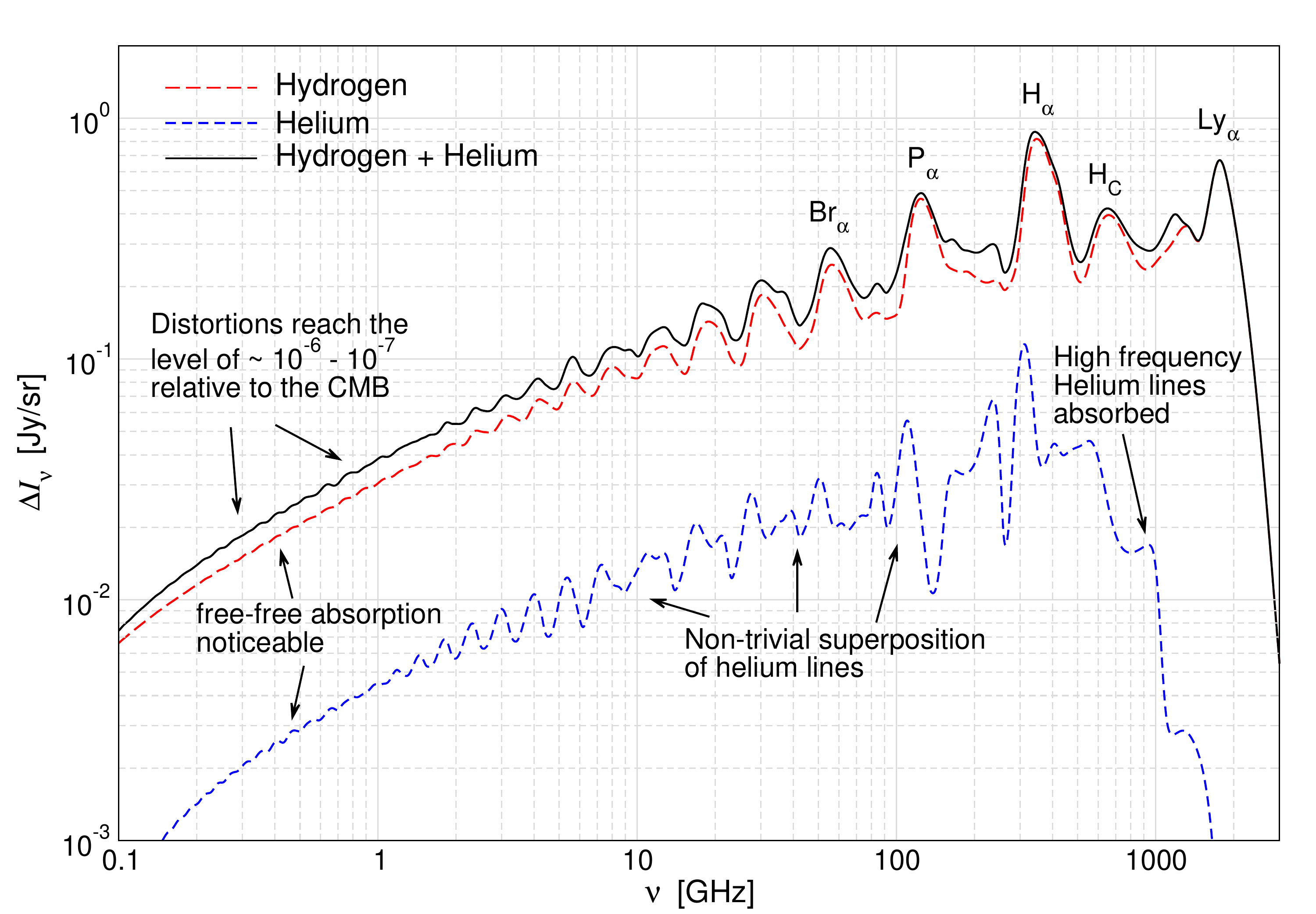}
\vspace{-2.0mm}
\caption{The cosmological recombination radiation arising from the hydrogen and helium components. The helium distortion {\it(blue)} is the net accumulation of the HeI and HeII emission along with other effects (i.e. helium absorption and collisions). The spectral series of hydrogen have also been highlighted. The unique spectral shapes given by the CRR would provide us with a revolutionary new cosmological probe of the atomic physics in the early Universe.}
\vspace{-1.0mm}
\label{fig:CRR}
\end{figure}

\vspace{-5mm}
\subsection{The Cosmological Recombination Radiation} 
\vspace{-3.5mm}
The cosmological recombination process causes another small but {\it inevitable distortion} of the CMB. Line emission from hydrogen and helium injects photons into the CMB, which after redshifting from $z\simeq 10^3$ are visible today as complex frequency structure in the microwave bands (Fig.~\ref{fig:CRR}) \citep{Dubrovich1975, RybickiDell94, DubroVlad95, Kholu2005, Wong2006, Jose2006, Chluba2006, Jose2008, Yacine2013RecSpec}.
The cosmological recombination radiation (CRR) has a simple dependence on cosmological parameters and the dynamics of recombination; since it includes not only hydrogen but also two helium recombinations, it probes eras well beyond the last-scattering surface observed by CMB anisotropies \citep{Chluba2008T0, Sunyaev2009, Chluba2016CosmoSpec}. Modern computations now include the bound-bound and free-bound contributions from hydrogen, neutral helium and hydrogenic helium and thus allow precise modeling of the total signal and its parameter dependences \citep{Chluba2016CosmoSpec}.

Cosmological recombination process is crucial for understanding and interpreting the CMB temperature and polarization anisotropies \citep{Zeldovich68, Peebles68, Sunyaev1970, Peebles1970}. It is thus critical to test our physical assumptions during this era \citep{Hu1995, Lewis2006, Jose2010, Shaw2011}. {\it The} CRR {\it provides one of the most direct methods to achieve this.}
It should enable a pristine measurement of the primordial helium abundance, long before the first stars have formed. On the other hand, it is fairly insensitive to the effective number of neutrino species and thus can help breaking the degeneracy with the primordial helium abundance \citep{Chluba2016CosmoSpec}. 

The CRR is also directly sensitive to new physics affecting the recombination dynamics. Decaying or annihilating particles could enhance the total emission caused by the primordial atoms \citep{Chluba2010}, leaving features that may help determining the time-dependence of the process through uncompensated atomic transitions in the pre-recombination era \citep[e.g.,][]{Chluba2008c}. The contributions from both helium recombinations furthermore arise from significantly earlier phases (redshifts $z\simeq 2000$ and $6000$, respectively; cf., Fig.~\ref{fig:stages}). This opens a new window to the primordial Universe that cannot be directly accessed in another way. 
Measurements of the CRR will also allow us to directly map the baryon density and other cosmological parameters at $z\simeq 10^3$ across the sky, providing an independent test of statistical isotropy \citep{Chluba2016CosmoSpec}. And finally, variations of fundamental constants should leave observable imprints in the shape of the CRR at otherwise inaccessible epochs \citep{Chluba2016CosmoSpec}.

Although the CRR is one of the smallest signals expected in standard $\Lambda$CDM (see Fig.~\ref{fig:future}), its detection is {\it on par} with the larger $\mu$-distortion from acoustic damping (see Sect.~\Secfgchall). This is because the CRR and its derivatives with respect to the cosmological parameters have many spectral features \citep[e.g., Fig.~\ref{fig:CRR} here and also Fig.~5 and 6 of][]{Chluba2016CosmoSpec}, making it easier to distinguish the signals from the much brighter but smoother foregrounds \citep{Vince2015, Mayuri2015}. 
For this reason, at very low frequencies ($\nu \simeq$ few~$\times$~GHz) the CRR could in principle be targeted from the ground using concepts like APSERa \citep{Mayuri2015}, however, to overcome atmospheric noise and access the more structured signal at high frequencies ($\nu \gtrsim 100$~GHz) a space mission will be required \citep{Vince2015}.
\SPIXIE could detect the distortion at the level of $\simeq 2\sigma$ (Sect.~\Secfgchall), opening a way to directly test our physical understanding of the $z=10^3$ Universe.

\vspace{-5mm}
\subsection{Reionization and Structure Formation}
\vspace{-3.5mm}
The epoch of reionization and the formation of cosmic structures mark additional important transitions in the evolution of our Universe. 
The largest all-sky spectral distortion signal is indeed caused by the reionization and structure-formation processes \citep{Sunyaev1972b, Hu1994pert, Cen1999, Refregier2000, Miniati2000, Oh2003}. Energy output from the first stars, accreting black holes, and gravitational shocks heats the baryons and electrons, which then up-scatter CMB photons to create a $y$-type distortion. The overall distortion is expected to reach $y\simeq$~few~$\times10^{-6}$ \citep{Refregier2000, Miniati2000, Khatri2015y, Hill2015}, about one order of magnitude below the current upper bound placed by \COBEF. Such a distortion {\it must} exist and provides a measure of the total thermal energy in (ionized) baryons in the Universe. Spectrometers like \PIXIE or \SPIXIE will measure this distortion to sub-percent precision (Sect.~\Secfgchall, Fig.~\ref{fig:future_errors}).
The low-redshift $y$-distortions from reionization and structure formation are furthermore anisotropic \citep[e.g.,][]{Refregier2000, Zhang2004, Pitrou2010, Alvarez2016} and thus opens new opportunities for cross-correlation studies (e.g., with CMB and 21 cm tomography).

\begin{figure}
\vspace{-6mm}
\includegraphics[width=0.68\columnwidth]{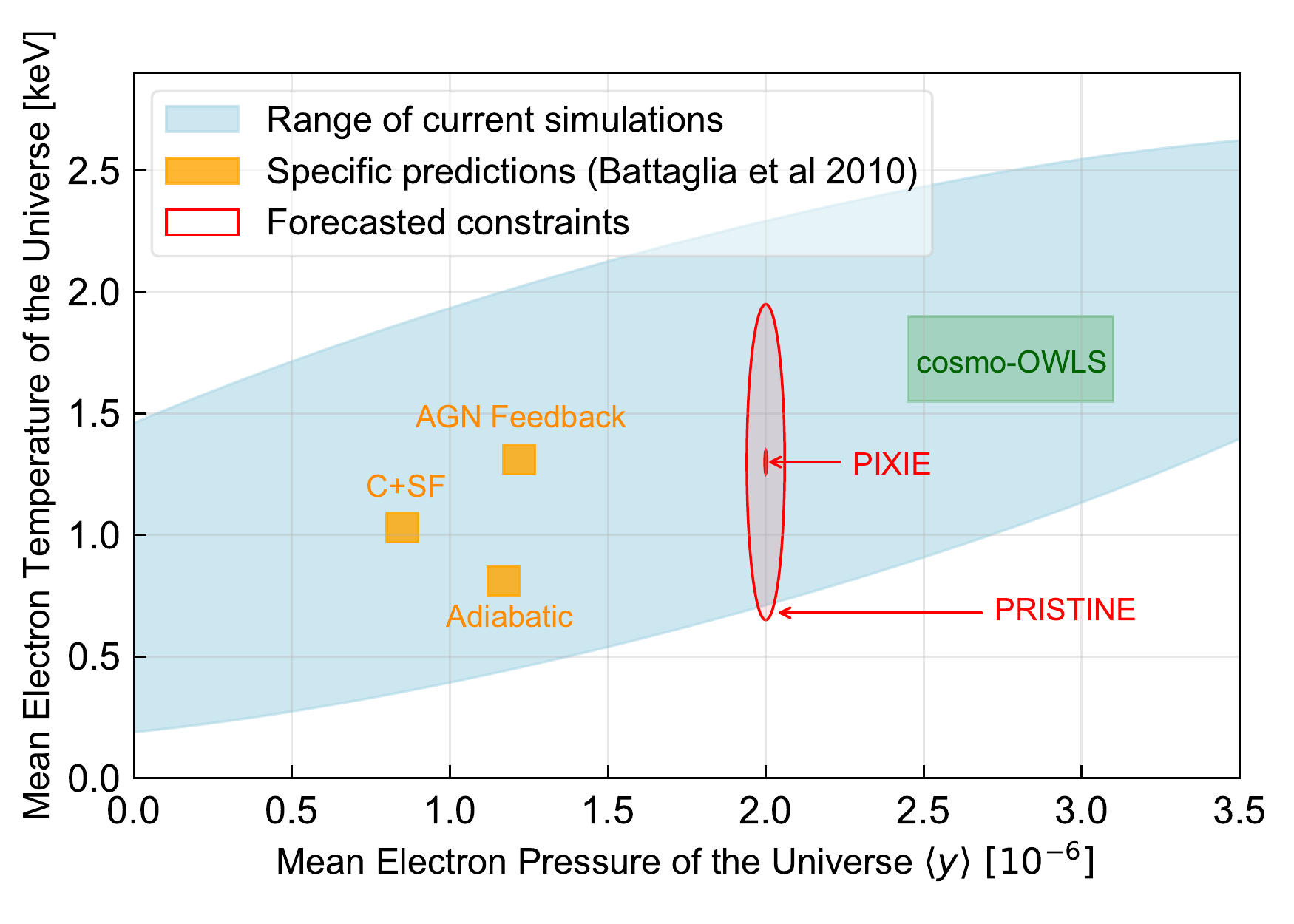}
   \vspace{-2mm}
   \caption{Theoretical predictions and forecasted constraints for the late-time Compton-$y$ and relativistic SZ spectral distortions due to structure formation and reionization, with $y=\pot{2}{-6}$ and $k\Te=1.3\,{\rm keV}$ as fiducial values \citep{Hill2015}. 
   The light blue ellipse encompasses the approximate range of several current predictions for these quantities.  Each of the labeled squares denotes a specific prediction from the simulations of~\cite{Battaglia2010}, where only the sub-grid feedback model is varied. The green rectangle indicates the range of results for the cosmo-OWLS simulations~\cite{LeBrun2014}. The red ellipses show the forecasted constraints on these quantities for \PRISTINE and \PIXIE (is hardly visible for \PIXIE); more powerful missions in the Voyage 2050 program would provide even tighter constraints. 
   }
   \label{fig:feedback}
\end{figure}

A large part of the low-redshift Compton-$y$ signal is due to halos with masses $M\simeq 10^{13}\,M_\odot$, which contain virialized gas with an electron temperature of $k T_{\rm e}\simeq 2-3$~keV. This causes a relativistic temperature correction (rSZ) \citep{Wright1979, Rephaeli1995, Sazonov1998, Itoh98, Challinor1998, ChlubaSZpack} to the $y$-distortion that can directly tell us about feedback mechanisms \citep{Hill2015}. In addition, both $y$ and the rSZ distortion depend directly on the shape and amplitude of the halo mass function, providing another cosmological measure of the growth of structure.
The level of the relativistic contributions is, however, uncertain and current estimates based in X-ray temperature-mass relations may underestimate the signal by a factor of $\simeq 1.5-2$ \citep{Kay2008, Lee2019}, thus further increasing the detectability of this signal. Nevertheless, with spectrometers like \PIXIE or \SPIXIE, the average relativistic temperature could be determined to tens of standard deviations (Sect.~\Secfgchall). 
With simultaneous high-precision measurements of both $y$ and rSZ, we will be able to place tight constraints on models of feedback in galaxy formation. This is illustrated in Fig.~\ref{fig:feedback}, which shows the range of current predictions for these quantities from state-of-the-art cosmological hydrodynamics simulations~\cite{Battaglia2010,LeBrun2014}, including precise predictions from different feedback implementations. The figure also shows forecasted constraints for \PRISTINE and \PIXIE, as illustrative spectral distortion missions.  It is clear that such measurements will strongly distinguish between current sub-grid feedback models, yielding significant breakthroughs in our understanding of galaxy formation.
A direct measurement of the average rSZ temperature would also shed new light on the 'missing baryon problem' \citep{Cen1999} without the need to resolve the warm-hot-intergalactic medium, a unique opportunity that we should make use of in the future.

The late-time $y$-distortion has an additional contribution at the level of $y\simeq \pot{{\rm few}}{-8}$ due to second-order Doppler terms from the large-scale velocity field \citep{Zeldovich1972, Hu1994pert}. This signal and the average distortion from the reionized $\simeq 10^4$~K gas could be accessed by masking resolved SZ clusters, or by isolating the latter signal through cross-correlations with galaxy and cluster catalogs. This procedure also reduces one of the largest primordial distortion foregrounds, the low-redshift $y$-distortion itself, and would therefore allow us to tighten the upper limits on early energy release occurring at $z\simeq 10^3-10^4$, a unique opportunity for combining CMB spectroscopy and imaging.
Measurements at $\nu\gtrsim 500$~GHz will furthermore probe the total cosmic ray energy density of the Universe through the non-thermal relativistic SZ effect \citep{Ensslin2000, Shimon2002, Colafrancesco2003, Colafrancesco06}.

\vspace{-3mm}
\subsection{Line Intensity Mapping}
\vspace{-3.5mm}
The measurement of the integrated Far-IR background \citep{1998ApJ...508..123F} was a significant legacy of the \COBEF mission. The amplitude of the Far-IR background suggests that half of the starlight in the Universe is absorbed and reprocessed through thermal dust emission. Similarly to the other spectral distortions, the extragalactic background light provides a synoptic view of energetic processes in all galaxies. The \COBEF measurement of integrated dust emission became a reference point for two decades of fruitful observations to resolve the sources of this emission into individual galaxies. The continuum radiation spectrum has no identifiers for the redshift of its emission, but cross-correlation with a galaxy redshift survey permits some dissection of the emission into its constituent redshifts \citep{2014A&A...570A..98S}. Future spectral surveys will be able to measure not only the dust continuum but also the integral of diffuse line radiation (namely CO ladder, [CII] and [NII]), which maps directly to redshift. This approach of Line Intensity Mapping has attracted significant attention in recent years as a probe for both galaxy evolution and fundamental cosmology \citep{2017arXiv170909066K, 2019BAAS...51c.101K}. Line emission traces cold, molecular gas (a precursor to star formation) and line emission excited by star formation \citep{2013ARA&A..51..105C}. \COBEF has insufficient sensitivity to extract this emission, and searches in the Planck data have hit fundamental limits \citep{2018MNRAS.478.1911P, 2019arXiv190401180Y} at a $3\sigma$ excess consistent with [CII] emission. New instruments are needed to constrain this signal~~\citep{2019MNRAS.tmp.1857P}.

A space-borne FTS is a unique platform for intensity mapping. It provides 1) access to the monopole of line emission \citep{2016MNRAS.458L..99M, 2016ApJ...833..153S}, 2) access to the largest modes of anisotropy on the sky, and 3) a highly precise passband calibration through differencing with a blackbody reference. Cross-correlation with a galaxy redshift survey allows the line signal to be extracted unambiguously from uncorrelated Milky Way foregrounds and may ultimately mitigate cosmic variance \citep{2017ApJ...838...82S}. This cross-correlation measures not only the line brightness, but also the SED of average galaxies as a function of frequency and time. \PIXIE and \SPIXIE will have a sufficient number of spectral channels to separate the correlated line and continuum emission \citep{2019ApJ...872...82S}. In each of its frequency channels, \PIXIE is expected to make a high significance detection of [CII] emission, from the present to $z{\sim}2$ \citep{2017ApJ...838...82S}, and detection of the CO $J$ ladder at $z<1$ (depending on poorly constrained emission models). Access to large volumes also permits probes of fundamental physics through searches for primordial non-Gaussianity. \citet{2019ApJ...872..126M} find that a 4-year \PIXIE survey could constrain $\sigma(f_{\rm NL}^{\rm loc}) = 2.1$, which is comparable to future goals of SKA and \SphereX \citep{Camera2015, Dore2014} and complements limits obtained with $\mu$-anisotropies (Sect.~\SecmuFNL). 
With $10\times$ the capability, \SPIXIE should reveal redshifted line emission monopole and anisotropy in all frequency channels. Much like for \COBEF, observations of integrated, redshifted line emission will provide a complement to efforts to catalog line emission from individual sources~\citep{2019arXiv190309164D}. 

\vspace{-3mm}
\subsection{Resonant Scattering Signals}
\vspace{-4mm}
Interactions of CMB photons with atoms can imprint additional frequency-dependent signals through resonant \citep{Loeb2001, Zaldarriaga2002, Kaustuv2004, Jose2005, Carlos2005, Carlos2006, Carlos2007, Carlos2017} and Rayleigh scattering effects \citep{Yu2001, Lewis2013}, or via collisional emission processes \citep{Righi2008b, Hernandezetal2007b, Gongetal2012, Carlos2017}, providing independent ways for learning about recombination, the dark ages and reionization. 
A detection of these frequency-dependent signals, even at large angular scales, is limited in general by sensitivity, foregrounds and especially by the accuracy of the inter-calibration between channels. The required level of precision will be naturally achieved by the proposed mission concepts for spectral distortions discussed here. 

Importantly, some of the signals can be detected using a spectrometer with moderate angular resolution ($\ell \lesssim 300$). For example, the resonant scattering of CMB photons in the H$\alpha$ line during cosmological recombination \citep{Jose2005,Carlos2007} is detectable with \PIXIE or \SPIXIE, providing a crucial demonstration of the methodology, which can be used for other lines (e.g. P$\alpha$) and novel polarization signals (i.e., $TE$ and $EE$ from H$\alpha$) with a \PRISM-like mission that hosts both a CMB spectrometer and high-resolution imager \citep{PRISM2013WPII}. The resonant scattering of CMB photons by the fine-structure lines of metals and heavy ions (i.e., OI, OIII, NII, NIII, CI, CII) produced by the first stars can also be observed at angular scales around the first Doppler peak. This effect causes a blurring of the original CMB anisotropy field on intermediate angular scales given by $\delta C_\ell\simeq -2\tau_{\rm X} C_\ell^{\rm CMB}$ \citep{Kaustuv2004,Carlos2006}, where $\tau_{\rm X}$ denotes the optical depth associated to a given transition ${\rm X}$ and $C_\ell^{\rm CMB}$ is the primordial CMB anisotropy angular power spectrum generated at $z\simeq 1100$. Typical frequencies involved in a few of the most relevant lines are $\nu_{\rm obs}\simeq 190\times 10/(1+z)$~GHz, $475\times 10/(1+z)$~GHz, and $206\times 10/(1+z)$~GHz for the [CII]~157.7~$\mu$m, [OI]~63~$\mu$m, and [OI]~145~$\mu$m transitions, respectively. Here, $z$ denotes the resonant scattering redshift. An FTS can provide relative calibration between different frequency channels at the level of few nK for $\nu<600$~GHz, thus enabling a sensitivity to values of $\tau_{\rm X}$ as low as $\simeq 10^{-5}$. As shown in Fig.~8 of \citep{Basu2019WP}, this level of inter-channel calibration uncertainty can shed light on the history of the metal pollution of the IGM during the end of the dark ages and the reonization epoch, thus constituting an alternative window to those cosmological times that is totally complementary to HI 21~cm observations.

Finally, the UV radiation field generated by stars at the end of the Dark Ages and during the reionization epoch influences the spin temperature associated with fine-structure transitions like [OI]~63\,$\mu$m, [OI]~145\,$\mu$m, and [CII]~157.7\,$\mu$m \citep{Hernandezetal2007b}. Through the Field-Wouthuysen effect, these transitions may be seen in absorption/emission against the CMB backlight, and thus will generate another type of distortion to the CMB blackbody spectrum at the $10^{-10}$--$10^{-7}$ level that is only reachable with a spectrometer in the ESA Voyage 2050 Program.

\vspace{-4mm}
\section{The path forward with CMB spectral distortions}
\label{sec:observ}
\vspace{-3mm}

\subsection{Technological challenges}
\vspace{-3.5mm}
The seminal measurements of the CMB blackbody spectrum by \COBEF in the early '90s cemented the Hot Big Bang model by ruling out any energy release greater than $\Delta \rho_\gamma / \rho_\gamma~\simeq~6~\times~10^{-5}$ (95\% c.l.) of the energy in CMB photons \citep{Mather1994, Fixsen1996, Fixsen2009}. Advances since then, in both detector technology and cryogenics, could improve this constraint by four orders of magnitude or more (e.g.,~with experimental concepts like \PIXIE \citep{Kogut2011PIXIE, Kogut2016SPIE}, \PRISM \citep{PRISM2013WPII}, \PRISTINE \citep{Nabila2018} or \SPIXIE \citep{Kogut2019WP}), opening an enormous discovery space for both predicted distortion signals and those caused by new physics. On the timescales relevant to the Voyage 2050 program we expect to go beyond, surpassing the crucial threshold of detecting the dissipation $\mu$-distortion at more than $3\sigma$.

\COBEF was not background limited; its sensitivity was set instead by phonon noise from the 1.4 K detector. Modern detectors, operating at $\simeq$~0.1~K, would have detector (dark) noise well below the intrinsic limit set by photon arrival statistics. The sensitivity of a background-limited instrument could be further improved by increasing its throughput or the integration time and, in a less trivial way, by modifying the mirror stroke (i.e., frequency-sampling) and reducing the optical load at high frequencies \citep{Kogut2019WP}. Combining replicas of the same telescope design can additionally enhance its capabilities. Modern blackbody calibrators now also reach sufficient thermal and spectral stability for the task. 
All these technological challenges can be overcome and it is possible to reach the required spectral sensitivities and coverage using FTS approaches that build on the legacy of \COBEF \citep{Kogut2011PIXIE, Kogut2016SPIE, PRISM2013WPII, Nabila2018, Kogut2019WP}.

As a point of comparison, it is worth noting that the raw instrumental sensitivities under consideration for these spectral distortion measurements are in the same range as recently proposed CMB imaging missions for the 2030s.  For example, the NASA-proposed Probe of Inflation and Cosmic Origins (PICO) aims to achieve an overall map-level sensitivity of $\approx 0.5$--$1$ $\mu$K-arcmin, after combining all of its 21 frequency channels (considering noise only, i.e., no foregrounds)~\cite{PICO2019}.  Averaging over the full sky, this corresponds to a monopole sensitivity of $\approx 0.02$--$0.04$ Jy/sr at a reference frequency of 150 GHz.  This is even below the range considered for an ESA Voyage 2050 spectral distortion mission concepts shown in Fig.~\ref{fig:future}.  Of course, more channels and absolute calibration are needed for the spectral distortion measurements in comparison to an imager, but nevertheless it is clear that the relevant raw sensitivities are entirely feasible \citep{Kogut2011PIXIE, Kogut2016SPIE, PRISM2013WPII, Nabila2018, Kogut2019WP}.

The designs of previously considered spectrometer concepts have evolved significantly in the past few years due to our improved understanding of the foreground challenge (Sect.~\Secfgchall). \PIXIE was proposed as a NASA mid-Ex mission (duration 4 years, resolution $\delta \theta\simeq 1.5^{\circ}$), while \PRISTINE was put forward as an F-class mission to ESA (duration 2 years, resolution $\delta \theta\simeq 0.75^{\circ}$). The \SPIXIE concept was described for the recent NASA Decadal Survey 2020 (duration 4-10 years, resolution $\delta \theta\simeq 0.5^{\circ}-2^\circ$ depending on the band) \citep{Kogut2019WP}. 
All these concepts used polarization-sensitive, absolutely-calibrated FTSs, with hundreds of spectral channels covering $\nu\simeq 10\,{\rm GHz}-{\rm few}\times {\rm THz}$.
The estimated sensitivity curves and channel distributions are illustrated in Fig.~\ref{fig:future} relative to the foreground and spectral signals. For \SPIXIE, the frequency range was split into three separate bands (each with many channels), to improve foreground mitigation. The anticipated spectral distortion constraints are summarized in Fig.~\ref{fig:future_errors}. \SPIXIE can reach $\sigma(y)\simeq \pot{1.6}{-9}$, $\sigma(k\Te) \simeq 0.02\,{\rm keV}$, $\sigma(\mu)\simeq \pot{7.7}{-9}$ and $\sigma(\Delta T)=10\,{\rm nK}$ in eight years of observation even after foreground marginalization (Sect.~\Secfgchall), and thus crosses the critical thresholds for detecting the dissipation $\mu$-distortion and CRR, embracing all expected distortions in the CSM.

\vspace{-3mm}
\subsection{Foreground Challenge for CMB Spectral Distortion}
\label{sec:foreground-challenge}
\begin{figure}
\vspace{-4mm}
   \includegraphics[width=0.8\columnwidth]{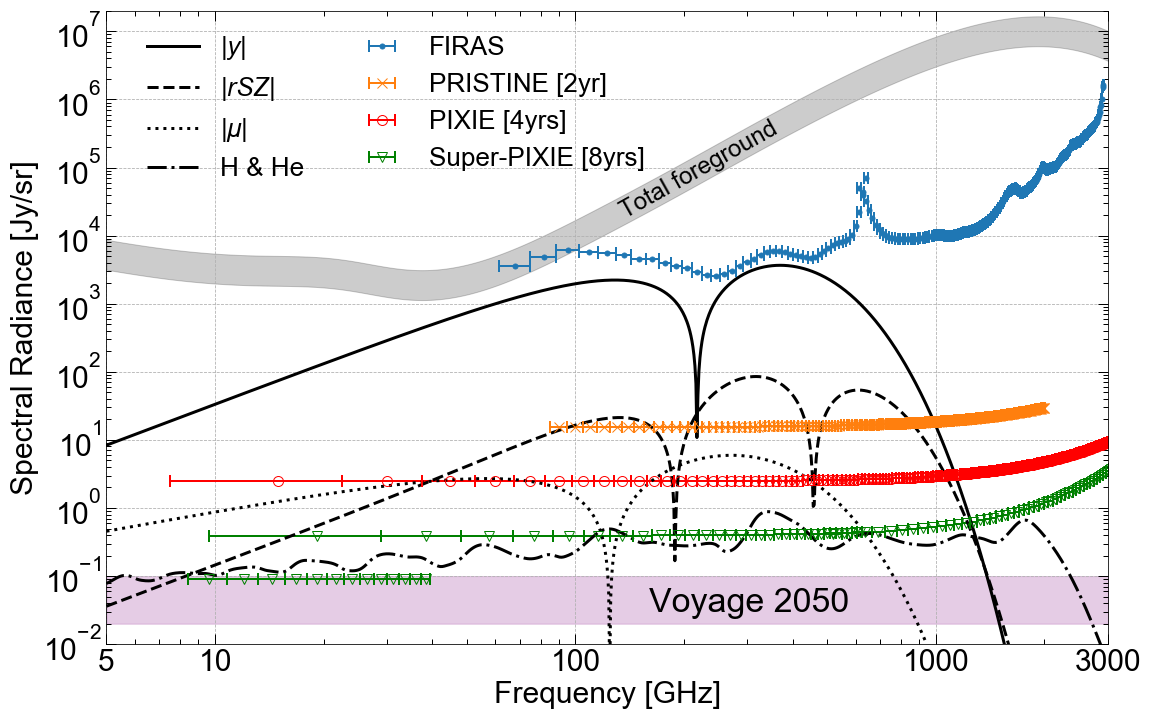}
   \caption{Level of the expected CSM spectral distortion signals and cumulative astrophysical foregrounds.
   The estimated sensitivities for various mission concepts are illustrated as well as their channel distribution. The \SPIXIE high and mid-frequency bands merged around $\nu \simeq 600\,{\rm GHz}$. Within the ESA Voyage 2050 program the $\simeq 0.01-0.1\,{\rm Jy/sr}$ level could be targeted, yielding clear detections of $\mu\simeq \pot{2}{-8}$ and the CRR.
   }
   \vspace{-2mm}
   \label{fig:future}
\end{figure}

\vspace{-3.8mm}
Robust detection of spectral distortion signals in the presence of bright astrophysical foregrounds requires observations over multiple decades in frequency, between $\simeq$~10~GHz and a few$\times$THz. Our current understanding of the intensity foregrounds comes primarily from \Planck, {\it WMAP} and assorted ground-based experiments. At the sensitivities of these observations, the intensity foregrounds could be modeled with sufficient accuracy using a limited set of parameters. We use this foreground parametrization to make spectral distortion forecasts \citep{abitbol_pixie}. 
Figure~\ref{fig:future} compares several predicted spectral distortions \citep[e.g.,][for overview]{Chluba2016} and the largest astrophysical foregrounds to the sensitivity of possible next-generation spectrometers. At high frequencies, the foregrounds are dominated by dust emission from the Milky Way and the cosmic infrared background, while at low frequencies Galactic synchrotron and free-free emission dominate. 

Pioneering steps towards $y\simeq 10^{-7}-10^{-6}$ and technology development may be possible from the ground and balloons, using concepts similar to COSMO, OLIMPO \citep{Masi2003, Schillaci2014}, ARCADE \citep{Kogut2006ARCADE, arcade2} and BISOU \citep{Bruno2019}. However, because the distortions peak at frequencies above 200~GHz, broad frequency coverage outside the atmospheric windows ultimately requires a space mission to detect $\mu\simeq 10^{-8}$ or the CRR \citep{Vince2015, Mayuri2015, 2016ApJ...833..153S, Chluba2017Moments, abitbol_pixie, Remazeilles2018mu}. 
To prepare for the analysis of CMB spectral distortions, we will be able to capitalize on existing analysis techniques \citep[e.g.,][]{Remazeilles2018} used in CMB anisotropy studies, although a new synergistic approach (combining multiple data sets) and observing strategy (e.g., small-patch vs. all-sky) have yet to be fully developed.

\begin{figure}
\vspace{-7mm}
   \includegraphics[width=0.88\columnwidth]{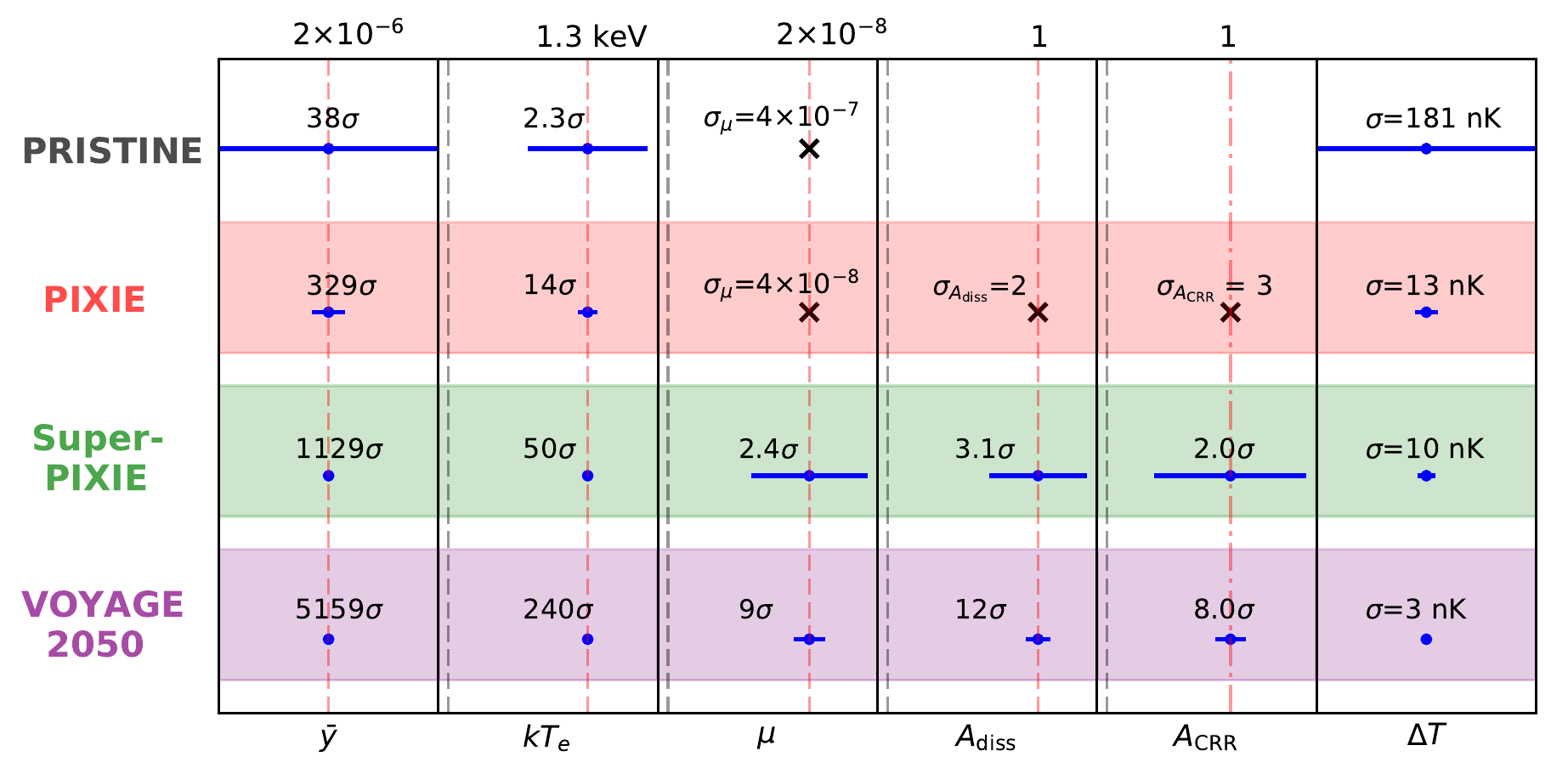}
   \vspace{-2mm}
   \caption{Anticipated distortion signal error estimates after foreground marginalization. Each panel depicts the evolution of measurement errors with improved sensitivity of the proposed experiments. The red dashed line marks the fiducial values for the parameters, while the black dashed line marks the null hypothesis. $A_{\rm diss}$ denotes amplitude for the spectral distortion template computed for the total distortion signal expected from damping of acoustic modes in $\Lambda$CDM. $A_{\rm CRR}$ denotes the amplitude for the spectral distortions induced by recombination of H and He (see Section~\SecCRR for details). For 'Voyage 2050', a scaled version of \SPIXIE was assumed with $\simeq 5$ times higher sensitivity.}
   \label{fig:future_errors}
\end{figure}

\begin{figure}
   \includegraphics[width=0.92\columnwidth]{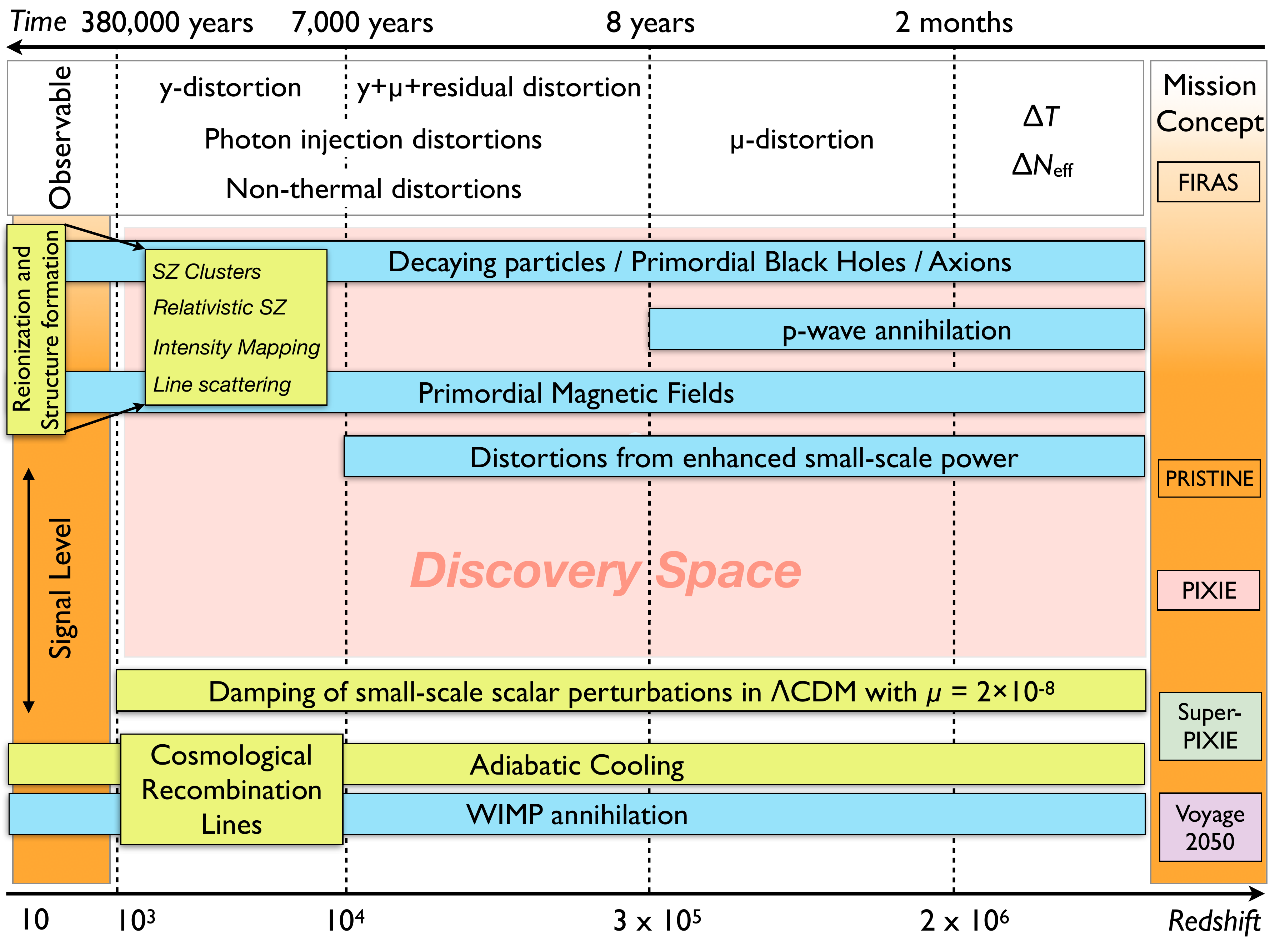}
   \vspace{-2mm}
   \caption{Science thresholds and mission concepts of increasing sensitivity. Guaranteed sources of distortions and their expected signal levels are shown in yellow), while non-standard processes with possible signal levels are presented in turquoise. Spectral distortions could open a new window to the pre-recombination Universe with a vast {\it discovery space} to new physics that is accessible on the path towards a detection and characterization of the $\mu$-distortion from the dissipation of small-scale acoustic modes set by inflation and the cosmological recombination radiation.}
   \label{fig:Missions}
\end{figure}

Using the known foreground signals, expected CMB spectral distortions, and realistic frequency coverage and sensitivity estimated from currently existing technologies (see Fig.~\ref{fig:future}), we produce forecasts for various spectrometer concepts, summarized in Fig.~\ref{fig:future_errors}. A detailed description of the forecasting method can be found in~\citet{abitbol_pixie}. The key points are as follows: 
a pathfinding concept like \PRISTINE could detect the relativistic SZ distortion at $\simeq 2\sigma$, measure the expected $y$-distortion at high significance and deliver an upper limit of $|\mu|<\pot{8}{-7}$ (95\% c.l.) using readily available technology with only 2 years of integration time. 
This would already yield important constraints on galactic feedback models (Fig.~\ref{fig:feedback}) and also provides us with invaluable information about distortion foregrounds. Should polarization sensitivity be included, these observations could also be used to obtain a cosmic-variance-limited measurement of $\tau$ and further mitigate foregrounds in planned $B$-mode polarization searches~\citep{Nabila2018}.

\PIXIE's extended low-frequency coverage and enhanced sensitivity produces strong improvements over \PRISTINE in detection significances for $k T_{\rm e}$ and $y$  (see Fig.~\ref{fig:future_errors}), while improving the upper limit on $\mu$-distortions by an additional order of magnitude to $|\mu|<\pot{8}{-8}$ (95\% c.l.). 
The \SPIXIE concept employs significantly more low-frequency sensitivity and could surpass the threshold for a detection of $\mu=\pot{2}{-8}$ at $3\sigma$. New information on the complexity of foregrounds could allow for better mission optimization in terms of angular resolution, scan strategy, sensitivity and frequency coverage. With sensitivity improvements over \SPIXIE by a factor of 5 (ignoring even an optimized frequency sampling from better understanding of the foreground complexity), a Voyage 2050 spectral distortion mission could target a $0.02\%$ measurement of $y$, a $0.3\%$ measurement of rSZ, a $10\%$ or better measurement of $\mu$, and a $10\%$ measurement of the cosmological recombination lines. This would cross all thresholds for characterization of the distortion expected in the CSM, while covering the full discovery space to new physics (see Fig.~\ref{fig:Missions}).

For these estimates, percent-level priors on the low-frequency foregrounds were imposed, anticipating external information from ground-based low-frequency observatories (e.g., C-BASS and S-PASS) to become available. For \PRISTINE, this has a significant impact on the forecasted errors, while the other cases are far less affected, suggesting that \PRISTINE's science capabilities could be significantly enhanced by adding channels below $\nu<90\,{\rm GHz}$.

\vspace{-5mm}
\subsection{Possible Mission Concepts and Experimental Roadmap}
\begin{figure}
   \vspace{-0mm}
   \includegraphics[width=1.0\columnwidth]{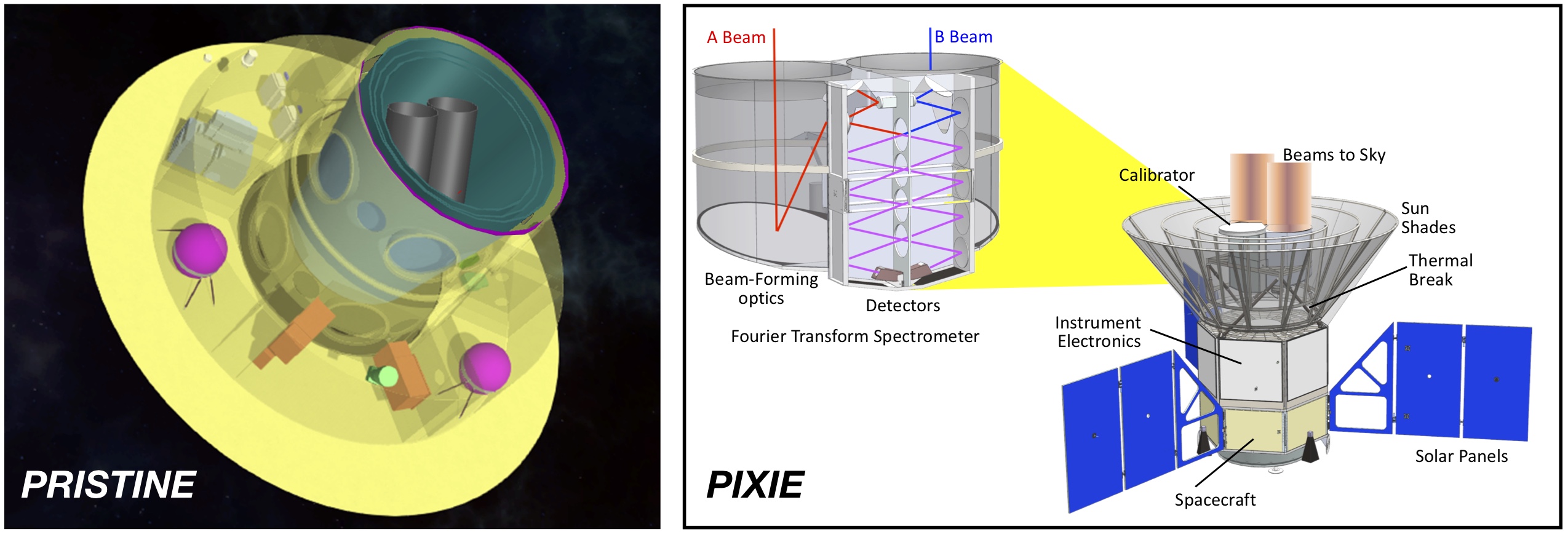}
   \vspace{-4mm}
   \caption{Illustrations of previous missions. The left panel depicts the observing platform envisioned for \PRISTINE, where the two telescopes (each with $\simeq 36\,{\rm cm}$ primary mirror) of the FTS are visible in gray~\citep{Nabila2018}. The right panel shows a zoomed in version of the FTS (primary mirrors $\simeq 55\,{\rm cm}$) alongside the \PIXIE spacecraft \citep{Kogut2016SPIE}. For \SPIXIE, multiple copies of the \PIXIE FTS are combined~\cite{Kogut2019WP}.}
   \label{fig:Mission-illustrations}
\end{figure}

\vspace{-3.5mm}
The next frontier in CMB spectroscopy is to detect the tiny departures from a perfect blackbody predicted in the current paradigm (see Fig.~\ref{fig:future}). This will open a completely new window on cosmology and particle physics, which is within reach of present-day technology, but requires a huge step forward in overall sensitivity from \COBEF\ -- ideally a factor of no less than $10^5$ (see Fig.~\ref{fig:Missions} for science thresholds). This sensitivity can be achieved using FTS concepts (for illustration see Fig.~\ref{fig:Mission-illustrations}) and we advocate for such a space mission in the ESA science program for 2035-2050.

However, while technically the required improvement of sensitivity seems within reach by 2035, and possibly even before that, it calls for a dedicated roadmap that builds on the heritage of \COBEF but minimizes the risks of too bold a single-step jump into the unknown. Foregrounds, in particular, are a source of concern. Past experience has shown that each new space mission operating at microwave and submillimeter wavelengths came with surprises concerning galactic and extragalactic foregrounds, resulting in substantial revisions to pre-existing models. 
So far, all obstacles could be overcome and we are optimistically looking towards future CMB anisotropy measurements to further our understanding of the Universe. In CMB spectroscopy, it will be crucial to understand how to best deal with those foregrounds, reducing residuals to 0.1~Jy/sr or better. This will be learnt from improved observations in the next decade.
Taking into account programmatic constraints at ESA and potential partners, we advocate for two possible paths forward to achieve the science goals in the ESA Voyage 2050 timeframe.

\vspace{-3mm}
\subsubsection{L-class mission with pathfinder}
\vspace{-3.5mm}
In the first scenario, we envision an ambitious L-class space mission based on a scaled and further optimized version of \SPIXIE \citep{Kogut2019WP} with $\simeq$ 2--5 times better sensitivity or more. 
This could allow a measurement of {\it all the expected primordial distortions}, in particular, a $>5\sigma$ detection of the $\mu$-distortion from dissipation and the CRR (see Fig.~\ref{fig:future_errors}), crossing the threshold to characterizing these signals. In this case, a low-cost {\it pathfinder}, similar to \PRISTINE \citep{Nabila2018} and consisting of a single FTS with a sensitivity improved by a factor $\simeq 10^2-10^{3}$ over \COBEF, could be flown in low-earth orbit around 2025-2030. 
Even a pathfinder like this would already cross important science thresholds (see Sect.~\Secfgchall, Fig.~\ref{fig:future_errors}), however, slightly enhanced low-frequency coverage and sensitivity may be desirable (see Sect.~\Secfgchall).
Based on the knowledge built in the analysis of the data from the pathfinder, we would further optimize the spectrometer for measuring spectral distortions in the presence of foreground emission. 
If polarization sensitivity is included, the pathfinder could furthermore complement ongoing searches for primordial $B$-modes, providing extra high frequency coverage and 
hence significant leverage for foreground removal at large angular scales if flown on until $\simeq$~2030.
This information could also be used to explore if polarization sensitivity can help with the cleaning of intensity foregrounds in spectral distortion studies.

In addition to optimizing the L-class spectrometer for foreground removal, one will have to understand how to best make use of data sets that will be available by then. Ground-based observations at low-frequencies $\nu \lesssim 5\,{\rm GHz}$, e.g., with C-BASS, S-PASS or the SKA, will provide valuable new information to constrain the fluctuating part of the low-frequency foregrounds. LSST, {\it Euclid} and DESI will have completed their galaxy surveys, allowing to build direct models for extra-galactic foreground signals that can be utilized. SO, CMB-S4 and \Litebird are also expected to have completed their observations and in combination with the spectrometer will again allow us to further model and subtract the fluctuating parts of the distortion foregrounds. Combined with the unprecedented control of systematics, frequency coverage and spectral sensitivity of the envisioned spectrometer this will provide us with the necessary tools to tackle various challenges.

While the required FTS technology has existed for several decades and is well understood, we may need to further explore alternatives to cover the low-frequency end ($\nu\simeq 5-50\,{\rm GHz}$) of the spectrum. This band yields the largest gains for measuring $\mu$-distortions \citep{abitbol_pixie} and spectrometer-on-chip or radiometer designs may perform better. In tandem with the pathfinder, an optimized L-class spectrometer should deliver breakthrough science encompassing all CSM distortions.

\vspace{-5mm}
\subsubsection{M-class mission within Voyage 2050 Program}
\vspace{-3.5mm}
The second option would be to fly, preferably by 2030-2035 and thus in collaboration with international partners, an intermediate (e.g., M-class) mission with a spectrometer design closely following \SPIXIE \citep{Kogut2019WP}. Depending on budgetary constraints, the number of FTS replicas can be adjusted, although the main cost driver stems from the cryogenic cooling chain.
We expect small changes to the design to be compensated by further optimization of the instrument, so that a mission like this should be able to detect the dissipation $\mu$-distortion at $\simeq 3\sigma$.
Although the $\mu$-detection significance could still suffer from foreground complexities, accurate measurements of the low-redshift $y$-distortion and SZ temperature correction are guaranteed. A vast discovery space would furthermore be explored, constraining many new physics examples with significantly improved bounds on $\mu$ (see Fig.~\ref{fig:Missions}).
In addition, there is a guaranteed scientific harvest from the delivered, absolutely-calibrated all-sky maps at hundreds of frequencies, not only for cosmology but also for many branches of astrophysics (see Sect.~3.4). 
Ultimate spectral distortion measurements would then be targeted with an independent future L-class space mission (potentially capitalizing on ideas to go back to the moon) beyond Voyage~2050.

\vspace{-4mm}
\subsection{Synergies}
\vspace{-3.5mm}
A CMB spectrometer will deliver many new constraints for cosmology (e.g., Fig.~\ref{fig:future_errors} and \ref{fig:Missions}). Here, we wish to emphasize the synergistic gains from combining it with future CMB imaging (ground- or space-based), 21 cm measurements and galaxy surveys. 
A CMB spectrometer will obtain low-resolution ($\delta \theta \simeq 1^\circ$), absolutely-calibrated maps of the full sky at hundreds of frequencies. These maps will allow us to calibrate high-resolution CMB imagers to unprecedented levels, opening many possibilities for studying CMB secondaries \citep{Basu2019WP}. They will also help with the cleaning of CMB foregrounds at the largest angular scales, providing unprecedented control of systematics and allowing us to explore significantly extended foreground parametrizations \citep[e.g.,][]{Chluba2017Moments}. 
If polarization sensitivity is included, as in all currently considered spectrometer designs, this would also allow us to obtain a cosmic-variance-limited measurement of $\tau$ and help in further improving constraints on primordial $B$-modes.
This, however, requires additional consideration, since it implies reduced spectral distortion sensitivity\footnote{The calibrator is stowed in polarization-mode, when both apertures directly observe the sky.}.

In return, high-resolution CMB imaging with upcoming or planned experiments (e.g., SO, CMB-S4 and \Litebird) and low-frequency ($\nu\lesssim 10\,{\rm GHz}$) observations by ground-based experiments (e.g., C-BASS, S-PASS and SKA) can provide vital information about the spatially-varying foreground components at small scales. LSST, {\it Euclid} and DESI will further help us to improve models of extra-galactic foreground signals (e.g., integrated CO emission \citep[e.g.,][]{Righi2008, Righi2008b, 2016MNRAS.458L..99M}). Combined with the CMB spectrometer, these will allow us to mitigate many of the foregrounds, enabling us to inch closer to the ultimate goals of detecting and characterizing $\mu$ and the CRR.

A CMB spectrometer will also open the path for many cross-correlation studies with future 21~cm measurements and galaxy surveys to further improve our understanding of the low-redshift universe at the largest angular scales \citep{Refregier2000, Zhang2004, Pajer2012, Ganc2012, Alvarez2016}. It can also be used to study the origin of large-scale CMB anomalies and isotropy of the Universe \citep{Dai2013, Planck2019STISO}.
All of the above provide unique scientific opportunities within the ESA Voyage 2050 Program that would further our understanding of astrophysics and cosmology.

\vspace{-3mm}
\section{Conclusions}
\vspace{-3.5mm}
CMB spectral distortions probe many processes throughout the history of the Universe. Precision spectroscopy, possible with existing technology, would provide key tests for processes expected within the CSM and open an enormous discovery space to new physics. This offers unique scientific opportunities for furthering our understanding of inflation, recombination, reionization and particle physics. Several experimental and theoretical challenges have to be overcome before we can fully exploit this new window to early- and late-universe physics. However, the potential gains are immense and the field is entering a phase of accelerated growth after decades of dormancy. 
With a coordinated approach, we could thus see first precision measurements of some of the fundamental observables of our Universe in the ESA Voyage 2050 Program.
World-wide this would be a unique initiative for advancing the long-standing legacy of \COBEF.

\theendnotes

\newpage

\bibliography{Lit}

\begin{thebibliography}{331}
\expandafter\ifx\csname natexlab\endcsname\relax\def\natexlab#1{#1}\fi
\expandafter\ifx\csname bibnamefont\endcsname\relax
  \def\bibnamefont#1{#1}\fi
\expandafter\ifx\csname bibfnamefont\endcsname\relax
  \def\bibfnamefont#1{#1}\fi
\expandafter\ifx\csname citenamefont\endcsname\relax
  \def\citenamefont#1{#1}\fi
\expandafter\ifx\csname url\endcsname\relax
  \def\url#1{\texttt{#1}}\fi
\expandafter\ifx\csname urlprefix\endcsname\relax\def\urlprefix{URL }\fi
\providecommand{\bibinfo}[2]{#2}
\providecommand{\eprint}[2][]{\url{#2}}

\bibitem[{\citenamefont{{Chluba} et~al.}(2019)}]{Chluba2019WPDEC}
\bibinfo{author}{\bibfnamefont{J.}~\bibnamefont{{Chluba}}}
  \bibnamefont{et~al.}, \bibinfo{journal}{\baas} \textbf{\bibinfo{volume}{51}},
  \bibinfo{eid}{184} (\bibinfo{year}{2019}), \eprint{1903.04218}.

\bibitem[{\citenamefont{{Mather} et~al.}(1994)}]{Mather1994}
\bibinfo{author}{\bibfnamefont{J.~C.} \bibnamefont{{Mather}}}
  \bibnamefont{et~al.}, \bibinfo{journal}{\apj} \textbf{\bibinfo{volume}{420}},
  \bibinfo{pages}{439} (\bibinfo{year}{1994}).

\bibitem[{\citenamefont{{Fixsen} et~al.}(1996)}]{Fixsen1996}
\bibinfo{author}{\bibfnamefont{D.~J.} \bibnamefont{{Fixsen}}}
  \bibnamefont{et~al.}, \bibinfo{journal}{\apj} \textbf{\bibinfo{volume}{473}},
  \bibinfo{pages}{576} (\bibinfo{year}{1996}), \eprint{arXiv:astro-ph/9605054}.

\bibitem[{\citenamefont{{Smoot} et~al.}(1992)}]{Smoot1992}
\bibinfo{author}{\bibfnamefont{G.~F.} \bibnamefont{{Smoot}}}
  \bibnamefont{et~al.}, \bibinfo{journal}{\apjl}
  \textbf{\bibinfo{volume}{396}}, \bibinfo{pages}{L1} (\bibinfo{year}{1992}).

\bibitem[{\citenamefont{{Bennett} et~al.}(2003)}]{WMAP_params}
\bibinfo{author}{\bibfnamefont{C.~L.} \bibnamefont{{Bennett}}}
  \bibnamefont{et~al.}, \bibinfo{journal}{\apjs}
  \textbf{\bibinfo{volume}{148}}, \bibinfo{pages}{1} (\bibinfo{year}{2003}).

\bibitem[{\citenamefont{{Planck Collaboration}}(2014)}]{Planck2013params}
\bibinfo{author}{\bibnamefont{{Planck Collaboration}}}, \bibinfo{journal}{\aap}
  \textbf{\bibinfo{volume}{571}}, \bibinfo{eid}{A16} (\bibinfo{year}{2014}).

\bibitem[{\citenamefont{{Fixsen}}(2009)}]{Fixsen2009}
\bibinfo{author}{\bibfnamefont{D.~J.} \bibnamefont{{Fixsen}}},
  \bibinfo{journal}{\apj} \textbf{\bibinfo{volume}{707}}, \bibinfo{pages}{916}
  (\bibinfo{year}{2009}), \eprint{0911.1955}.

\bibitem[{\citenamefont{{Silk} and {Chluba}}(2014)}]{Silk2014Science}
\bibinfo{author}{\bibfnamefont{J.}~\bibnamefont{{Silk}}} \bibnamefont{and}
  \bibinfo{author}{\bibfnamefont{J.}~\bibnamefont{{Chluba}}},
  \bibinfo{journal}{Science} \textbf{\bibinfo{volume}{344}},
  \bibinfo{pages}{586} (\bibinfo{year}{2014}).

\bibitem[{\citenamefont{{Zeldovich} and {Sunyaev}}(1969)}]{Zeldovich1969}
\bibinfo{author}{\bibfnamefont{Y.~B.} \bibnamefont{{Zeldovich}}}
  \bibnamefont{and} \bibinfo{author}{\bibfnamefont{R.~A.}
  \bibnamefont{{Sunyaev}}}, \bibinfo{journal}{\apss}
  \textbf{\bibinfo{volume}{4}}, \bibinfo{pages}{301} (\bibinfo{year}{1969}).

\bibitem[{\citenamefont{{Sunyaev} and
  {Zeldovich}}(1970{\natexlab{a}})}]{Sunyaev1970mu}
\bibinfo{author}{\bibfnamefont{R.~A.} \bibnamefont{{Sunyaev}}}
  \bibnamefont{and} \bibinfo{author}{\bibfnamefont{Y.~B.}
  \bibnamefont{{Zeldovich}}}, \bibinfo{journal}{\apss}
  \textbf{\bibinfo{volume}{7}}, \bibinfo{pages}{20}
  (\bibinfo{year}{1970}{\natexlab{a}}).

\bibitem[{\citenamefont{{Illarionov} and {Sunyaev}}(1975)}]{Illarionov1975b}
\bibinfo{author}{\bibfnamefont{A.~F.} \bibnamefont{{Illarionov}}}
  \bibnamefont{and} \bibinfo{author}{\bibfnamefont{R.~A.}
  \bibnamefont{{Sunyaev}}}, \bibinfo{journal}{Soviet Astronomy}
  \textbf{\bibinfo{volume}{18}}, \bibinfo{pages}{691} (\bibinfo{year}{1975}).

\bibitem[{\citenamefont{{Burigana} et~al.}(1991)\citenamefont{{Burigana},
  {Danese}, and {de Zotti}}}]{Burigana1991}
\bibinfo{author}{\bibfnamefont{C.}~\bibnamefont{{Burigana}}},
  \bibinfo{author}{\bibfnamefont{L.}~\bibnamefont{{Danese}}}, \bibnamefont{and}
  \bibinfo{author}{\bibfnamefont{G.}~\bibnamefont{{de Zotti}}},
  \bibinfo{journal}{\aap} \textbf{\bibinfo{volume}{246}}, \bibinfo{pages}{49}
  (\bibinfo{year}{1991}).

\bibitem[{\citenamefont{{Hu} and {Silk}}(1993{\natexlab{a}})}]{Hu1993}
\bibinfo{author}{\bibfnamefont{W.}~\bibnamefont{{Hu}}} \bibnamefont{and}
  \bibinfo{author}{\bibfnamefont{J.}~\bibnamefont{{Silk}}},
  \bibinfo{journal}{\prd} \textbf{\bibinfo{volume}{48}}, \bibinfo{pages}{485}
  (\bibinfo{year}{1993}{\natexlab{a}}).

\bibitem[{\citenamefont{{Chluba} and {Sunyaev}}(2012)}]{Chluba2011therm}
\bibinfo{author}{\bibfnamefont{J.}~\bibnamefont{{Chluba}}} \bibnamefont{and}
  \bibinfo{author}{\bibfnamefont{R.~A.} \bibnamefont{{Sunyaev}}},
  \bibinfo{journal}{\mnras} \textbf{\bibinfo{volume}{419}},
  \bibinfo{pages}{1294} (\bibinfo{year}{2012}), \eprint{1109.6552}.

\bibitem[{\citenamefont{{Khatri} and {Sunyaev}}(2012)}]{Khatri2012mix}
\bibinfo{author}{\bibfnamefont{R.}~\bibnamefont{{Khatri}}} \bibnamefont{and}
  \bibinfo{author}{\bibfnamefont{R.~A.} \bibnamefont{{Sunyaev}}},
  \bibinfo{journal}{\jcap} \textbf{\bibinfo{volume}{9}}, \bibinfo{eid}{016}
  (\bibinfo{year}{2012}), \eprint{1207.6654}.

\bibitem[{\citenamefont{{Chluba}}(2013{\natexlab{a}})}]{Chluba2013Green}
\bibinfo{author}{\bibfnamefont{J.}~\bibnamefont{{Chluba}}},
  \bibinfo{journal}{\mnras} \textbf{\bibinfo{volume}{434}},
  \bibinfo{pages}{352} (\bibinfo{year}{2013}{\natexlab{a}}),
  \eprint{1304.6120}.

\bibitem[{\citenamefont{{Chluba}}(2013{\natexlab{b}})}]{Chluba2013fore}
\bibinfo{author}{\bibfnamefont{J.}~\bibnamefont{{Chluba}}},
  \bibinfo{journal}{\mnras} \textbf{\bibinfo{volume}{436}},
  \bibinfo{pages}{2232} (\bibinfo{year}{2013}{\natexlab{b}}),
  \eprint{1304.6121}.

\bibitem[{\citenamefont{{Chluba} and {Jeong}}(2014)}]{Chluba2013PCA}
\bibinfo{author}{\bibfnamefont{J.}~\bibnamefont{{Chluba}}} \bibnamefont{and}
  \bibinfo{author}{\bibfnamefont{D.}~\bibnamefont{{Jeong}}},
  \bibinfo{journal}{\mnras} \textbf{\bibinfo{volume}{438}},
  \bibinfo{pages}{2065} (\bibinfo{year}{2014}), \eprint{1306.5751}.

\bibitem[{\citenamefont{{Chluba}}(2015)}]{Chluba2015GreensII}
\bibinfo{author}{\bibfnamefont{J.}~\bibnamefont{{Chluba}}},
  \bibinfo{journal}{\mnras} \textbf{\bibinfo{volume}{454}},
  \bibinfo{pages}{4182} (\bibinfo{year}{2015}), \eprint{1506.06582}.

\bibitem[{\citenamefont{{Dubrovich}}(1975)}]{Dubrovich1975}
\bibinfo{author}{\bibfnamefont{V.~K.} \bibnamefont{{Dubrovich}}},
  \bibinfo{journal}{Soviet Astronomy Letters} \textbf{\bibinfo{volume}{1}},
  \bibinfo{pages}{196} (\bibinfo{year}{1975}).

\bibitem[{\citenamefont{{Sunyaev} and {Chluba}}(2009)}]{Sunyaev2009}
\bibinfo{author}{\bibfnamefont{R.~A.} \bibnamefont{{Sunyaev}}}
  \bibnamefont{and} \bibinfo{author}{\bibfnamefont{J.}~\bibnamefont{{Chluba}}},
  \bibinfo{journal}{Astronomische Nachrichten} \textbf{\bibinfo{volume}{330}},
  \bibinfo{pages}{657} (\bibinfo{year}{2009}), \eprint{0908.0435}.

\bibitem[{\citenamefont{{Chluba} and
  {Ali-Ha{\"i}moud}}(2016)}]{Chluba2016CosmoSpec}
\bibinfo{author}{\bibfnamefont{J.}~\bibnamefont{{Chluba}}} \bibnamefont{and}
  \bibinfo{author}{\bibfnamefont{Y.}~\bibnamefont{{Ali-Ha{\"i}moud}}},
  \bibinfo{journal}{\mnras} \textbf{\bibinfo{volume}{456}},
  \bibinfo{pages}{3494} (\bibinfo{year}{2016}), \eprint{1510.03877}.

\bibitem[{\citenamefont{{Lyubarsky} and {Sunyaev}}(1983)}]{Liubarskii83}
\bibinfo{author}{\bibfnamefont{Y.~E.} \bibnamefont{{Lyubarsky}}}
  \bibnamefont{and} \bibinfo{author}{\bibfnamefont{R.~A.}
  \bibnamefont{{Sunyaev}}}, \bibinfo{journal}{\aap}
  \textbf{\bibinfo{volume}{123}}, \bibinfo{pages}{171} (\bibinfo{year}{1983}).

\bibitem[{\citenamefont{{Chluba} and {Sunyaev}}(2009)}]{Chluba2008c}
\bibinfo{author}{\bibfnamefont{J.}~\bibnamefont{{Chluba}}} \bibnamefont{and}
  \bibinfo{author}{\bibfnamefont{R.~A.} \bibnamefont{{Sunyaev}}},
  \bibinfo{journal}{\aap} \textbf{\bibinfo{volume}{501}}, \bibinfo{pages}{29}
  (\bibinfo{year}{2009}), \eprint{0803.3584}.

\bibitem[{\citenamefont{{Chluba}}(2010)}]{Chluba2010a}
\bibinfo{author}{\bibfnamefont{J.}~\bibnamefont{{Chluba}}},
  \bibinfo{journal}{\mnras} \textbf{\bibinfo{volume}{402}},
  \bibinfo{pages}{1195} (\bibinfo{year}{2010}), \eprint{0910.3663}.

\bibitem[{\citenamefont{{En{\ss}lin} and {Kaiser}}(2000)}]{Ensslin2000}
\bibinfo{author}{\bibfnamefont{T.~A.} \bibnamefont{{En{\ss}lin}}}
  \bibnamefont{and} \bibinfo{author}{\bibfnamefont{C.~R.}
  \bibnamefont{{Kaiser}}}, \bibinfo{journal}{\aap}
  \textbf{\bibinfo{volume}{360}}, \bibinfo{pages}{417} (\bibinfo{year}{2000}),
  \eprint{arXiv:astro-ph/0001429}.

\bibitem[{\citenamefont{{Slatyer}}(2016)}]{Slatyer2015}
\bibinfo{author}{\bibfnamefont{T.~R.} \bibnamefont{{Slatyer}}},
  \bibinfo{journal}{Phys.~Rev.} \textbf{\bibinfo{volume}{D93}},
  \bibinfo{eid}{023521} (\bibinfo{year}{2016}), \eprint{1506.03812}.

\bibitem[{\citenamefont{{Acharya} and {Khatri}}(2019)}]{Acharya2018}
\bibinfo{author}{\bibfnamefont{S.~K.} \bibnamefont{{Acharya}}}
  \bibnamefont{and} \bibinfo{author}{\bibfnamefont{R.}~\bibnamefont{{Khatri}}},
  \bibinfo{journal}{\prd} \textbf{\bibinfo{volume}{99}}, \bibinfo{eid}{043520}
  (\bibinfo{year}{2019}), \eprint{1808.02897}.

\bibitem[{\citenamefont{{Sunyaev} and
  {Zeldovich}}(1970{\natexlab{b}})}]{Sunyaev1970SPEC}
\bibinfo{author}{\bibfnamefont{R.~A.} \bibnamefont{{Sunyaev}}}
  \bibnamefont{and} \bibinfo{author}{\bibfnamefont{Y.~B.}
  \bibnamefont{{Zeldovich}}}, \bibinfo{journal}{Comments on Astrophysics and
  Space Physics} \textbf{\bibinfo{volume}{2}}, \bibinfo{pages}{66}
  (\bibinfo{year}{1970}{\natexlab{b}}).

\bibitem[{\citenamefont{{Sunyaev} and {Zeldovich}}(1980)}]{Sunyaev1980ARAA}
\bibinfo{author}{\bibfnamefont{R.~A.} \bibnamefont{{Sunyaev}}}
  \bibnamefont{and} \bibinfo{author}{\bibfnamefont{I.~B.}
  \bibnamefont{{Zeldovich}}}, \bibinfo{journal}{\araa}
  \textbf{\bibinfo{volume}{18}}, \bibinfo{pages}{537} (\bibinfo{year}{1980}).

\bibitem[{\citenamefont{{Sunyaev} and {Khatri}}(2013)}]{Sunyaev2013}
\bibinfo{author}{\bibfnamefont{R.~A.} \bibnamefont{{Sunyaev}}}
  \bibnamefont{and} \bibinfo{author}{\bibfnamefont{R.}~\bibnamefont{{Khatri}}},
  \bibinfo{journal}{IJMPD} \textbf{\bibinfo{volume}{22}},
  \bibinfo{eid}{1330014} (\bibinfo{year}{2013}), \eprint{1302.6553}.

\bibitem[{\citenamefont{{Andr{\'e}} et~al.}(2014)}]{PRISM2013WPII}
\bibinfo{author}{\bibfnamefont{P.}~\bibnamefont{{Andr{\'e}}}}
  \bibnamefont{et~al.}, \bibinfo{journal}{\jcap} \textbf{\bibinfo{volume}{2}},
  \bibinfo{eid}{006} (\bibinfo{year}{2014}), \eprint{1310.1554}.

\bibitem[{\citenamefont{{Chluba}}(2014)}]{Chluba2014Moriond}
\bibinfo{author}{\bibfnamefont{J.}~\bibnamefont{{Chluba}}},
  \bibinfo{journal}{ArXiv e-prints}  (\bibinfo{year}{2014}),
  \eprint{1405.6938}.

\bibitem[{\citenamefont{{Tashiro}}(2014)}]{Tashiro2014}
\bibinfo{author}{\bibfnamefont{H.}~\bibnamefont{{Tashiro}}},
  \bibinfo{journal}{Prog. of Theo. and Exp. Physics}
  \textbf{\bibinfo{volume}{2014}}, \bibinfo{eid}{06B107}
  (\bibinfo{year}{2014}).

\bibitem[{\citenamefont{{De Zotti} et~al.}(2016)\citenamefont{{De Zotti},
  {Negrello}, {Castex}, {Lapi}, and {Bonato}}}]{deZotti2015}
\bibinfo{author}{\bibfnamefont{G.}~\bibnamefont{{De Zotti}}},
  \bibinfo{author}{\bibfnamefont{M.}~\bibnamefont{{Negrello}}},
  \bibinfo{author}{\bibfnamefont{G.}~\bibnamefont{{Castex}}},
  \bibinfo{author}{\bibfnamefont{A.}~\bibnamefont{{Lapi}}}, \bibnamefont{and}
  \bibinfo{author}{\bibfnamefont{M.}~\bibnamefont{{Bonato}}},
  \bibinfo{journal}{\jcap} \textbf{\bibinfo{volume}{3}}, \bibinfo{eid}{047}
  (\bibinfo{year}{2016}).

\bibitem[{\citenamefont{{Chluba}}(2016)}]{Chluba2016}
\bibinfo{author}{\bibfnamefont{J.}~\bibnamefont{{Chluba}}},
  \bibinfo{journal}{\mnras} \textbf{\bibinfo{volume}{460}},
  \bibinfo{pages}{227} (\bibinfo{year}{2016}), \eprint{1603.02496}.

\bibitem[{\citenamefont{{Chluba}}(2018)}]{Chluba2018}
\bibinfo{author}{\bibfnamefont{J.}~\bibnamefont{{Chluba}}},
  \bibinfo{journal}{ArXiv e-prints}  (\bibinfo{year}{2018}),
  \eprint{1806.02915}.

\bibitem[{\citenamefont{{Kogut} et~al.}(2011{\natexlab{a}})}]{Kogut2011}
\bibinfo{author}{\bibfnamefont{A.}~\bibnamefont{{Kogut}}} \bibnamefont{et~al.},
  \bibinfo{journal}{\apj} \textbf{\bibinfo{volume}{734}}, \bibinfo{pages}{4}
  (\bibinfo{year}{2011}{\natexlab{a}}), \eprint{0901.0562}.

\bibitem[{\citenamefont{{Mukherjee}
  et~al.}(2018{\natexlab{a}})\citenamefont{{Mukherjee}, {Silk}, and
  {Wandelt}}}]{Mukherjee2018FSD}
\bibinfo{author}{\bibfnamefont{S.}~\bibnamefont{{Mukherjee}}},
  \bibinfo{author}{\bibfnamefont{J.}~\bibnamefont{{Silk}}}, \bibnamefont{and}
  \bibinfo{author}{\bibfnamefont{B.~D.} \bibnamefont{{Wandelt}}},
  \bibinfo{journal}{\mnras} \textbf{\bibinfo{volume}{477}},
  \bibinfo{pages}{4473} (\bibinfo{year}{2018}{\natexlab{a}}),
  \eprint{1801.05120}.

\bibitem[{\citenamefont{{Kogut} et~al.}(2011{\natexlab{b}})}]{Kogut2011PIXIE}
\bibinfo{author}{\bibfnamefont{A.}~\bibnamefont{{Kogut}}} \bibnamefont{et~al.},
  \bibinfo{journal}{\jcap} \textbf{\bibinfo{volume}{7}}, \bibinfo{pages}{25}
  (\bibinfo{year}{2011}{\natexlab{b}}), \eprint{1105.2044}.

\bibitem[{\citenamefont{{Kogut} et~al.}(2016)}]{Kogut2016SPIE}
\bibinfo{author}{\bibfnamefont{A.}~\bibnamefont{{Kogut}}} \bibnamefont{et~al.},
  in \emph{\bibinfo{booktitle}{SPIE Conference Series}} (\bibinfo{year}{2016}),
  vol. \bibinfo{volume}{9904} of \emph{\bibinfo{series}{Proc.SPIE}}, p.
  \bibinfo{pages}{99040W}.

\bibitem[{\citenamefont{{Sunyaev} and
  {Zeldovich}}(1972{\natexlab{a}})}]{Sunyaev1972CoASP}
\bibinfo{author}{\bibfnamefont{R.~A.} \bibnamefont{{Sunyaev}}}
  \bibnamefont{and} \bibinfo{author}{\bibfnamefont{Y.~B.}
  \bibnamefont{{Zeldovich}}}, \bibinfo{journal}{Comments on Astrophysics and
  Space Physics} \textbf{\bibinfo{volume}{4}}, \bibinfo{pages}{173}
  (\bibinfo{year}{1972}{\natexlab{a}}).

\bibitem[{\citenamefont{{Carlstrom} et~al.}(2002)\citenamefont{{Carlstrom},
  {Holder}, and {Reese}}}]{Carlstrom2002}
\bibinfo{author}{\bibfnamefont{J.~E.} \bibnamefont{{Carlstrom}}},
  \bibinfo{author}{\bibfnamefont{G.~P.} \bibnamefont{{Holder}}},
  \bibnamefont{and} \bibinfo{author}{\bibfnamefont{E.~D.}
  \bibnamefont{{Reese}}}, \bibinfo{journal}{\araa}
  \textbf{\bibinfo{volume}{40}}, \bibinfo{pages}{643} (\bibinfo{year}{2002}),
  \eprint{arXiv:astro-ph/0208192}.

\bibitem[{\citenamefont{{Pajer} and {Zaldarriaga}}(2012)}]{Pajer2012}
\bibinfo{author}{\bibfnamefont{E.}~\bibnamefont{{Pajer}}} \bibnamefont{and}
  \bibinfo{author}{\bibfnamefont{M.}~\bibnamefont{{Zaldarriaga}}},
  \bibinfo{journal}{\prl} \textbf{\bibinfo{volume}{109}}, \bibinfo{eid}{021302}
  (\bibinfo{year}{2012}), \eprint{1201.5375}.

\bibitem[{\citenamefont{{Ganc} and {Komatsu}}(2012)}]{Ganc2012}
\bibinfo{author}{\bibfnamefont{J.}~\bibnamefont{{Ganc}}} \bibnamefont{and}
  \bibinfo{author}{\bibfnamefont{E.}~\bibnamefont{{Komatsu}}},
  \bibinfo{journal}{\prd} \textbf{\bibinfo{volume}{86}}, \bibinfo{eid}{023518}
  (\bibinfo{year}{2012}), \eprint{1204.4241}.

\bibitem[{\citenamefont{Emami et~al.}(2015)}]{Emami:2015xqa}
\bibinfo{author}{\bibfnamefont{R.}~\bibnamefont{Emami}} \bibnamefont{et~al.},
  \bibinfo{journal}{Phys. Rev.} \textbf{\bibinfo{volume}{D91}},
  \bibinfo{pages}{123531} (\bibinfo{year}{2015}), \eprint{1504.00675}.

\bibitem[{\citenamefont{{Ota}}(2016)}]{Ota2016muE}
\bibinfo{author}{\bibfnamefont{A.}~\bibnamefont{{Ota}}},
  \bibinfo{journal}{Phys.~Rev.} \textbf{\bibinfo{volume}{D94}},
  \bibinfo{eid}{103520} (\bibinfo{year}{2016}).

\bibitem[{\citenamefont{{Danese} and {de Zotti}}(1981)}]{Danese1981}
\bibinfo{author}{\bibfnamefont{L.}~\bibnamefont{{Danese}}} \bibnamefont{and}
  \bibinfo{author}{\bibfnamefont{G.}~\bibnamefont{{de Zotti}}},
  \bibinfo{journal}{\aap} \textbf{\bibinfo{volume}{94}}, \bibinfo{pages}{L33}
  (\bibinfo{year}{1981}).

\bibitem[{\citenamefont{{Balashev} et~al.}(2015)}]{Balashev2015}
\bibinfo{author}{\bibfnamefont{S.~A.} \bibnamefont{{Balashev}}}
  \bibnamefont{et~al.}, \bibinfo{journal}{\apj} \textbf{\bibinfo{volume}{810}},
  \bibinfo{eid}{131} (\bibinfo{year}{2015}), \eprint{1505.06028}.

\bibitem[{\citenamefont{{Burigana} et~al.}(2018)}]{Burigana2018CORE}
\bibinfo{author}{\bibfnamefont{C.}~\bibnamefont{{Burigana}}}
  \bibnamefont{et~al.}, \bibinfo{journal}{\jcap}
  \textbf{\bibinfo{volume}{2018}}, \bibinfo{eid}{021} (\bibinfo{year}{2018}),
  \eprint{1704.05764}.

\bibitem[{\citenamefont{{Refregier} et~al.}(2000)\citenamefont{{Refregier},
  {Komatsu}, {Spergel}, and {Pen}}}]{Refregier2000}
\bibinfo{author}{\bibfnamefont{A.}~\bibnamefont{{Refregier}}},
  \bibinfo{author}{\bibfnamefont{E.}~\bibnamefont{{Komatsu}}},
  \bibinfo{author}{\bibfnamefont{D.~N.} \bibnamefont{{Spergel}}},
  \bibnamefont{and} \bibinfo{author}{\bibfnamefont{U.-L.} \bibnamefont{{Pen}}},
  \bibinfo{journal}{\prd} \textbf{\bibinfo{volume}{61}}, \bibinfo{eid}{123001}
  (\bibinfo{year}{2000}).

\bibitem[{\citenamefont{{Zhang} et~al.}(2004)\citenamefont{{Zhang}, {Pen}, and
  {Trac}}}]{Zhang2004}
\bibinfo{author}{\bibfnamefont{P.}~\bibnamefont{{Zhang}}},
  \bibinfo{author}{\bibfnamefont{U.-L.} \bibnamefont{{Pen}}}, \bibnamefont{and}
  \bibinfo{author}{\bibfnamefont{H.}~\bibnamefont{{Trac}}},
  \bibinfo{journal}{\mnras} \textbf{\bibinfo{volume}{355}},
  \bibinfo{pages}{451} (\bibinfo{year}{2004}), \eprint{arXiv:astro-ph/0402115}.

\bibitem[{\citenamefont{{Pitrou} et~al.}(2010)\citenamefont{{Pitrou},
  {Bernardeau}, and {Uzan}}}]{Pitrou2010}
\bibinfo{author}{\bibfnamefont{C.}~\bibnamefont{{Pitrou}}},
  \bibinfo{author}{\bibfnamefont{F.}~\bibnamefont{{Bernardeau}}},
  \bibnamefont{and} \bibinfo{author}{\bibfnamefont{J.-P.}
  \bibnamefont{{Uzan}}}, \bibinfo{journal}{\jcap} \textbf{\bibinfo{volume}{7}},
  \bibinfo{pages}{19} (\bibinfo{year}{2010}), \eprint{0912.3655}.

\bibitem[{\citenamefont{{Hill} and {Pajer}}(2013)}]{Hill2013}
\bibinfo{author}{\bibfnamefont{J.~C.} \bibnamefont{{Hill}}} \bibnamefont{and}
  \bibinfo{author}{\bibfnamefont{E.}~\bibnamefont{{Pajer}}},
  \bibinfo{journal}{Phys.~Rev.} \textbf{\bibinfo{volume}{D88}},
  \bibinfo{eid}{063526} (\bibinfo{year}{2013}), \eprint{1303.4726}.

\bibitem[{\citenamefont{{Alvarez}}(2016)}]{Alvarez2016}
\bibinfo{author}{\bibfnamefont{M.~A.} \bibnamefont{{Alvarez}}},
  \bibinfo{journal}{\apj} \textbf{\bibinfo{volume}{824}}, \bibinfo{eid}{118}
  (\bibinfo{year}{2016}), \eprint{1511.02846}.

\bibitem[{\citenamefont{{Dai} et~al.}(2013)\citenamefont{{Dai}, {Jeong},
  {Kamionkowski}, and {Chluba}}}]{Dai2013}
\bibinfo{author}{\bibfnamefont{L.}~\bibnamefont{{Dai}}},
  \bibinfo{author}{\bibfnamefont{D.}~\bibnamefont{{Jeong}}},
  \bibinfo{author}{\bibfnamefont{M.}~\bibnamefont{{Kamionkowski}}},
  \bibnamefont{and} \bibinfo{author}{\bibfnamefont{J.}~\bibnamefont{{Chluba}}},
  \bibinfo{journal}{\prd} \textbf{\bibinfo{volume}{87}}, \bibinfo{eid}{123005}
  (\bibinfo{year}{2013}), \eprint{1303.6949}.

\bibitem[{\citenamefont{{Kogut} et~al.}(2019)}]{Kogut2019WP}
\bibinfo{author}{\bibfnamefont{A.}~\bibnamefont{{Kogut}}} \bibnamefont{et~al.},
  \bibinfo{journal}{arXiv e-prints}  (\bibinfo{year}{2019}),
  \eprint{1907.13195}.

\bibitem[{\citenamefont{{Basu} et~al.}(2019)}]{Basu2019WP}
\bibinfo{author}{\bibfnamefont{K.}~\bibnamefont{{Basu}}} \bibnamefont{et~al.},
  \bibinfo{journal}{Voyage 2050 Survey}  (\bibinfo{year}{2019}).

\bibitem[{\citenamefont{{Calabrese} et~al.}(2017)\citenamefont{{Calabrese},
  {Alonso}, and {Dunkley}}}]{Calabrese2017}
\bibinfo{author}{\bibfnamefont{E.}~\bibnamefont{{Calabrese}}},
  \bibinfo{author}{\bibfnamefont{D.}~\bibnamefont{{Alonso}}}, \bibnamefont{and}
  \bibinfo{author}{\bibfnamefont{J.}~\bibnamefont{{Dunkley}}},
  \bibinfo{journal}{\prd} \textbf{\bibinfo{volume}{95}}, \bibinfo{eid}{063504}
  (\bibinfo{year}{2017}), \eprint{1611.10269}.

\bibitem[{\citenamefont{Collaboration}(2018)}]{Planck2018params}
\bibinfo{author}{\bibfnamefont{P.}~\bibnamefont{Collaboration}},
  \bibinfo{journal}{ArXiv e-prints}  (\bibinfo{year}{2018}),
  \eprint{1807.06209}.

\bibitem[{\citenamefont{Starobinsky}(1980)}]{Starobinsky:1980te}
\bibinfo{author}{\bibfnamefont{A.~A.} \bibnamefont{Starobinsky}},
  \bibinfo{journal}{Phys. Lett.} \textbf{\bibinfo{volume}{B91}},
  \bibinfo{pages}{99} (\bibinfo{year}{1980}).

\bibitem[{\citenamefont{Guth}(1981)}]{Guth:1980zm}
\bibinfo{author}{\bibfnamefont{A.~H.} \bibnamefont{Guth}},
  \bibinfo{journal}{Phys. Rev.} \textbf{\bibinfo{volume}{D23}},
  \bibinfo{pages}{347} (\bibinfo{year}{1981}).

\bibitem[{\citenamefont{Linde}(1982)}]{Linde:1981mu}
\bibinfo{author}{\bibfnamefont{A.~D.} \bibnamefont{Linde}},
  \bibinfo{journal}{Phys. Lett.} \textbf{\bibinfo{volume}{108B}},
  \bibinfo{pages}{389} (\bibinfo{year}{1982}).

\bibitem[{\citenamefont{Albrecht and Steinhardt}(1982)}]{Albrecht:1982wi}
\bibinfo{author}{\bibfnamefont{A.}~\bibnamefont{Albrecht}} \bibnamefont{and}
  \bibinfo{author}{\bibfnamefont{P.~J.} \bibnamefont{Steinhardt}},
  \bibinfo{journal}{\prl} \textbf{\bibinfo{volume}{48}}, \bibinfo{pages}{1220}
  (\bibinfo{year}{1982}).

\bibitem[{\citenamefont{{Mukhanov} and {Chibisov}}(1981)}]{mukhanov}
\bibinfo{author}{\bibfnamefont{V.~F.} \bibnamefont{{Mukhanov}}}
  \bibnamefont{and} \bibinfo{author}{\bibfnamefont{G.~V.}
  \bibnamefont{{Chibisov}}}, \bibinfo{journal}{SJETP Letters}
  \textbf{\bibinfo{volume}{33}}, \bibinfo{pages}{532} (\bibinfo{year}{1981}).

\bibitem[{\citenamefont{Hawking}(1982)}]{Hawking:1982cz}
\bibinfo{author}{\bibfnamefont{S.~W.} \bibnamefont{Hawking}},
  \bibinfo{journal}{Phys. Lett.} \textbf{\bibinfo{volume}{115B}},
  \bibinfo{pages}{295} (\bibinfo{year}{1982}).

\bibitem[{\citenamefont{Starobinsky}(1979)}]{Starobinsky:1979ty}
\bibinfo{author}{\bibfnamefont{A.~A.} \bibnamefont{Starobinsky}},
  \bibinfo{journal}{JETP Lett.} \textbf{\bibinfo{volume}{30}},
  \bibinfo{pages}{682} (\bibinfo{year}{1979}).

\bibitem[{\citenamefont{Guth and Pi}(1982)}]{Guth:1982ec}
\bibinfo{author}{\bibfnamefont{A.~H.} \bibnamefont{Guth}} \bibnamefont{and}
  \bibinfo{author}{\bibfnamefont{S.~Y.} \bibnamefont{Pi}},
  \bibinfo{journal}{\prl} \textbf{\bibinfo{volume}{49}}, \bibinfo{pages}{1110}
  (\bibinfo{year}{1982}).

\bibitem[{\citenamefont{Gasperini and Veneziano}(1993)}]{Gasperini:1992em}
\bibinfo{author}{\bibfnamefont{M.}~\bibnamefont{Gasperini}} \bibnamefont{and}
  \bibinfo{author}{\bibfnamefont{G.}~\bibnamefont{Veneziano}},
  \bibinfo{journal}{Astropart. Phys.} \textbf{\bibinfo{volume}{1}},
  \bibinfo{pages}{317} (\bibinfo{year}{1993}), \eprint{hep-th/9211021}.

\bibitem[{\citenamefont{Wands}(1999)}]{Wands:1998yp}
\bibinfo{author}{\bibfnamefont{D.}~\bibnamefont{Wands}},
  \bibinfo{journal}{Phys. Rev.} \textbf{\bibinfo{volume}{D60}},
  \bibinfo{pages}{023507} (\bibinfo{year}{1999}), \eprint{gr-qc/9809062}.

\bibitem[{\citenamefont{Khoury et~al.}(2002)\citenamefont{Khoury, Ovrut,
  Seiberg, Steinhardt, and Turok}}]{Khoury:2001bz}
\bibinfo{author}{\bibfnamefont{J.}~\bibnamefont{Khoury}},
  \bibinfo{author}{\bibfnamefont{B.~A.} \bibnamefont{Ovrut}},
  \bibinfo{author}{\bibfnamefont{N.}~\bibnamefont{Seiberg}},
  \bibinfo{author}{\bibfnamefont{P.~J.} \bibnamefont{Steinhardt}},
  \bibnamefont{and} \bibinfo{author}{\bibfnamefont{N.}~\bibnamefont{Turok}},
  \bibinfo{journal}{Phys. Rev.} \textbf{\bibinfo{volume}{D65}},
  \bibinfo{pages}{086007} (\bibinfo{year}{2002}).

\bibitem[{\citenamefont{Hollands and Wald}(2002)}]{Hollands:2002yb}
\bibinfo{author}{\bibfnamefont{S.}~\bibnamefont{Hollands}} \bibnamefont{and}
  \bibinfo{author}{\bibfnamefont{R.~M.} \bibnamefont{Wald}},
  \bibinfo{journal}{Gen. Rel. Grav.} \textbf{\bibinfo{volume}{34}},
  \bibinfo{pages}{2043} (\bibinfo{year}{2002}), \eprint{gr-qc/0205058}.

\bibitem[{\citenamefont{Brandenberger et~al.}(2007)\citenamefont{Brandenberger,
  Nayeri, Patil, and Vafa}}]{Brandenberger:2006vv}
\bibinfo{author}{\bibfnamefont{R.~H.} \bibnamefont{Brandenberger}},
  \bibinfo{author}{\bibfnamefont{A.}~\bibnamefont{Nayeri}},
  \bibinfo{author}{\bibfnamefont{S.~P.} \bibnamefont{Patil}}, \bibnamefont{and}
  \bibinfo{author}{\bibfnamefont{C.}~\bibnamefont{Vafa}},
  \bibinfo{journal}{Int. J. Mod. Phys.} \textbf{\bibinfo{volume}{A22}},
  \bibinfo{pages}{3621} (\bibinfo{year}{2007}).

\bibitem[{\citenamefont{Craps et~al.}(2012)\citenamefont{Craps, Hertog, and
  Turok}}]{Craps:2007ch}
\bibinfo{author}{\bibfnamefont{B.}~\bibnamefont{Craps}},
  \bibinfo{author}{\bibfnamefont{T.}~\bibnamefont{Hertog}}, \bibnamefont{and}
  \bibinfo{author}{\bibfnamefont{N.}~\bibnamefont{Turok}},
  \bibinfo{journal}{Phys. Rev.} \textbf{\bibinfo{volume}{D86}},
  \bibinfo{pages}{043513} (\bibinfo{year}{2012}), \eprint{0712.4180}.

\bibitem[{\citenamefont{Creminelli et~al.}(2010)\citenamefont{Creminelli,
  Nicolis, and Trincherini}}]{Creminelli:2010ba}
\bibinfo{author}{\bibfnamefont{P.}~\bibnamefont{Creminelli}},
  \bibinfo{author}{\bibfnamefont{A.}~\bibnamefont{Nicolis}}, \bibnamefont{and}
  \bibinfo{author}{\bibfnamefont{E.}~\bibnamefont{Trincherini}},
  \bibinfo{journal}{JCAP} \textbf{\bibinfo{volume}{1011}}, \bibinfo{pages}{021}
  (\bibinfo{year}{2010}), \eprint{1007.0027}.

\bibitem[{\citenamefont{Ijjas and Steinhardt}(2016)}]{Ijjas:2016tpn}
\bibinfo{author}{\bibfnamefont{A.}~\bibnamefont{Ijjas}} \bibnamefont{and}
  \bibinfo{author}{\bibfnamefont{P.~J.} \bibnamefont{Steinhardt}},
  \bibinfo{journal}{\prl} \textbf{\bibinfo{volume}{117}},
  \bibinfo{pages}{121304} (\bibinfo{year}{2016}), \eprint{1606.08880}.

\bibitem[{\citenamefont{{Dobre} and otheres}(2018)}]{Dobre2018}
\bibinfo{author}{\bibfnamefont{D.~A.} \bibnamefont{{Dobre}}} \bibnamefont{and}
  \bibinfo{author}{\bibnamefont{otheres}}, \bibinfo{journal}{\jcap}
  \textbf{\bibinfo{volume}{2018}}, \bibinfo{eid}{020} (\bibinfo{year}{2018}),
  \eprint{1712.10272}.

\bibitem[{\citenamefont{Ade et~al.}(2018)}]{Ade:2018gkx}
\bibinfo{author}{\bibfnamefont{P.~A.~R.} \bibnamefont{Ade}}
  \bibnamefont{et~al.} (\bibinfo{collaboration}{BICEP2, Keck Array}),
  \bibinfo{journal}{Submitted to: \prl}  (\bibinfo{year}{2018}),
  \eprint{1810.05216}.

\bibitem[{\citenamefont{{Bezrukov} and {Shaposhnikov}}(2008)}]{Bezrukov2008}
\bibinfo{author}{\bibfnamefont{F.}~\bibnamefont{{Bezrukov}}} \bibnamefont{and}
  \bibinfo{author}{\bibfnamefont{M.}~\bibnamefont{{Shaposhnikov}}},
  \bibinfo{journal}{Phys. Lett.} \textbf{\bibinfo{volume}{B659}},
  \bibinfo{pages}{703} (\bibinfo{year}{2008}), \eprint{0710.3755}.

\bibitem[{\citenamefont{Byrnes et~al.}(2018)\citenamefont{Byrnes, Cole, and
  Patil}}]{Byrnes:2018txb}
\bibinfo{author}{\bibfnamefont{C.~T.} \bibnamefont{Byrnes}},
  \bibinfo{author}{\bibfnamefont{P.~S.} \bibnamefont{Cole}}, \bibnamefont{and}
  \bibinfo{author}{\bibfnamefont{S.~P.} \bibnamefont{Patil}}
  (\bibinfo{year}{2018}), \eprint{1811.11158}.

\bibitem[{\citenamefont{{Sunyaev} and
  {Zeldovich}}(1970{\natexlab{c}})}]{Sunyaev1970diss}
\bibinfo{author}{\bibfnamefont{R.~A.} \bibnamefont{{Sunyaev}}}
  \bibnamefont{and} \bibinfo{author}{\bibfnamefont{Y.~B.}
  \bibnamefont{{Zeldovich}}}, \bibinfo{journal}{\apss}
  \textbf{\bibinfo{volume}{9}}, \bibinfo{pages}{368}
  (\bibinfo{year}{1970}{\natexlab{c}}).

\bibitem[{\citenamefont{{Daly}}(1991)}]{Daly1991}
\bibinfo{author}{\bibfnamefont{R.~A.} \bibnamefont{{Daly}}},
  \bibinfo{journal}{\apj} \textbf{\bibinfo{volume}{371}}, \bibinfo{pages}{14}
  (\bibinfo{year}{1991}).

\bibitem[{\citenamefont{{Hu} et~al.}(1994{\natexlab{a}})\citenamefont{{Hu},
  {Scott}, and {Silk}}}]{Hu1994}
\bibinfo{author}{\bibfnamefont{W.}~\bibnamefont{{Hu}}},
  \bibinfo{author}{\bibfnamefont{D.}~\bibnamefont{{Scott}}}, \bibnamefont{and}
  \bibinfo{author}{\bibfnamefont{J.}~\bibnamefont{{Silk}}},
  \bibinfo{journal}{\apjl} \textbf{\bibinfo{volume}{430}}, \bibinfo{pages}{L5}
  (\bibinfo{year}{1994}{\natexlab{a}}), \eprint{arXiv:astro-ph/9402045}.

\bibitem[{\citenamefont{{Chluba}
  et~al.}(2012{\natexlab{a}})\citenamefont{{Chluba}, {Khatri}, and
  {Sunyaev}}}]{Chluba2012}
\bibinfo{author}{\bibfnamefont{J.}~\bibnamefont{{Chluba}}},
  \bibinfo{author}{\bibfnamefont{R.}~\bibnamefont{{Khatri}}}, \bibnamefont{and}
  \bibinfo{author}{\bibfnamefont{R.~A.} \bibnamefont{{Sunyaev}}},
  \bibinfo{journal}{\mnras} \textbf{\bibinfo{volume}{425}},
  \bibinfo{pages}{1129} (\bibinfo{year}{2012}{\natexlab{a}}),
  \eprint{1202.0057}.

\bibitem[{\citenamefont{{Khatri} et~al.}(2012)\citenamefont{{Khatri},
  {Sunyaev}, and {Chluba}}}]{Khatri2012short2x2}
\bibinfo{author}{\bibfnamefont{R.}~\bibnamefont{{Khatri}}},
  \bibinfo{author}{\bibfnamefont{R.~A.} \bibnamefont{{Sunyaev}}},
  \bibnamefont{and} \bibinfo{author}{\bibfnamefont{J.}~\bibnamefont{{Chluba}}},
  \bibinfo{journal}{\aap} \textbf{\bibinfo{volume}{543}}, \bibinfo{eid}{A136}
  (\bibinfo{year}{2012}), \eprint{1205.2871}.

\bibitem[{\citenamefont{{Chluba}
  et~al.}(2012{\natexlab{b}})\citenamefont{{Chluba}, {Erickcek}, and
  {Ben-Dayan}}}]{Chluba2012inflaton}
\bibinfo{author}{\bibfnamefont{J.}~\bibnamefont{{Chluba}}},
  \bibinfo{author}{\bibfnamefont{A.~L.} \bibnamefont{{Erickcek}}},
  \bibnamefont{and}
  \bibinfo{author}{\bibfnamefont{I.}~\bibnamefont{{Ben-Dayan}}},
  \bibinfo{journal}{\apj} \textbf{\bibinfo{volume}{758}}, \bibinfo{eid}{76}
  (\bibinfo{year}{2012}{\natexlab{b}}), \eprint{1203.2681}.

\bibitem[{\citenamefont{{Chluba}
  et~al.}(2015{\natexlab{a}})\citenamefont{{Chluba}, {Hamann}, and
  {Patil}}}]{Chluba2015IJMPD}
\bibinfo{author}{\bibfnamefont{J.}~\bibnamefont{{Chluba}}},
  \bibinfo{author}{\bibfnamefont{J.}~\bibnamefont{{Hamann}}}, \bibnamefont{and}
  \bibinfo{author}{\bibfnamefont{S.~P.} \bibnamefont{{Patil}}},
  \bibinfo{journal}{IJMP D} \textbf{\bibinfo{volume}{24}},
  \bibinfo{eid}{1530023} (\bibinfo{year}{2015}{\natexlab{a}}),
  \eprint{1505.01834}.

\bibitem[{\citenamefont{{Cabass}
  et~al.}(2016{\natexlab{a}})\citenamefont{{Cabass}, {Melchiorri}, and
  {Pajer}}}]{Cabass2016}
\bibinfo{author}{\bibfnamefont{G.}~\bibnamefont{{Cabass}}},
  \bibinfo{author}{\bibfnamefont{A.}~\bibnamefont{{Melchiorri}}},
  \bibnamefont{and} \bibinfo{author}{\bibfnamefont{E.}~\bibnamefont{{Pajer}}},
  \bibinfo{journal}{Phys.~Rev.} \textbf{\bibinfo{volume}{D93}},
  \bibinfo{eid}{083515} (\bibinfo{year}{2016}{\natexlab{a}}),
  \eprint{1602.05578}.

\bibitem[{\citenamefont{{Pajer} and {Zaldarriaga}}(2013)}]{Pajer2012b}
\bibinfo{author}{\bibfnamefont{E.}~\bibnamefont{{Pajer}}} \bibnamefont{and}
  \bibinfo{author}{\bibfnamefont{M.}~\bibnamefont{{Zaldarriaga}}},
  \bibinfo{journal}{\jcap} \textbf{\bibinfo{volume}{2}}, \bibinfo{eid}{036}
  (\bibinfo{year}{2013}), \eprint{1206.4479}.

\bibitem[{\citenamefont{{Khatri} and {Sunyaev}}(2013)}]{Khatri2013forecast}
\bibinfo{author}{\bibfnamefont{R.}~\bibnamefont{{Khatri}}} \bibnamefont{and}
  \bibinfo{author}{\bibfnamefont{R.~A.} \bibnamefont{{Sunyaev}}},
  \bibinfo{journal}{\jcap} \textbf{\bibinfo{volume}{6}}, \bibinfo{eid}{026}
  (\bibinfo{year}{2013}), \eprint{1303.7212}.

\bibitem[{\citenamefont{{Powell}}(2012)}]{Powell2012}
\bibinfo{author}{\bibfnamefont{B.~A.} \bibnamefont{{Powell}}},
  \bibinfo{journal}{ArXiv e-prints}  (\bibinfo{year}{2012}),
  \eprint{1209.2024}.

\bibitem[{\citenamefont{{Clesse} et~al.}(2014)\citenamefont{{Clesse},
  {Garbrecht}, and {Zhu}}}]{Clesse2014}
\bibinfo{author}{\bibfnamefont{S.}~\bibnamefont{{Clesse}}},
  \bibinfo{author}{\bibfnamefont{B.}~\bibnamefont{{Garbrecht}}},
  \bibnamefont{and} \bibinfo{author}{\bibfnamefont{Y.}~\bibnamefont{{Zhu}}},
  \bibinfo{journal}{\jcap} \textbf{\bibinfo{volume}{2014}}, \bibinfo{eid}{046}
  (\bibinfo{year}{2014}), \eprint{1402.2257}.

\bibitem[{\citenamefont{{Cabass}
  et~al.}(2016{\natexlab{b}})\citenamefont{{Cabass}, {Di Valentino},
  {Melchiorri}, {Pajer}, and {Silk}}}]{Cabass2016SRparams}
\bibinfo{author}{\bibfnamefont{G.}~\bibnamefont{{Cabass}}},
  \bibinfo{author}{\bibfnamefont{E.}~\bibnamefont{{Di Valentino}}},
  \bibinfo{author}{\bibfnamefont{A.}~\bibnamefont{{Melchiorri}}},
  \bibinfo{author}{\bibfnamefont{E.}~\bibnamefont{{Pajer}}}, \bibnamefont{and}
  \bibinfo{author}{\bibfnamefont{J.}~\bibnamefont{{Silk}}},
  \bibinfo{journal}{Phys.~Rev.} \textbf{\bibinfo{volume}{D94}},
  \bibinfo{eid}{023523} (\bibinfo{year}{2016}{\natexlab{b}}).

\bibitem[{\citenamefont{Starobinsky}(1992)}]{Starobinsky:1992ts}
\bibinfo{author}{\bibfnamefont{A.~A.} \bibnamefont{Starobinsky}},
  \bibinfo{journal}{JETP Lett.} \textbf{\bibinfo{volume}{55}},
  \bibinfo{pages}{489} (\bibinfo{year}{1992}), \bibinfo{note}{[Pisma Zh. Eksp.
  Teor. Fiz.55,477(1992)]}.

\bibitem[{\citenamefont{Adams et~al.}(2001)\citenamefont{Adams, Cresswell, and
  Easther}}]{Adams:2001vc}
\bibinfo{author}{\bibfnamefont{J.~A.} \bibnamefont{Adams}},
  \bibinfo{author}{\bibfnamefont{B.}~\bibnamefont{Cresswell}},
  \bibnamefont{and} \bibinfo{author}{\bibfnamefont{R.}~\bibnamefont{Easther}},
  \bibinfo{journal}{Phys.~Rev.} \textbf{\bibinfo{volume}{D64}},
  \bibinfo{pages}{123514} (\bibinfo{year}{2001}).

\bibitem[{\citenamefont{Hazra et~al.}(2014)\citenamefont{Hazra, Shafieloo,
  Smoot, and Starobinsky}}]{Hazra:2014jka}
\bibinfo{author}{\bibfnamefont{D.~K.} \bibnamefont{Hazra}},
  \bibinfo{author}{\bibfnamefont{A.}~\bibnamefont{Shafieloo}},
  \bibinfo{author}{\bibfnamefont{G.~F.} \bibnamefont{Smoot}}, \bibnamefont{and}
  \bibinfo{author}{\bibfnamefont{A.~A.} \bibnamefont{Starobinsky}},
  \bibinfo{journal}{\prl} \textbf{\bibinfo{volume}{113}},
  \bibinfo{pages}{071301} (\bibinfo{year}{2014}).

\bibitem[{\citenamefont{Polnarev and Musco}(2007)}]{Polnarev:2006aa}
\bibinfo{author}{\bibfnamefont{A.~G.} \bibnamefont{Polnarev}} \bibnamefont{and}
  \bibinfo{author}{\bibfnamefont{I.}~\bibnamefont{Musco}},
  \bibinfo{journal}{Class. Quant. Grav.} \textbf{\bibinfo{volume}{24}},
  \bibinfo{pages}{1405} (\bibinfo{year}{2007}), \eprint{gr-qc/0605122}.

\bibitem[{\citenamefont{{Kohri} et~al.}(2008)\citenamefont{{Kohri}, {Lyth}, and
  {Melchiorri}}}]{Kohri:2007qn}
\bibinfo{author}{\bibfnamefont{K.}~\bibnamefont{{Kohri}}},
  \bibinfo{author}{\bibfnamefont{D.~H.} \bibnamefont{{Lyth}}},
  \bibnamefont{and}
  \bibinfo{author}{\bibfnamefont{A.}~\bibnamefont{{Melchiorri}}},
  \bibinfo{journal}{\jcap} \textbf{\bibinfo{volume}{2008}}, \bibinfo{eid}{038}
  (\bibinfo{year}{2008}), \eprint{0711.5006}.

\bibitem[{\citenamefont{{Ben-Dayan} and {Brustein}}(2010)}]{Ido2010}
\bibinfo{author}{\bibfnamefont{I.}~\bibnamefont{{Ben-Dayan}}} \bibnamefont{and}
  \bibinfo{author}{\bibfnamefont{R.}~\bibnamefont{{Brustein}}},
  \bibinfo{journal}{\jcap} \textbf{\bibinfo{volume}{2010}}, \bibinfo{eid}{007}
  (\bibinfo{year}{2010}), \eprint{0907.2384}.

\bibitem[{\citenamefont{Choudhury and Mazumdar}(2014)}]{Choudhury:2013woa}
\bibinfo{author}{\bibfnamefont{S.}~\bibnamefont{Choudhury}} \bibnamefont{and}
  \bibinfo{author}{\bibfnamefont{A.}~\bibnamefont{Mazumdar}},
  \bibinfo{journal}{Phys. Lett.} \textbf{\bibinfo{volume}{B733}},
  \bibinfo{pages}{270} (\bibinfo{year}{2014}), \eprint{1307.5119}.

\bibitem[{\citenamefont{Clesse and García-Bellido}(2015)}]{Clesse:2015wea}
\bibinfo{author}{\bibfnamefont{S.}~\bibnamefont{Clesse}} \bibnamefont{and}
  \bibinfo{author}{\bibfnamefont{J.}~\bibnamefont{García-Bellido}},
  \bibinfo{journal}{Phys. Rev.} \textbf{\bibinfo{volume}{D92}},
  \bibinfo{pages}{023524} (\bibinfo{year}{2015}), \eprint{1501.07565}.

\bibitem[{\citenamefont{Germani and Prokopec}(2017)}]{Germani:2017bcs}
\bibinfo{author}{\bibfnamefont{C.}~\bibnamefont{Germani}} \bibnamefont{and}
  \bibinfo{author}{\bibfnamefont{T.}~\bibnamefont{Prokopec}},
  \bibinfo{journal}{Phys. Dark Univ.} \textbf{\bibinfo{volume}{18}},
  \bibinfo{pages}{6} (\bibinfo{year}{2017}), \eprint{1706.04226}.

\bibitem[{\citenamefont{Barnaby and Huang}(2009)}]{Barnaby:2009dd}
\bibinfo{author}{\bibfnamefont{N.}~\bibnamefont{Barnaby}} \bibnamefont{and}
  \bibinfo{author}{\bibfnamefont{Z.}~\bibnamefont{Huang}},
  \bibinfo{journal}{Phys. Rev.} \textbf{\bibinfo{volume}{D80}},
  \bibinfo{pages}{126018} (\bibinfo{year}{2009}), \eprint{0909.0751}.

\bibitem[{\citenamefont{Cook and Sorbo}(2012)}]{Cook:2011hg}
\bibinfo{author}{\bibfnamefont{J.~L.} \bibnamefont{Cook}} \bibnamefont{and}
  \bibinfo{author}{\bibfnamefont{L.}~\bibnamefont{Sorbo}},
  \bibinfo{journal}{Phys. Rev.} \textbf{\bibinfo{volume}{D85}},
  \bibinfo{pages}{023534} (\bibinfo{year}{2012}), \eprint{1109.0022}.

\bibitem[{\citenamefont{Battefeld et~al.}(2014)\citenamefont{Battefeld,
  Battefeld, and Fiene}}]{Battefeld:2013bfl}
\bibinfo{author}{\bibfnamefont{D.}~\bibnamefont{Battefeld}},
  \bibinfo{author}{\bibfnamefont{T.}~\bibnamefont{Battefeld}},
  \bibnamefont{and} \bibinfo{author}{\bibfnamefont{D.}~\bibnamefont{Fiene}},
  \bibinfo{journal}{Phys. Rev.} \textbf{\bibinfo{volume}{D89}},
  \bibinfo{pages}{123523} (\bibinfo{year}{2014}), \eprint{1309.4082}.

\bibitem[{\citenamefont{{Dimastrogiovanni}
  et~al.}(2017)\citenamefont{{Dimastrogiovanni}, {Fasiello}, and
  {Fujita}}}]{Dimastrogiovanni2017}
\bibinfo{author}{\bibfnamefont{E.}~\bibnamefont{{Dimastrogiovanni}}},
  \bibinfo{author}{\bibfnamefont{M.}~\bibnamefont{{Fasiello}}},
  \bibnamefont{and} \bibinfo{author}{\bibfnamefont{T.}~\bibnamefont{{Fujita}}},
  \bibinfo{journal}{\jcap} \textbf{\bibinfo{volume}{2017}}, \bibinfo{eid}{019}
  (\bibinfo{year}{2017}), \eprint{1608.04216}.

\bibitem[{\citenamefont{Domcke and Mukaida}(2018)}]{Domcke:2018eki}
\bibinfo{author}{\bibfnamefont{V.}~\bibnamefont{Domcke}} \bibnamefont{and}
  \bibinfo{author}{\bibfnamefont{K.}~\bibnamefont{Mukaida}},
  \bibinfo{journal}{JCAP} \textbf{\bibinfo{volume}{1811}}, \bibinfo{pages}{020}
  (\bibinfo{year}{2018}), \eprint{1806.08769}.

\bibitem[{\citenamefont{Linde}(1994)}]{Linde:1993cn}
\bibinfo{author}{\bibfnamefont{A.~D.} \bibnamefont{Linde}},
  \bibinfo{journal}{Phys. Rev.} \textbf{\bibinfo{volume}{D49}},
  \bibinfo{pages}{748} (\bibinfo{year}{1994}), \eprint{astro-ph/9307002}.

\bibitem[{\citenamefont{Lyth and Stewart}(1996)}]{Lyth:1996kt}
\bibinfo{author}{\bibfnamefont{D.~H.} \bibnamefont{Lyth}} \bibnamefont{and}
  \bibinfo{author}{\bibfnamefont{E.~D.} \bibnamefont{Stewart}},
  \bibinfo{journal}{Phys. Rev.} \textbf{\bibinfo{volume}{D54}},
  \bibinfo{pages}{7186} (\bibinfo{year}{1996}), \eprint{hep-ph/9606412}.

\bibitem[{\citenamefont{{Garc{\'\i}a-Bellido}
  et~al.}(1996)\citenamefont{{Garc{\'\i}a-Bellido}, {Linde}, and
  {Wands}}}]{Juan1996PhRvD}
\bibinfo{author}{\bibfnamefont{J.}~\bibnamefont{{Garc{\'\i}a-Bellido}}},
  \bibinfo{author}{\bibfnamefont{A.}~\bibnamefont{{Linde}}}, \bibnamefont{and}
  \bibinfo{author}{\bibfnamefont{D.}~\bibnamefont{{Wands}}},
  \bibinfo{journal}{Phys.~Rev.} \textbf{\bibinfo{volume}{D54}},
  \bibinfo{pages}{6040} (\bibinfo{year}{1996}).

\bibitem[{\citenamefont{Abolhasani et~al.}(2011)\citenamefont{Abolhasani,
  Firouzjahi, and Namjoo}}]{Abolhasani:2010kn}
\bibinfo{author}{\bibfnamefont{A.~A.} \bibnamefont{Abolhasani}},
  \bibinfo{author}{\bibfnamefont{H.}~\bibnamefont{Firouzjahi}},
  \bibnamefont{and} \bibinfo{author}{\bibfnamefont{M.~H.}
  \bibnamefont{Namjoo}}, \bibinfo{journal}{Class. Quant. Grav.}
  \textbf{\bibinfo{volume}{28}}, \bibinfo{pages}{075009}
  (\bibinfo{year}{2011}).

\bibitem[{\citenamefont{Clesse et~al.}(2014)\citenamefont{Clesse, Garbrecht,
  and Zhu}}]{Clesse:2014pna}
\bibinfo{author}{\bibfnamefont{S.}~\bibnamefont{Clesse}},
  \bibinfo{author}{\bibfnamefont{B.}~\bibnamefont{Garbrecht}},
  \bibnamefont{and} \bibinfo{author}{\bibfnamefont{Y.}~\bibnamefont{Zhu}},
  \bibinfo{journal}{JCAP} \textbf{\bibinfo{volume}{1410}}, \bibinfo{pages}{046}
  (\bibinfo{year}{2014}), \eprint{1402.2257}.

\bibitem[{\citenamefont{{Barnaby} and {Peloso}}(2011)}]{Barnaby2011}
\bibinfo{author}{\bibfnamefont{N.}~\bibnamefont{{Barnaby}}} \bibnamefont{and}
  \bibinfo{author}{\bibfnamefont{M.}~\bibnamefont{{Peloso}}},
  \bibinfo{journal}{\prl} \textbf{\bibinfo{volume}{106}}, \bibinfo{eid}{181301}
  (\bibinfo{year}{2011}), \eprint{1011.1500}.

\bibitem[{\citenamefont{{Barnaby} et~al.}(2012)\citenamefont{{Barnaby},
  {Pajer}, and {Peloso}}}]{Barnaby2012}
\bibinfo{author}{\bibfnamefont{N.}~\bibnamefont{{Barnaby}}},
  \bibinfo{author}{\bibfnamefont{E.}~\bibnamefont{{Pajer}}}, \bibnamefont{and}
  \bibinfo{author}{\bibfnamefont{M.}~\bibnamefont{{Peloso}}},
  \bibinfo{journal}{Phys.~Rev.} \textbf{\bibinfo{volume}{D85}},
  \bibinfo{eid}{023525} (\bibinfo{year}{2012}), \eprint{1110.3327}.

\bibitem[{\citenamefont{Meerburg and Pajer}(2013)}]{Meerburg:2012id}
\bibinfo{author}{\bibfnamefont{P.~D.} \bibnamefont{Meerburg}} \bibnamefont{and}
  \bibinfo{author}{\bibfnamefont{E.}~\bibnamefont{Pajer}},
  \bibinfo{journal}{JCAP} \textbf{\bibinfo{volume}{1302}}, \bibinfo{pages}{017}
  (\bibinfo{year}{2013}), \eprint{1203.6076}.

\bibitem[{\citenamefont{{Nakama} et~al.}(2017)\citenamefont{{Nakama}, {Chluba},
  and {Kamionkowski}}}]{Nakama2017}
\bibinfo{author}{\bibfnamefont{T.}~\bibnamefont{{Nakama}}},
  \bibinfo{author}{\bibfnamefont{J.}~\bibnamefont{{Chluba}}}, \bibnamefont{and}
  \bibinfo{author}{\bibfnamefont{M.}~\bibnamefont{{Kamionkowski}}},
  \bibinfo{journal}{Phys.~Rev.} \textbf{\bibinfo{volume}{D95}},
  \bibinfo{eid}{121302} (\bibinfo{year}{2017}), \eprint{1703.10559}.

\bibitem[{\citenamefont{{Cho} et~al.}(2017)\citenamefont{{Cho}, {Hong},
  {Stewart}, and {Zoe}}}]{Cho2017}
\bibinfo{author}{\bibfnamefont{K.}~\bibnamefont{{Cho}}},
  \bibinfo{author}{\bibfnamefont{S.~E.} \bibnamefont{{Hong}}},
  \bibinfo{author}{\bibfnamefont{E.~D.} \bibnamefont{{Stewart}}},
  \bibnamefont{and} \bibinfo{author}{\bibfnamefont{H.}~\bibnamefont{{Zoe}}},
  \bibinfo{journal}{\jcap} \textbf{\bibinfo{volume}{2017}}, \bibinfo{eid}{002}
  (\bibinfo{year}{2017}), \eprint{1705.02741}.

\bibitem[{\citenamefont{{Ota} et~al.}(2014)\citenamefont{{Ota}, {Takahashi},
  {Tashiro}, and {Yamaguchi}}}]{Ota2014}
\bibinfo{author}{\bibfnamefont{A.}~\bibnamefont{{Ota}}},
  \bibinfo{author}{\bibfnamefont{T.}~\bibnamefont{{Takahashi}}},
  \bibinfo{author}{\bibfnamefont{H.}~\bibnamefont{{Tashiro}}},
  \bibnamefont{and}
  \bibinfo{author}{\bibfnamefont{M.}~\bibnamefont{{Yamaguchi}}},
  \bibinfo{journal}{\jcap} \textbf{\bibinfo{volume}{10}}, \bibinfo{eid}{029}
  (\bibinfo{year}{2014}), \eprint{1406.0451}.

\bibitem[{\citenamefont{{Chluba}
  et~al.}(2015{\natexlab{b}})\citenamefont{{Chluba}, {Dai}, {Grin}, {Amin}, and
  {Kamionkowski}}}]{Chluba2015}
\bibinfo{author}{\bibfnamefont{J.}~\bibnamefont{{Chluba}}},
  \bibinfo{author}{\bibfnamefont{L.}~\bibnamefont{{Dai}}},
  \bibinfo{author}{\bibfnamefont{D.}~\bibnamefont{{Grin}}},
  \bibinfo{author}{\bibfnamefont{M.~A.} \bibnamefont{{Amin}}},
  \bibnamefont{and}
  \bibinfo{author}{\bibfnamefont{M.}~\bibnamefont{{Kamionkowski}}},
  \bibinfo{journal}{\mnras} \textbf{\bibinfo{volume}{446}},
  \bibinfo{pages}{2871} (\bibinfo{year}{2015}{\natexlab{b}}).

\bibitem[{\citenamefont{{Hu} and {Sugiyama}}(1994)}]{Hu1994isocurv}
\bibinfo{author}{\bibfnamefont{W.}~\bibnamefont{{Hu}}} \bibnamefont{and}
  \bibinfo{author}{\bibfnamefont{N.}~\bibnamefont{{Sugiyama}}},
  \bibinfo{journal}{\apj} \textbf{\bibinfo{volume}{436}}, \bibinfo{pages}{456}
  (\bibinfo{year}{1994}), \eprint{arXiv:astro-ph/9403031}.

\bibitem[{\citenamefont{{Dent} et~al.}(2012)\citenamefont{{Dent}, {Easson}, and
  {Tashiro}}}]{Dent2012}
\bibinfo{author}{\bibfnamefont{J.~B.} \bibnamefont{{Dent}}},
  \bibinfo{author}{\bibfnamefont{D.~A.} \bibnamefont{{Easson}}},
  \bibnamefont{and}
  \bibinfo{author}{\bibfnamefont{H.}~\bibnamefont{{Tashiro}}},
  \bibinfo{journal}{Phys.~Rev.} \textbf{\bibinfo{volume}{D86}},
  \bibinfo{eid}{023514} (\bibinfo{year}{2012}), \eprint{1202.6066}.

\bibitem[{\citenamefont{{Chluba} and {Grin}}(2013)}]{Chluba2013iso}
\bibinfo{author}{\bibfnamefont{J.}~\bibnamefont{{Chluba}}} \bibnamefont{and}
  \bibinfo{author}{\bibfnamefont{D.}~\bibnamefont{{Grin}}},
  \bibinfo{journal}{\mnras} \textbf{\bibinfo{volume}{434}},
  \bibinfo{pages}{1619} (\bibinfo{year}{2013}), \eprint{1304.4596}.

\bibitem[{\citenamefont{{Haga} et~al.}(2018)\citenamefont{{Haga}, {Inomata},
  {Ota}, and {Ravenni}}}]{Haga2018}
\bibinfo{author}{\bibfnamefont{T.}~\bibnamefont{{Haga}}},
  \bibinfo{author}{\bibfnamefont{K.}~\bibnamefont{{Inomata}}},
  \bibinfo{author}{\bibfnamefont{A.}~\bibnamefont{{Ota}}}, \bibnamefont{and}
  \bibinfo{author}{\bibfnamefont{A.}~\bibnamefont{{Ravenni}}},
  \bibinfo{journal}{\jcap} \textbf{\bibinfo{volume}{2018}}, \bibinfo{eid}{036}
  (\bibinfo{year}{2018}).

\bibitem[{\citenamefont{{Ade} et~al.}(2019)}]{SOWP2018}
\bibinfo{author}{\bibfnamefont{P.}~\bibnamefont{{Ade}}} \bibnamefont{et~al.},
  \bibinfo{journal}{\jcap} \textbf{\bibinfo{volume}{2019}}, \bibinfo{eid}{056}
  (\bibinfo{year}{2019}), \eprint{1808.07445}.

\bibitem[{\citenamefont{{Suzuki} et~al.}(2018)}]{Suzuki2018}
\bibinfo{author}{\bibfnamefont{A.}~\bibnamefont{{Suzuki}}}
  \bibnamefont{et~al.}, \bibinfo{journal}{Journal of Low Temperature Physics}
  (\bibinfo{year}{2018}), \eprint{1801.06987}.

\bibitem[{\citenamefont{{Biagetti} et~al.}(2013)\citenamefont{{Biagetti},
  {Perrier}, {Riotto}, and {Desjacques}}}]{Biagetti2013}
\bibinfo{author}{\bibfnamefont{M.}~\bibnamefont{{Biagetti}}},
  \bibinfo{author}{\bibfnamefont{H.}~\bibnamefont{{Perrier}}},
  \bibinfo{author}{\bibfnamefont{A.}~\bibnamefont{{Riotto}}}, \bibnamefont{and}
  \bibinfo{author}{\bibfnamefont{V.}~\bibnamefont{{Desjacques}}},
  \bibinfo{journal}{Phys.~Rev.} \textbf{\bibinfo{volume}{D87}},
  \bibinfo{eid}{063521} (\bibinfo{year}{2013}).

\bibitem[{\citenamefont{{Ota} et~al.}(2015)\citenamefont{{Ota}, {Sekiguchi},
  {Tada}, and {Yokoyama}}}]{Ota2015aniso_iso}
\bibinfo{author}{\bibfnamefont{A.}~\bibnamefont{{Ota}}},
  \bibinfo{author}{\bibfnamefont{T.}~\bibnamefont{{Sekiguchi}}},
  \bibinfo{author}{\bibfnamefont{Y.}~\bibnamefont{{Tada}}}, \bibnamefont{and}
  \bibinfo{author}{\bibfnamefont{S.}~\bibnamefont{{Yokoyama}}},
  \bibinfo{journal}{\jcap} \textbf{\bibinfo{volume}{2015}}, \bibinfo{eid}{013}
  (\bibinfo{year}{2015}), \eprint{1412.4517}.

\bibitem[{\citenamefont{{Khatri} and
  {Sunyaev}}(2015{\natexlab{a}})}]{Khatri2015aniso}
\bibinfo{author}{\bibfnamefont{R.}~\bibnamefont{{Khatri}}} \bibnamefont{and}
  \bibinfo{author}{\bibfnamefont{R.}~\bibnamefont{{Sunyaev}}},
  \bibinfo{journal}{\jcap} \textbf{\bibinfo{volume}{9}}, \bibinfo{eid}{026}
  (\bibinfo{year}{2015}{\natexlab{a}}), \eprint{1507.05615}.

\bibitem[{\citenamefont{{Chluba}
  et~al.}(2017{\natexlab{a}})\citenamefont{{Chluba}, {Dimastrogiovanni},
  {Amin}, and {Kamionkowski}}}]{Chluba2017}
\bibinfo{author}{\bibfnamefont{J.}~\bibnamefont{{Chluba}}},
  \bibinfo{author}{\bibfnamefont{E.}~\bibnamefont{{Dimastrogiovanni}}},
  \bibinfo{author}{\bibfnamefont{M.~A.} \bibnamefont{{Amin}}},
  \bibnamefont{and}
  \bibinfo{author}{\bibfnamefont{M.}~\bibnamefont{{Kamionkowski}}},
  \bibinfo{journal}{\mnras} \textbf{\bibinfo{volume}{466}},
  \bibinfo{pages}{2390} (\bibinfo{year}{2017}{\natexlab{a}}).

\bibitem[{\citenamefont{Ota}(2016)}]{Ota:2016mqd}
\bibinfo{author}{\bibfnamefont{A.}~\bibnamefont{Ota}}, \bibinfo{journal}{Phys.
  Rev.} \textbf{\bibinfo{volume}{D94}}, \bibinfo{pages}{103520}
  (\bibinfo{year}{2016}), \eprint{1607.00212}.

\bibitem[{\citenamefont{{Ravenni} et~al.}(2017)\citenamefont{{Ravenni},
  {Liguori}, {Bartolo}, and {Shiraishi}}}]{Ravenni2017}
\bibinfo{author}{\bibfnamefont{A.}~\bibnamefont{{Ravenni}}},
  \bibinfo{author}{\bibfnamefont{M.}~\bibnamefont{{Liguori}}},
  \bibinfo{author}{\bibfnamefont{N.}~\bibnamefont{{Bartolo}}},
  \bibnamefont{and}
  \bibinfo{author}{\bibfnamefont{M.}~\bibnamefont{{Shiraishi}}},
  \bibinfo{journal}{\jcap} \textbf{\bibinfo{volume}{9}}, \bibinfo{eid}{042}
  (\bibinfo{year}{2017}), \eprint{1707.04759}.

\bibitem[{\citenamefont{Cabass et~al.}(2018)\citenamefont{Cabass, Pajer, and
  van~der Woude}}]{Cabass:2018jgj}
\bibinfo{author}{\bibfnamefont{G.}~\bibnamefont{Cabass}},
  \bibinfo{author}{\bibfnamefont{E.}~\bibnamefont{Pajer}}, \bibnamefont{and}
  \bibinfo{author}{\bibfnamefont{D.}~\bibnamefont{van~der Woude}},
  \bibinfo{journal}{JCAP} \textbf{\bibinfo{volume}{1808}}, \bibinfo{pages}{050}
  (\bibinfo{year}{2018}), \eprint{1805.08775}.

\bibitem[{\citenamefont{{Dimastrogiovanni} and
  {Emami}}(2016)}]{Dimastrogiovanni2016}
\bibinfo{author}{\bibfnamefont{E.}~\bibnamefont{{Dimastrogiovanni}}}
  \bibnamefont{and} \bibinfo{author}{\bibfnamefont{R.}~\bibnamefont{{Emami}}},
  \bibinfo{journal}{\jcap} \textbf{\bibinfo{volume}{12}}, \bibinfo{eid}{015}
  (\bibinfo{year}{2016}), \eprint{1606.04286}.

\bibitem[{\citenamefont{Akrami et~al.}(2019)}]{Akrami:2019izv}
\bibinfo{author}{\bibfnamefont{Y.}~\bibnamefont{Akrami}} \bibnamefont{et~al.}
  (\bibinfo{collaboration}{Planck}) (\bibinfo{year}{2019}),
  \eprint{1905.05697}.

\bibitem[{\citenamefont{Bartolo et~al.}(2016)\citenamefont{Bartolo, Liguori,
  and Shiraishi}}]{Bartolo:2015fqz}
\bibinfo{author}{\bibfnamefont{N.}~\bibnamefont{Bartolo}},
  \bibinfo{author}{\bibfnamefont{M.}~\bibnamefont{Liguori}}, \bibnamefont{and}
  \bibinfo{author}{\bibfnamefont{M.}~\bibnamefont{Shiraishi}},
  \bibinfo{journal}{JCAP} \textbf{\bibinfo{volume}{1603}}, \bibinfo{pages}{029}
  (\bibinfo{year}{2016}), \eprint{1511.01474}.

\bibitem[{\citenamefont{Becker and Huterer}(2012)}]{Becker:2012je}
\bibinfo{author}{\bibfnamefont{A.}~\bibnamefont{Becker}} \bibnamefont{and}
  \bibinfo{author}{\bibfnamefont{D.}~\bibnamefont{Huterer}},
  \bibinfo{journal}{Phys. Rev. Lett.} \textbf{\bibinfo{volume}{109}},
  \bibinfo{pages}{121302} (\bibinfo{year}{2012}), \eprint{1207.5788}.

\bibitem[{\citenamefont{Oppizzi et~al.}(2018)\citenamefont{Oppizzi, Liguori,
  Renzi, Arroja, and Bartolo}}]{Oppizzi:2017nfy}
\bibinfo{author}{\bibfnamefont{F.}~\bibnamefont{Oppizzi}},
  \bibinfo{author}{\bibfnamefont{M.}~\bibnamefont{Liguori}},
  \bibinfo{author}{\bibfnamefont{A.}~\bibnamefont{Renzi}},
  \bibinfo{author}{\bibfnamefont{F.}~\bibnamefont{Arroja}}, \bibnamefont{and}
  \bibinfo{author}{\bibfnamefont{N.}~\bibnamefont{Bartolo}},
  \bibinfo{journal}{JCAP} \textbf{\bibinfo{volume}{1805}}, \bibinfo{pages}{045}
  (\bibinfo{year}{2018}), \eprint{1711.08286}.

\bibitem[{\citenamefont{{Remazeilles} and
  {Chluba}}(2018{\natexlab{a}})}]{Remazeilles2018:mu}
\bibinfo{author}{\bibfnamefont{M.}~\bibnamefont{{Remazeilles}}}
  \bibnamefont{and} \bibinfo{author}{\bibfnamefont{J.}~\bibnamefont{{Chluba}}},
  \bibinfo{journal}{\mnras} \textbf{\bibinfo{volume}{478}},
  \bibinfo{pages}{807} (\bibinfo{year}{2018}{\natexlab{a}}),
  \eprint{1802.10101}.

\bibitem[{\citenamefont{{Hanany} et~al.}(2019)}]{PICO2019}
\bibinfo{author}{\bibfnamefont{S.}~\bibnamefont{{Hanany}}}
  \bibnamefont{et~al.}, \bibinfo{eid}{arXiv:1902.10541} (\bibinfo{year}{2019}),
  \eprint{1902.10541}.

\bibitem[{\citenamefont{{Remazeilles} et~al.}(2011)\citenamefont{{Remazeilles},
  {Delabrouille}, and {Cardoso}}}]{Remazeilles2011a}
\bibinfo{author}{\bibfnamefont{M.}~\bibnamefont{{Remazeilles}}},
  \bibinfo{author}{\bibfnamefont{J.}~\bibnamefont{{Delabrouille}}},
  \bibnamefont{and} \bibinfo{author}{\bibfnamefont{J.-F.}
  \bibnamefont{{Cardoso}}}, \bibinfo{journal}{\mnras}
  \textbf{\bibinfo{volume}{410}}, \bibinfo{pages}{2481} (\bibinfo{year}{2011}),
  \eprint{1006.5599}.

\bibitem[{\citenamefont{{Camera} et~al.}(2015)\citenamefont{{Camera}, {Santos},
  and {Maartens}}}]{Camera2015}
\bibinfo{author}{\bibfnamefont{S.}~\bibnamefont{{Camera}}},
  \bibinfo{author}{\bibfnamefont{M.~G.} \bibnamefont{{Santos}}},
  \bibnamefont{and}
  \bibinfo{author}{\bibfnamefont{R.}~\bibnamefont{{Maartens}}},
  \bibinfo{journal}{\mnras} \textbf{\bibinfo{volume}{448}},
  \bibinfo{pages}{1035} (\bibinfo{year}{2015}), \eprint{1409.8286}.

\bibitem[{\citenamefont{{Dor{\'e}} et~al.}(2014)}]{Dore2014}
\bibinfo{author}{\bibfnamefont{O.}~\bibnamefont{{Dor{\'e}}}}
  \bibnamefont{et~al.}, \bibinfo{journal}{arXiv e-prints}
  \bibinfo{eid}{arXiv:1412.4872} (\bibinfo{year}{2014}), \eprint{1412.4872}.

\bibitem[{\citenamefont{Dalal et~al.}(2008)\citenamefont{Dalal, Dore, Huterer,
  and Shirokov}}]{Dalal:2007cu}
\bibinfo{author}{\bibfnamefont{N.}~\bibnamefont{Dalal}},
  \bibinfo{author}{\bibfnamefont{O.}~\bibnamefont{Dore}},
  \bibinfo{author}{\bibfnamefont{D.}~\bibnamefont{Huterer}}, \bibnamefont{and}
  \bibinfo{author}{\bibfnamefont{A.}~\bibnamefont{Shirokov}},
  \bibinfo{journal}{Phys. Rev.} \textbf{\bibinfo{volume}{D77}},
  \bibinfo{pages}{123514} (\bibinfo{year}{2008}), \eprint{0710.4560}.

\bibitem[{\citenamefont{Matarrese and Verde}(2008)}]{Matarrese:2008nc}
\bibinfo{author}{\bibfnamefont{S.}~\bibnamefont{Matarrese}} \bibnamefont{and}
  \bibinfo{author}{\bibfnamefont{L.}~\bibnamefont{Verde}},
  \bibinfo{journal}{Astrophys. J.} \textbf{\bibinfo{volume}{677}},
  \bibinfo{pages}{L77} (\bibinfo{year}{2008}), \eprint{0801.4826}.

\bibitem[{\citenamefont{Maldacena}(2003)}]{Maldacena:2002vr}
\bibinfo{author}{\bibfnamefont{J.~M.} \bibnamefont{Maldacena}},
  \bibinfo{journal}{JHEP} \textbf{\bibinfo{volume}{05}}, \bibinfo{pages}{013}
  (\bibinfo{year}{2003}), \eprint{astro-ph/0210603}.

\bibitem[{\citenamefont{{Kawasaki} et~al.}(2005)\citenamefont{{Kawasaki},
  {Kohri}, and {Moroi}}}]{Kawasaki2005}
\bibinfo{author}{\bibfnamefont{M.}~\bibnamefont{{Kawasaki}}},
  \bibinfo{author}{\bibfnamefont{K.}~\bibnamefont{{Kohri}}}, \bibnamefont{and}
  \bibinfo{author}{\bibfnamefont{T.}~\bibnamefont{{Moroi}}},
  \bibinfo{journal}{Phys.~Rev.} \textbf{\bibinfo{volume}{D71}},
  \bibinfo{eid}{083502} (\bibinfo{year}{2005}).

\bibitem[{\citenamefont{{Kawasaki} et~al.}(2018)\citenamefont{{Kawasaki},
  {Kohri}, {Moroi}, and {Takaesu}}}]{Kawasaki2018}
\bibinfo{author}{\bibfnamefont{M.}~\bibnamefont{{Kawasaki}}},
  \bibinfo{author}{\bibfnamefont{K.}~\bibnamefont{{Kohri}}},
  \bibinfo{author}{\bibfnamefont{T.}~\bibnamefont{{Moroi}}}, \bibnamefont{and}
  \bibinfo{author}{\bibfnamefont{Y.}~\bibnamefont{{Takaesu}}},
  \bibinfo{journal}{Phys.~Rev.} \textbf{\bibinfo{volume}{D97}},
  \bibinfo{eid}{023502} (\bibinfo{year}{2018}).

\bibitem[{\citenamefont{Ahmed et~al.}(2010)}]{Ahmed:2009zw}
\bibinfo{author}{\bibfnamefont{Z.}~\bibnamefont{Ahmed}} \bibnamefont{et~al.}
  (\bibinfo{collaboration}{CDMS-II}), \bibinfo{journal}{Science}
  \textbf{\bibinfo{volume}{327}}, \bibinfo{pages}{1619} (\bibinfo{year}{2010}),
  \eprint{0912.3592}.

\bibitem[{\citenamefont{Aprile et~al.}(2012)}]{Aprile:2012nq}
\bibinfo{author}{\bibfnamefont{E.}~\bibnamefont{Aprile}} \bibnamefont{et~al.}
  (\bibinfo{collaboration}{XENON100}), \bibinfo{journal}{\prl}
  \textbf{\bibinfo{volume}{109}}, \bibinfo{pages}{181301}
  (\bibinfo{year}{2012}), \eprint{1207.5988}.

\bibitem[{\citenamefont{Angloher et~al.}(2016)}]{Angloher:2015ewa}
\bibinfo{author}{\bibfnamefont{G.}~\bibnamefont{Angloher}} \bibnamefont{et~al.}
  (\bibinfo{collaboration}{CRESST}), \bibinfo{journal}{Eur. Phys. J.}
  \textbf{\bibinfo{volume}{C76}}, \bibinfo{pages}{25} (\bibinfo{year}{2016}),
  \eprint{1509.01515}.

\bibitem[{\citenamefont{Agnese et~al.}(2016)}]{Agnese:2015nto}
\bibinfo{author}{\bibfnamefont{R.}~\bibnamefont{Agnese}} \bibnamefont{et~al.}
  (\bibinfo{collaboration}{SuperCDMS}), \bibinfo{journal}{\prl}
  \textbf{\bibinfo{volume}{116}}, \bibinfo{pages}{071301}
  (\bibinfo{year}{2016}), \eprint{1509.02448}.

\bibitem[{\citenamefont{Tan et~al.}(2016)}]{Tan:2016zwf}
\bibinfo{author}{\bibfnamefont{A.}~\bibnamefont{Tan}} \bibnamefont{et~al.}
  (\bibinfo{collaboration}{PandaX-II}), \bibinfo{journal}{\prl}
  \textbf{\bibinfo{volume}{117}}, \bibinfo{pages}{121303}
  (\bibinfo{year}{2016}), \eprint{1607.07400}.

\bibitem[{\citenamefont{Akerib et~al.}(2017)}]{Akerib:2016vxi}
\bibinfo{author}{\bibfnamefont{D.~S.} \bibnamefont{Akerib}}
  \bibnamefont{et~al.} (\bibinfo{collaboration}{LUX}), \bibinfo{journal}{\prl}
  \textbf{\bibinfo{volume}{118}}, \bibinfo{pages}{021303}
  (\bibinfo{year}{2017}), \eprint{1608.07648}.

\bibitem[{\citenamefont{{Jungman} et~al.}(1996)\citenamefont{{Jungman},
  {Kamionkowski}, and {Griest}}}]{Jungman1996}
\bibinfo{author}{\bibfnamefont{G.}~\bibnamefont{{Jungman}}},
  \bibinfo{author}{\bibfnamefont{M.}~\bibnamefont{{Kamionkowski}}},
  \bibnamefont{and} \bibinfo{author}{\bibfnamefont{K.}~\bibnamefont{{Griest}}},
  \bibinfo{journal}{\physrep} \textbf{\bibinfo{volume}{267}},
  \bibinfo{pages}{195} (\bibinfo{year}{1996}), \eprint{hep-ph/9506380}.

\bibitem[{\citenamefont{{Feng} et~al.}(2003{\natexlab{a}})\citenamefont{{Feng},
  {Rajaraman}, and {Takayama}}}]{Feng2003PhRvL}
\bibinfo{author}{\bibfnamefont{J.~L.} \bibnamefont{{Feng}}},
  \bibinfo{author}{\bibfnamefont{A.}~\bibnamefont{{Rajaraman}}},
  \bibnamefont{and}
  \bibinfo{author}{\bibfnamefont{F.}~\bibnamefont{{Takayama}}},
  \bibinfo{journal}{\prl} \textbf{\bibinfo{volume}{91}}, \bibinfo{eid}{011302}
  (\bibinfo{year}{2003}{\natexlab{a}}).

\bibitem[{\citenamefont{{Feng} et~al.}(2003{\natexlab{b}})\citenamefont{{Feng},
  {Rajaraman}, and {Takayama}}}]{Feng2003}
\bibinfo{author}{\bibfnamefont{J.~L.} \bibnamefont{{Feng}}},
  \bibinfo{author}{\bibfnamefont{A.}~\bibnamefont{{Rajaraman}}},
  \bibnamefont{and}
  \bibinfo{author}{\bibfnamefont{F.}~\bibnamefont{{Takayama}}},
  \bibinfo{journal}{Phys.~Rev.} \textbf{\bibinfo{volume}{D68}},
  \bibinfo{eid}{063504} (\bibinfo{year}{2003}{\natexlab{b}}).

\bibitem[{\citenamefont{{Kusenko}}(2009)}]{Kusenko2009}
\bibinfo{author}{\bibfnamefont{A.}~\bibnamefont{{Kusenko}}},
  \bibinfo{journal}{\physrep} \textbf{\bibinfo{volume}{481}},
  \bibinfo{pages}{1} (\bibinfo{year}{2009}), \eprint{0906.2968}.

\bibitem[{\citenamefont{{Feng}}(2010)}]{Feng2010}
\bibinfo{author}{\bibfnamefont{J.~L.} \bibnamefont{{Feng}}},
  \bibinfo{journal}{\araa} \textbf{\bibinfo{volume}{48}}, \bibinfo{pages}{495}
  (\bibinfo{year}{2010}), \eprint{1003.0904}.

\bibitem[{\citenamefont{{Carr} et~al.}(2010)\citenamefont{{Carr}, {Kohri},
  {Sendouda}, and {Yokoyama}}}]{Carr2010}
\bibinfo{author}{\bibfnamefont{B.~J.} \bibnamefont{{Carr}}},
  \bibinfo{author}{\bibfnamefont{K.}~\bibnamefont{{Kohri}}},
  \bibinfo{author}{\bibfnamefont{Y.}~\bibnamefont{{Sendouda}}},
  \bibnamefont{and}
  \bibinfo{author}{\bibfnamefont{J.}~\bibnamefont{{Yokoyama}}},
  \bibinfo{journal}{Phys.~Rev.} \textbf{\bibinfo{volume}{D81}},
  \bibinfo{pages}{104019} (\bibinfo{year}{2010}).

\bibitem[{\citenamefont{{Marsh}}(2016)}]{Marsh2016Rev}
\bibinfo{author}{\bibfnamefont{D.~J.~E.} \bibnamefont{{Marsh}}},
  \bibinfo{journal}{\physrep} \textbf{\bibinfo{volume}{643}},
  \bibinfo{pages}{1} (\bibinfo{year}{2016}), \eprint{1510.07633}.

\bibitem[{\citenamefont{{Ellis} et~al.}(1992)\citenamefont{{Ellis}, {Gelmini},
  {Lopez}, {Nanopoulos}, and {Sarkar}}}]{Ellis1992}
\bibinfo{author}{\bibfnamefont{J.}~\bibnamefont{{Ellis}}},
  \bibinfo{author}{\bibfnamefont{G.~B.} \bibnamefont{{Gelmini}}},
  \bibinfo{author}{\bibfnamefont{J.~L.} \bibnamefont{{Lopez}}},
  \bibinfo{author}{\bibfnamefont{D.~V.} \bibnamefont{{Nanopoulos}}},
  \bibnamefont{and} \bibinfo{author}{\bibfnamefont{S.}~\bibnamefont{{Sarkar}}},
  \bibinfo{journal}{Nu. Phys. B} \textbf{\bibinfo{volume}{373}},
  \bibinfo{pages}{399} (\bibinfo{year}{1992}).

\bibitem[{\citenamefont{Adams et~al.}(1998)\citenamefont{Adams, Sarkar, and
  Sciama}}]{Adams1998nr}
\bibinfo{author}{\bibfnamefont{J.~A.} \bibnamefont{Adams}},
  \bibinfo{author}{\bibfnamefont{S.}~\bibnamefont{Sarkar}}, \bibnamefont{and}
  \bibinfo{author}{\bibfnamefont{D.~W.} \bibnamefont{Sciama}},
  \bibinfo{journal}{\mnras} \textbf{\bibinfo{volume}{301}},
  \bibinfo{pages}{210} (\bibinfo{year}{1998}), \eprint{astro-ph/9805108}.

\bibitem[{\citenamefont{{Chen} and {Kamionkowski}}(2004)}]{Chen2004}
\bibinfo{author}{\bibfnamefont{X.}~\bibnamefont{{Chen}}} \bibnamefont{and}
  \bibinfo{author}{\bibfnamefont{M.}~\bibnamefont{{Kamionkowski}}},
  \bibinfo{journal}{Phys.~Rev.} \textbf{\bibinfo{volume}{D70}},
  \bibinfo{pages}{043502} (\bibinfo{year}{2004}),
  \eprint{arXiv:astro-ph/0310473}.

\bibitem[{\citenamefont{{Padmanabhan} and
  {Finkbeiner}}(2005)}]{Padmanabhan2005}
\bibinfo{author}{\bibfnamefont{N.}~\bibnamefont{{Padmanabhan}}}
  \bibnamefont{and} \bibinfo{author}{\bibfnamefont{D.~P.}
  \bibnamefont{{Finkbeiner}}}, \bibinfo{journal}{Phys.~Rev.}
  \textbf{\bibinfo{volume}{D72}}, \bibinfo{pages}{023508}
  (\bibinfo{year}{2005}).

\bibitem[{\citenamefont{{Galli} et~al.}(2009)\citenamefont{{Galli}, {Iocco},
  {Bertone}, and {Melchiorri}}}]{Galli2009}
\bibinfo{author}{\bibfnamefont{S.}~\bibnamefont{{Galli}}},
  \bibinfo{author}{\bibfnamefont{F.}~\bibnamefont{{Iocco}}},
  \bibinfo{author}{\bibfnamefont{G.}~\bibnamefont{{Bertone}}},
  \bibnamefont{and}
  \bibinfo{author}{\bibfnamefont{A.}~\bibnamefont{{Melchiorri}}},
  \bibinfo{journal}{\prd} \textbf{\bibinfo{volume}{80}},
  \bibinfo{pages}{023505} (\bibinfo{year}{2009}), \eprint{0905.0003}.

\bibitem[{\citenamefont{Slatyer et~al.}(2009)\citenamefont{Slatyer,
  Padmanabhan, and Finkbeiner}}]{Slatyer2009}
\bibinfo{author}{\bibfnamefont{T.~R.} \bibnamefont{Slatyer}},
  \bibinfo{author}{\bibfnamefont{N.}~\bibnamefont{Padmanabhan}},
  \bibnamefont{and} \bibinfo{author}{\bibfnamefont{D.~P.}
  \bibnamefont{Finkbeiner}}, \bibinfo{journal}{Phys.~Rev.}
  \textbf{\bibinfo{volume}{D80}}, \bibinfo{eid}{043526} (\bibinfo{year}{2009}).

\bibitem[{\citenamefont{{Slatyer} and {Wu}}(2017)}]{Slatyer2017}
\bibinfo{author}{\bibfnamefont{T.~R.} \bibnamefont{{Slatyer}}}
  \bibnamefont{and} \bibinfo{author}{\bibfnamefont{C.-L.} \bibnamefont{{Wu}}},
  \bibinfo{journal}{Phys.~Rev.} \textbf{\bibinfo{volume}{D95}},
  \bibinfo{eid}{023010} (\bibinfo{year}{2017}), \eprint{1610.06933}.

\bibitem[{\citenamefont{{Poulin} et~al.}(2017)\citenamefont{{Poulin},
  {Lesgourgues}, and {Serpico}}}]{Poulin2017}
\bibinfo{author}{\bibfnamefont{V.}~\bibnamefont{{Poulin}}},
  \bibinfo{author}{\bibfnamefont{J.}~\bibnamefont{{Lesgourgues}}},
  \bibnamefont{and} \bibinfo{author}{\bibfnamefont{P.~D.}
  \bibnamefont{{Serpico}}}, \bibinfo{journal}{\jcap}
  \textbf{\bibinfo{volume}{3}}, \bibinfo{eid}{043} (\bibinfo{year}{2017}),
  \eprint{1610.10051}.

\bibitem[{\citenamefont{{Wilkinson}
  et~al.}(2014{\natexlab{a}})\citenamefont{{Wilkinson}, {B{\oe}hm}, and
  {Lesgourgues}}}]{Wilkinson2014}
\bibinfo{author}{\bibfnamefont{R.~J.} \bibnamefont{{Wilkinson}}},
  \bibinfo{author}{\bibfnamefont{C.}~\bibnamefont{{B{\oe}hm}}},
  \bibnamefont{and}
  \bibinfo{author}{\bibfnamefont{J.}~\bibnamefont{{Lesgourgues}}},
  \bibinfo{journal}{\jcap} \textbf{\bibinfo{volume}{2014}}, \bibinfo{eid}{011}
  (\bibinfo{year}{2014}{\natexlab{a}}), \eprint{1401.7597}.

\bibitem[{\citenamefont{{Dvorkin} et~al.}(2014)\citenamefont{{Dvorkin}, {Blum},
  and {Kamionkowski}}}]{Dvorkin2014}
\bibinfo{author}{\bibfnamefont{C.}~\bibnamefont{{Dvorkin}}},
  \bibinfo{author}{\bibfnamefont{K.}~\bibnamefont{{Blum}}}, \bibnamefont{and}
  \bibinfo{author}{\bibfnamefont{M.}~\bibnamefont{{Kamionkowski}}},
  \bibinfo{journal}{Phys.~Rev.} \textbf{\bibinfo{volume}{D89}},
  \bibinfo{eid}{023519} (\bibinfo{year}{2014}), \eprint{1311.2937}.

\bibitem[{\citenamefont{{Wilkinson}
  et~al.}(2014{\natexlab{b}})\citenamefont{{Wilkinson}, {Lesgourgues}, and
  {B{\oe}hm}}}]{Wilkinson2014b}
\bibinfo{author}{\bibfnamefont{R.~J.} \bibnamefont{{Wilkinson}}},
  \bibinfo{author}{\bibfnamefont{J.}~\bibnamefont{{Lesgourgues}}},
  \bibnamefont{and}
  \bibinfo{author}{\bibfnamefont{C.}~\bibnamefont{{B{\oe}hm}}},
  \bibinfo{journal}{\jcap} \textbf{\bibinfo{volume}{2014}}, \bibinfo{eid}{026}
  (\bibinfo{year}{2014}{\natexlab{b}}), \eprint{1309.7588}.

\bibitem[{\citenamefont{{Gluscevic} and {Boddy}}(2018)}]{Gluscevic2018}
\bibinfo{author}{\bibfnamefont{V.}~\bibnamefont{{Gluscevic}}} \bibnamefont{and}
  \bibinfo{author}{\bibfnamefont{K.~K.} \bibnamefont{{Boddy}}},
  \bibinfo{journal}{\prl} \textbf{\bibinfo{volume}{121}}, \bibinfo{eid}{081301}
  (\bibinfo{year}{2018}), \eprint{1712.07133}.

\bibitem[{\citenamefont{{Boddy} and {Gluscevic}}(2018)}]{Boddy2018}
\bibinfo{author}{\bibfnamefont{K.~K.} \bibnamefont{{Boddy}}} \bibnamefont{and}
  \bibinfo{author}{\bibfnamefont{V.}~\bibnamefont{{Gluscevic}}},
  \bibinfo{journal}{Phys.~Rev.} \textbf{\bibinfo{volume}{D98}},
  \bibinfo{eid}{083510} (\bibinfo{year}{2018}), \eprint{1801.08609}.

\bibitem[{\citenamefont{{Jedamzik}}(2008)}]{Jedamzik2008}
\bibinfo{author}{\bibfnamefont{K.}~\bibnamefont{{Jedamzik}}},
  \bibinfo{journal}{\jcap} \textbf{\bibinfo{volume}{3}}, \bibinfo{eid}{008}
  (\bibinfo{year}{2008}), \eprint{0710.5153}.

\bibitem[{\citenamefont{{Poulin} and {Serpico}}(2015)}]{Poulin2015Loop}
\bibinfo{author}{\bibfnamefont{V.}~\bibnamefont{{Poulin}}} \bibnamefont{and}
  \bibinfo{author}{\bibfnamefont{P.~D.} \bibnamefont{{Serpico}}},
  \bibinfo{journal}{\prl} \textbf{\bibinfo{volume}{114}}, \bibinfo{eid}{091101}
  (\bibinfo{year}{2015}), \eprint{1502.01250}.

\bibitem[{\citenamefont{{Sarkar} and {Cooper}}(1984)}]{Sarkar1984}
\bibinfo{author}{\bibfnamefont{S.}~\bibnamefont{{Sarkar}}} \bibnamefont{and}
  \bibinfo{author}{\bibfnamefont{A.~M.} \bibnamefont{{Cooper}}},
  \bibinfo{journal}{Physics Letters B} \textbf{\bibinfo{volume}{148}},
  \bibinfo{pages}{347} (\bibinfo{year}{1984}).

\bibitem[{\citenamefont{{Ellis} et~al.}(1985)\citenamefont{{Ellis},
  {Nanopoulos}, and {Sarkar}}}]{Ellis1985}
\bibinfo{author}{\bibfnamefont{J.}~\bibnamefont{{Ellis}}},
  \bibinfo{author}{\bibfnamefont{D.~V.} \bibnamefont{{Nanopoulos}}},
  \bibnamefont{and} \bibinfo{author}{\bibfnamefont{S.}~\bibnamefont{{Sarkar}}},
  \bibinfo{journal}{Nuclear Physics B} \textbf{\bibinfo{volume}{259}},
  \bibinfo{pages}{175} (\bibinfo{year}{1985}).

\bibitem[{\citenamefont{{Kawasaki} and {Sato}}(1986)}]{Kawasaki1986}
\bibinfo{author}{\bibfnamefont{M.}~\bibnamefont{{Kawasaki}}} \bibnamefont{and}
  \bibinfo{author}{\bibfnamefont{K.}~\bibnamefont{{Sato}}},
  \bibinfo{journal}{Physics Letters B} \textbf{\bibinfo{volume}{169}},
  \bibinfo{pages}{280} (\bibinfo{year}{1986}).

\bibitem[{\citenamefont{{Hu} and {Silk}}(1993{\natexlab{b}})}]{Hu1993b}
\bibinfo{author}{\bibfnamefont{W.}~\bibnamefont{{Hu}}} \bibnamefont{and}
  \bibinfo{author}{\bibfnamefont{J.}~\bibnamefont{{Silk}}},
  \bibinfo{journal}{\prl} \textbf{\bibinfo{volume}{70}}, \bibinfo{pages}{2661}
  (\bibinfo{year}{1993}{\natexlab{b}}).

\bibitem[{\citenamefont{{Aalberts} et~al.}(2018)}]{Aalberts2018}
\bibinfo{author}{\bibfnamefont{J.~L.} \bibnamefont{{Aalberts}}}
  \bibnamefont{et~al.}, \bibinfo{journal}{Phys.~Rev.}
  \textbf{\bibinfo{volume}{D98}}, \bibinfo{eid}{023001} (\bibinfo{year}{2018}).

\bibitem[{\citenamefont{{McDonald} et~al.}(2001)\citenamefont{{McDonald},
  {Scherrer}, and {Walker}}}]{McDonald2001}
\bibinfo{author}{\bibfnamefont{P.}~\bibnamefont{{McDonald}}},
  \bibinfo{author}{\bibfnamefont{R.~J.} \bibnamefont{{Scherrer}}},
  \bibnamefont{and} \bibinfo{author}{\bibfnamefont{T.~P.}
  \bibnamefont{{Walker}}}, \bibinfo{journal}{Phys.~Rev.}
  \textbf{\bibinfo{volume}{D63}}, \bibinfo{pages}{023001}
  (\bibinfo{year}{2001}).

\bibitem[{\citenamefont{{Tashiro}
  et~al.}(2013{\natexlab{a}})\citenamefont{{Tashiro}, {Silk}, and
  {Marsh}}}]{Tashiro2013}
\bibinfo{author}{\bibfnamefont{H.}~\bibnamefont{{Tashiro}}},
  \bibinfo{author}{\bibfnamefont{J.}~\bibnamefont{{Silk}}}, \bibnamefont{and}
  \bibinfo{author}{\bibfnamefont{D.~J.~E.} \bibnamefont{{Marsh}}},
  \bibinfo{journal}{Phys.~Rev.} \textbf{\bibinfo{volume}{D88}},
  \bibinfo{eid}{125024} (\bibinfo{year}{2013}{\natexlab{a}}),
  \eprint{1308.0314}.

\bibitem[{\citenamefont{{Ejlli} and {Dolgov}}(2014)}]{Ejlli2013}
\bibinfo{author}{\bibfnamefont{D.}~\bibnamefont{{Ejlli}}} \bibnamefont{and}
  \bibinfo{author}{\bibfnamefont{A.~D.} \bibnamefont{{Dolgov}}},
  \bibinfo{journal}{Phys.~Rev.} \textbf{\bibinfo{volume}{D90}},
  \bibinfo{eid}{063514} (\bibinfo{year}{2014}), \eprint{1312.3558}.

\bibitem[{\citenamefont{{Mukherjee}
  et~al.}(2018{\natexlab{b}})\citenamefont{{Mukherjee}, {Khatri}, and
  {Wandelt}}}]{Mukherjee2018}
\bibinfo{author}{\bibfnamefont{S.}~\bibnamefont{{Mukherjee}}},
  \bibinfo{author}{\bibfnamefont{R.}~\bibnamefont{{Khatri}}}, \bibnamefont{and}
  \bibinfo{author}{\bibfnamefont{B.~D.} \bibnamefont{{Wandelt}}},
  \bibinfo{journal}{\jcap} \textbf{\bibinfo{volume}{4}}, \bibinfo{eid}{045}
  (\bibinfo{year}{2018}{\natexlab{b}}), \eprint{1801.09701}.

\bibitem[{\citenamefont{{Dimastrogiovanni}
  et~al.}(2016)\citenamefont{{Dimastrogiovanni}, {Krauss}, and
  {Chluba}}}]{Dimastrogiovanni2015}
\bibinfo{author}{\bibfnamefont{E.}~\bibnamefont{{Dimastrogiovanni}}},
  \bibinfo{author}{\bibfnamefont{L.~M.} \bibnamefont{{Krauss}}},
  \bibnamefont{and} \bibinfo{author}{\bibfnamefont{J.}~\bibnamefont{{Chluba}}},
  \bibinfo{journal}{Phys.~Rev.} \textbf{\bibinfo{volume}{D94}},
  \bibinfo{eid}{023518} (\bibinfo{year}{2016}).

\bibitem[{\citenamefont{{Ostriker} and {Thompson}}(1987)}]{Ostriker1987}
\bibinfo{author}{\bibfnamefont{J.~P.} \bibnamefont{{Ostriker}}}
  \bibnamefont{and}
  \bibinfo{author}{\bibfnamefont{C.}~\bibnamefont{{Thompson}}},
  \bibinfo{journal}{\apjl} \textbf{\bibinfo{volume}{323}}, \bibinfo{pages}{L97}
  (\bibinfo{year}{1987}).

\bibitem[{\citenamefont{{Tashiro}
  et~al.}(2013{\natexlab{b}})\citenamefont{{Tashiro}, {Sabancilar}, and
  {Vachaspati}}}]{Tashiro2012b}
\bibinfo{author}{\bibfnamefont{H.}~\bibnamefont{{Tashiro}}},
  \bibinfo{author}{\bibfnamefont{E.}~\bibnamefont{{Sabancilar}}},
  \bibnamefont{and}
  \bibinfo{author}{\bibfnamefont{T.}~\bibnamefont{{Vachaspati}}},
  \bibinfo{journal}{\jcap} \textbf{\bibinfo{volume}{8}}, \bibinfo{eid}{035}
  (\bibinfo{year}{2013}{\natexlab{b}}), \eprint{1212.3283}.

\bibitem[{\citenamefont{{Ali-Ha{\"\i}moud}
  et~al.}(2015)\citenamefont{{Ali-Ha{\"\i}moud}, {Chluba}, and
  {Kamionkowski}}}]{Yacine2015DM}
\bibinfo{author}{\bibfnamefont{Y.}~\bibnamefont{{Ali-Ha{\"\i}moud}}},
  \bibinfo{author}{\bibfnamefont{J.}~\bibnamefont{{Chluba}}}, \bibnamefont{and}
  \bibinfo{author}{\bibfnamefont{M.}~\bibnamefont{{Kamionkowski}}},
  \bibinfo{journal}{Phys.~Rev.} \textbf{\bibinfo{volume}{D115}},
  \bibinfo{eid}{071304} (\bibinfo{year}{2015}).

\bibitem[{\citenamefont{{Diacoumis} and {Wong}}(2017)}]{Diacoumis2017}
\bibinfo{author}{\bibfnamefont{J.~A.~D.} \bibnamefont{{Diacoumis}}}
  \bibnamefont{and} \bibinfo{author}{\bibfnamefont{Y.~Y.~Y.}
  \bibnamefont{{Wong}}}, \bibinfo{journal}{\jcap} \textbf{\bibinfo{volume}{9}},
  \bibinfo{eid}{011} (\bibinfo{year}{2017}), \eprint{1707.07050}.

\bibitem[{\citenamefont{{Kumar} et~al.}(2019)}]{Kumar2018}
\bibinfo{author}{\bibfnamefont{S.}~\bibnamefont{{Kumar}}} \bibnamefont{et~al.},
  \bibinfo{journal}{\prd} \textbf{\bibinfo{volume}{99}}, \bibinfo{eid}{023521}
  (\bibinfo{year}{2019}), \eprint{1804.08601}.

\bibitem[{\citenamefont{{Jedamzik} et~al.}(2000)\citenamefont{{Jedamzik},
  {Katalini{\'c}}, and {Olinto}}}]{Jedamzik2000}
\bibinfo{author}{\bibfnamefont{K.}~\bibnamefont{{Jedamzik}}},
  \bibinfo{author}{\bibfnamefont{V.}~\bibnamefont{{Katalini{\'c}}}},
  \bibnamefont{and} \bibinfo{author}{\bibfnamefont{A.~V.}
  \bibnamefont{{Olinto}}}, \bibinfo{journal}{PRL}
  \textbf{\bibinfo{volume}{85}}, \bibinfo{pages}{700} (\bibinfo{year}{2000}),
  \eprint{arXiv:astro-ph/9911100}.

\bibitem[{\citenamefont{{Sethi} and {Subramanian}}(2005)}]{Sethi2005}
\bibinfo{author}{\bibfnamefont{S.~K.} \bibnamefont{{Sethi}}} \bibnamefont{and}
  \bibinfo{author}{\bibfnamefont{K.}~\bibnamefont{{Subramanian}}},
  \bibinfo{journal}{\mnras} \textbf{\bibinfo{volume}{356}},
  \bibinfo{pages}{778} (\bibinfo{year}{2005}), \eprint{astro-ph/0405413}.

\bibitem[{\citenamefont{{Kunze} and {Komatsu}}(2014)}]{Kunze2014}
\bibinfo{author}{\bibfnamefont{K.~E.} \bibnamefont{{Kunze}}} \bibnamefont{and}
  \bibinfo{author}{\bibfnamefont{E.}~\bibnamefont{{Komatsu}}},
  \bibinfo{journal}{\jcap} \textbf{\bibinfo{volume}{1}}, \bibinfo{eid}{009}
  (\bibinfo{year}{2014}), \eprint{1309.7994}.

\bibitem[{\citenamefont{{Wagstaff} and {Banerjee}}(2015)}]{Wagstaff2015}
\bibinfo{author}{\bibfnamefont{J.~M.} \bibnamefont{{Wagstaff}}}
  \bibnamefont{and}
  \bibinfo{author}{\bibfnamefont{R.}~\bibnamefont{{Banerjee}}},
  \bibinfo{journal}{Phys.~Rev.} \textbf{\bibinfo{volume}{D92}},
  \bibinfo{eid}{123004} (\bibinfo{year}{2015}), \eprint{1508.01683}.

\bibitem[{\citenamefont{{Pani} and {Loeb}}(2013)}]{Pani2013}
\bibinfo{author}{\bibfnamefont{P.}~\bibnamefont{{Pani}}} \bibnamefont{and}
  \bibinfo{author}{\bibfnamefont{A.}~\bibnamefont{{Loeb}}},
  \bibinfo{journal}{Phys.~Rev.} \textbf{\bibinfo{volume}{D88}},
  \bibinfo{eid}{041301} (\bibinfo{year}{2013}), \eprint{1307.5176}.

\bibitem[{\citenamefont{{Clesse} and
  {Garc{\'{\i}}a-Bellido}}(2015)}]{Clesse2015PBH}
\bibinfo{author}{\bibfnamefont{S.}~\bibnamefont{{Clesse}}} \bibnamefont{and}
  \bibinfo{author}{\bibfnamefont{J.}~\bibnamefont{{Garc{\'{\i}}a-Bellido}}},
  \bibinfo{journal}{Phys.~Rev.} \textbf{\bibinfo{volume}{D92}},
  \bibinfo{eid}{023524} (\bibinfo{year}{2015}), \eprint{1501.07565}.

\bibitem[{\citenamefont{Nakama et~al.}(2018{\natexlab{a}})\citenamefont{Nakama,
  Carr, and Silk}}]{Nakama2017xvq}
\bibinfo{author}{\bibfnamefont{T.}~\bibnamefont{Nakama}},
  \bibinfo{author}{\bibfnamefont{B.}~\bibnamefont{Carr}}, \bibnamefont{and}
  \bibinfo{author}{\bibfnamefont{J.}~\bibnamefont{Silk}},
  \bibinfo{journal}{\prd} \textbf{\bibinfo{volume}{D97}},
  \bibinfo{pages}{043525} (\bibinfo{year}{2018}{\natexlab{a}}),
  \eprint{1710.06945}.

\bibitem[{\citenamefont{{Bird} et~al.}(2016)}]{Bird2016}
\bibinfo{author}{\bibfnamefont{S.}~\bibnamefont{{Bird}}} \bibnamefont{et~al.},
  \bibinfo{journal}{\prl} \textbf{\bibinfo{volume}{116}}, \bibinfo{eid}{201301}
  (\bibinfo{year}{2016}), \eprint{1603.00464}.

\bibitem[{\citenamefont{Sasaki et~al.}(2016)}]{Sasaki:2016jop}
\bibinfo{author}{\bibfnamefont{M.}~\bibnamefont{Sasaki}} \bibnamefont{et~al.},
  \bibinfo{journal}{Phys. Rev. Lett.} \textbf{\bibinfo{volume}{117}},
  \bibinfo{pages}{061101} (\bibinfo{year}{2016}), \eprint{1603.08338}.

\bibitem[{\citenamefont{Sasaki et~al.}(2018)}]{Sasaki:2018dmp}
\bibinfo{author}{\bibfnamefont{M.}~\bibnamefont{Sasaki}} \bibnamefont{et~al.},
  \bibinfo{journal}{Class. Quant. Grav.} \textbf{\bibinfo{volume}{35}},
  \bibinfo{pages}{063001} (\bibinfo{year}{2018}), \eprint{1801.05235}.

\bibitem[{\citenamefont{Abbott et~al.}(2016)}]{Abbott:2016blz}
\bibinfo{author}{\bibfnamefont{B.~P.} \bibnamefont{Abbott}}
  \bibnamefont{et~al.} (\bibinfo{collaboration}{LIGO Scientific, Virgo}),
  \bibinfo{journal}{Phys. Rev. Lett.} \textbf{\bibinfo{volume}{116}},
  \bibinfo{pages}{061102} (\bibinfo{year}{2016}), \eprint{1602.03837}.

\bibitem[{\citenamefont{Niikura et~al.}(2019)}]{Niikura:2017zjd}
\bibinfo{author}{\bibfnamefont{H.}~\bibnamefont{Niikura}} \bibnamefont{et~al.},
  \bibinfo{journal}{Nat. Astron.} \textbf{\bibinfo{volume}{3}},
  \bibinfo{pages}{524} (\bibinfo{year}{2019}), \eprint{1701.02151}.

\bibitem[{\citenamefont{Carr et~al.}(2016)\citenamefont{Carr, Kuhnel, and
  Sandstad}}]{Carr:2016drx}
\bibinfo{author}{\bibfnamefont{B.}~\bibnamefont{Carr}},
  \bibinfo{author}{\bibfnamefont{F.}~\bibnamefont{Kuhnel}}, \bibnamefont{and}
  \bibinfo{author}{\bibfnamefont{M.}~\bibnamefont{Sandstad}},
  \bibinfo{journal}{Phys. Rev.} \textbf{\bibinfo{volume}{D94}},
  \bibinfo{pages}{083504} (\bibinfo{year}{2016}), \eprint{1607.06077}.

\bibitem[{\citenamefont{Kawasaki et~al.}(2012)\citenamefont{Kawasaki, Kusenko,
  and Yanagida}}]{Kawasaki:2012kn}
\bibinfo{author}{\bibfnamefont{M.}~\bibnamefont{Kawasaki}},
  \bibinfo{author}{\bibfnamefont{A.}~\bibnamefont{Kusenko}}, \bibnamefont{and}
  \bibinfo{author}{\bibfnamefont{T.~T.} \bibnamefont{Yanagida}},
  \bibinfo{journal}{Phys. Lett.} \textbf{\bibinfo{volume}{B711}},
  \bibinfo{pages}{1} (\bibinfo{year}{2012}), \eprint{1202.3848}.

\bibitem[{\citenamefont{{Kohri} et~al.}(2014)\citenamefont{{Kohri}, {Nakama},
  and {Suyama}}}]{Kohri2014SMBH}
\bibinfo{author}{\bibfnamefont{K.}~\bibnamefont{{Kohri}}},
  \bibinfo{author}{\bibfnamefont{T.}~\bibnamefont{{Nakama}}}, \bibnamefont{and}
  \bibinfo{author}{\bibfnamefont{T.}~\bibnamefont{{Suyama}}},
  \bibinfo{journal}{Phys.~Rev.} \textbf{\bibinfo{volume}{D90}},
  \bibinfo{eid}{083514} (\bibinfo{year}{2014}), \eprint{1405.5999}.

\bibitem[{\citenamefont{Kawasaki and Murai}(2019)}]{Kawasaki:2019iis}
\bibinfo{author}{\bibfnamefont{M.}~\bibnamefont{Kawasaki}} \bibnamefont{and}
  \bibinfo{author}{\bibfnamefont{K.}~\bibnamefont{Murai}}
  (\bibinfo{year}{2019}), \eprint{1907.02273}.

\bibitem[{\citenamefont{Carr}(1975)}]{Carr:1975qj}
\bibinfo{author}{\bibfnamefont{B.~J.} \bibnamefont{Carr}},
  \bibinfo{journal}{Astrophys. J.} \textbf{\bibinfo{volume}{201}},
  \bibinfo{pages}{1} (\bibinfo{year}{1975}).

\bibitem[{\citenamefont{Harada et~al.}(2013)}]{Harada:2013epa}
\bibinfo{author}{\bibfnamefont{T.}~\bibnamefont{Harada}} \bibnamefont{et~al.},
  \bibinfo{journal}{\prd} \textbf{\bibinfo{volume}{88}},
  \bibinfo{pages}{084051} (\bibinfo{year}{2013}), \eprint{1309.4201}.

\bibitem[{\citenamefont{Harada et~al.}(2017)}]{Harada:2017fjm}
\bibinfo{author}{\bibfnamefont{T.}~\bibnamefont{Harada}} \bibnamefont{et~al.},
  \bibinfo{journal}{Phys. Rev.} \textbf{\bibinfo{volume}{D96}},
  \bibinfo{pages}{083517} (\bibinfo{year}{2017}), \eprint{1707.03595}.

\bibitem[{\citenamefont{{Inomata} et~al.}(2018)}]{Inomata:2017vxo}
\bibinfo{author}{\bibfnamefont{K.}~\bibnamefont{{Inomata}}}
  \bibnamefont{et~al.}, \bibinfo{journal}{\prd} \textbf{\bibinfo{volume}{97}},
  \bibinfo{eid}{043514} (\bibinfo{year}{2018}), \eprint{1711.06129}.

\bibitem[{\citenamefont{{Lyth}}(2011)}]{Lyth:2011kj}
\bibinfo{author}{\bibfnamefont{D.~H.} \bibnamefont{{Lyth}}},
  \bibinfo{journal}{arXiv e-prints} \bibinfo{eid}{arXiv:1107.1681}
  (\bibinfo{year}{2011}), \eprint{1107.1681}.

\bibitem[{\citenamefont{Kawasaki et~al.}(2013)\citenamefont{Kawasaki, Kitajima,
  and Yanagida}}]{Kawasaki:2012wr}
\bibinfo{author}{\bibfnamefont{M.}~\bibnamefont{Kawasaki}},
  \bibinfo{author}{\bibfnamefont{N.}~\bibnamefont{Kitajima}}, \bibnamefont{and}
  \bibinfo{author}{\bibfnamefont{T.~T.} \bibnamefont{Yanagida}},
  \bibinfo{journal}{Phys. Rev.} \textbf{\bibinfo{volume}{D87}},
  \bibinfo{pages}{063519} (\bibinfo{year}{2013}), \eprint{1207.2550}.

\bibitem[{\citenamefont{Kohri et~al.}(2013)\citenamefont{Kohri, Lin, and
  Matsuda}}]{Kohri:2012yw}
\bibinfo{author}{\bibfnamefont{K.}~\bibnamefont{Kohri}},
  \bibinfo{author}{\bibfnamefont{C.}~\bibnamefont{Lin}}, \bibnamefont{and}
  \bibinfo{author}{\bibfnamefont{T.}~\bibnamefont{Matsuda}},
  \bibinfo{journal}{Phys. Rev.} \textbf{\bibinfo{volume}{D87}},
  \bibinfo{pages}{103527} (\bibinfo{year}{2013}), \eprint{1211.2371}.

\bibitem[{\citenamefont{Frampton et~al.}(2010)\citenamefont{Frampton, Kawasaki,
  Takahashi, and Yanagida}}]{Frampton:2010sw}
\bibinfo{author}{\bibfnamefont{P.~H.} \bibnamefont{Frampton}},
  \bibinfo{author}{\bibfnamefont{M.}~\bibnamefont{Kawasaki}},
  \bibinfo{author}{\bibfnamefont{F.}~\bibnamefont{Takahashi}},
  \bibnamefont{and} \bibinfo{author}{\bibfnamefont{T.~T.}
  \bibnamefont{Yanagida}}, \bibinfo{journal}{JCAP}
  \textbf{\bibinfo{volume}{1004}}, \bibinfo{pages}{023} (\bibinfo{year}{2010}),
  \eprint{1001.2308}.

\bibitem[{\citenamefont{Josan et~al.}(2009)\citenamefont{Josan, Green, and
  Malik}}]{Josan:2009qn}
\bibinfo{author}{\bibfnamefont{A.~S.} \bibnamefont{Josan}},
  \bibinfo{author}{\bibfnamefont{A.~M.} \bibnamefont{Green}}, \bibnamefont{and}
  \bibinfo{author}{\bibfnamefont{K.~A.} \bibnamefont{Malik}},
  \bibinfo{journal}{Phys. Rev.} \textbf{\bibinfo{volume}{D79}},
  \bibinfo{pages}{103520} (\bibinfo{year}{2009}), \eprint{0903.3184}.

\bibitem[{\citenamefont{St{\"o}cker et~al.}(2018)\citenamefont{St{\"o}cker,
  Kr{\"a}mer, Lesgourgues, and Poulin}}]{Stocker:2018avm}
\bibinfo{author}{\bibfnamefont{P.}~\bibnamefont{St{\"o}cker}},
  \bibinfo{author}{\bibfnamefont{M.}~\bibnamefont{Kr{\"a}mer}},
  \bibinfo{author}{\bibfnamefont{J.}~\bibnamefont{Lesgourgues}},
  \bibnamefont{and} \bibinfo{author}{\bibfnamefont{V.}~\bibnamefont{Poulin}},
  \bibinfo{journal}{JCAP} \textbf{\bibinfo{volume}{1803}}, \bibinfo{pages}{018}
  (\bibinfo{year}{2018}), \eprint{1801.01871}.

\bibitem[{\citenamefont{Ali-Haïmoud and
  Kamionkowski}(2017)}]{Ali-Haimoud:2016mbv}
\bibinfo{author}{\bibfnamefont{Y.}~\bibnamefont{Ali-Haïmoud}}
  \bibnamefont{and}
  \bibinfo{author}{\bibfnamefont{M.}~\bibnamefont{Kamionkowski}},
  \bibinfo{journal}{Phys. Rev.} \textbf{\bibinfo{volume}{D95}},
  \bibinfo{pages}{043534} (\bibinfo{year}{2017}), \eprint{1612.05644}.

\bibitem[{\citenamefont{Poulin et~al.}(2017)}]{Poulin:2017bwe}
\bibinfo{author}{\bibfnamefont{V.}~\bibnamefont{Poulin}} \bibnamefont{et~al.},
  \bibinfo{journal}{Phys. Rev.} \textbf{\bibinfo{volume}{D96}},
  \bibinfo{pages}{083524} (\bibinfo{year}{2017}), \eprint{1707.04206}.

\bibitem[{\citenamefont{{Jeong} et~al.}(2014)}]{Jeong2014}
\bibinfo{author}{\bibfnamefont{D.}~\bibnamefont{{Jeong}}} \bibnamefont{et~al.},
  \bibinfo{journal}{\prl} \textbf{\bibinfo{volume}{113}}, \bibinfo{eid}{061301}
  (\bibinfo{year}{2014}), \eprint{1403.3697}.

\bibitem[{\citenamefont{Nakama et~al.}(2014)\citenamefont{Nakama, Suyama, and
  Yokoyama}}]{Nakama:2014vla}
\bibinfo{author}{\bibfnamefont{T.}~\bibnamefont{Nakama}},
  \bibinfo{author}{\bibfnamefont{T.}~\bibnamefont{Suyama}}, \bibnamefont{and}
  \bibinfo{author}{\bibfnamefont{J.}~\bibnamefont{Yokoyama}},
  \bibinfo{journal}{Phys. Rev. Lett.} \textbf{\bibinfo{volume}{113}},
  \bibinfo{pages}{061302} (\bibinfo{year}{2014}), \eprint{1403.5407}.

\bibitem[{\citenamefont{Inomata et~al.}(2016)\citenamefont{Inomata, Kawasaki,
  and Tada}}]{Inomata:2016uip}
\bibinfo{author}{\bibfnamefont{K.}~\bibnamefont{Inomata}},
  \bibinfo{author}{\bibfnamefont{M.}~\bibnamefont{Kawasaki}}, \bibnamefont{and}
  \bibinfo{author}{\bibfnamefont{Y.}~\bibnamefont{Tada}},
  \bibinfo{journal}{Phys. Rev.} \textbf{\bibinfo{volume}{D94}},
  \bibinfo{pages}{043527} (\bibinfo{year}{2016}), \eprint{1605.04646}.

\bibitem[{\citenamefont{Nakama et~al.}(2018{\natexlab{b}})\citenamefont{Nakama,
  Carr, and Silk}}]{Nakama:2017xvq}
\bibinfo{author}{\bibfnamefont{T.}~\bibnamefont{Nakama}},
  \bibinfo{author}{\bibfnamefont{B.}~\bibnamefont{Carr}}, \bibnamefont{and}
  \bibinfo{author}{\bibfnamefont{J.}~\bibnamefont{Silk}},
  \bibinfo{journal}{Phys. Rev.} \textbf{\bibinfo{volume}{D97}},
  \bibinfo{pages}{043525} (\bibinfo{year}{2018}{\natexlab{b}}),
  \eprint{1710.06945}.

\bibitem[{\citenamefont{{Inman} and {Ali-Ha{\"\i}moud}}(2019)}]{Inman2019}
\bibinfo{author}{\bibfnamefont{D.}~\bibnamefont{{Inman}}} \bibnamefont{and}
  \bibinfo{author}{\bibfnamefont{Y.}~\bibnamefont{{Ali-Ha{\"\i}moud}}},
  \bibinfo{journal}{arXiv e-prints} \bibinfo{eid}{arXiv:1907.08129}
  (\bibinfo{year}{2019}), \eprint{1907.08129}.

\bibitem[{\citenamefont{Tashiro and Sugiyama}(2008)}]{Tashiro:2008sf}
\bibinfo{author}{\bibfnamefont{H.}~\bibnamefont{Tashiro}} \bibnamefont{and}
  \bibinfo{author}{\bibfnamefont{N.}~\bibnamefont{Sugiyama}},
  \bibinfo{journal}{Phys. Rev.} \textbf{\bibinfo{volume}{D78}},
  \bibinfo{pages}{023004} (\bibinfo{year}{2008}), \eprint{0801.3172}.

\bibitem[{\citenamefont{Peccei and Quinn}(1977)}]{PhysRevLett.38.1440}
\bibinfo{author}{\bibfnamefont{R.~D.} \bibnamefont{Peccei}} \bibnamefont{and}
  \bibinfo{author}{\bibfnamefont{H.~R.} \bibnamefont{Quinn}},
  \bibinfo{journal}{\prl} \textbf{\bibinfo{volume}{38}}, \bibinfo{pages}{1440}
  (\bibinfo{year}{1977}).

\bibitem[{\citenamefont{Weinberg}(1978)}]{PhysRevLett.40.223}
\bibinfo{author}{\bibfnamefont{S.}~\bibnamefont{Weinberg}},
  \bibinfo{journal}{\prl} \textbf{\bibinfo{volume}{40}}, \bibinfo{pages}{223}
  (\bibinfo{year}{1978}).

\bibitem[{\citenamefont{Wilczek}(1978)}]{PhysRevLett.40.279}
\bibinfo{author}{\bibfnamefont{F.}~\bibnamefont{Wilczek}},
  \bibinfo{journal}{\prl} \textbf{\bibinfo{volume}{40}}, \bibinfo{pages}{279}
  (\bibinfo{year}{1978}).

\bibitem[{\citenamefont{Svrcek and Witten}(2006)}]{Svrcek:2006yi}
\bibinfo{author}{\bibfnamefont{P.}~\bibnamefont{Svrcek}} \bibnamefont{and}
  \bibinfo{author}{\bibfnamefont{E.}~\bibnamefont{Witten}},
  \bibinfo{journal}{JHEP} \textbf{\bibinfo{volume}{06}}, \bibinfo{pages}{051}
  (\bibinfo{year}{2006}), \eprint{hep-th/0605206}.

\bibitem[{\citenamefont{Arvanitaki et~al.}(2010)}]{Arvanitaki:2009fg}
\bibinfo{author}{\bibfnamefont{A.}~\bibnamefont{Arvanitaki}}
  \bibnamefont{et~al.}, \bibinfo{journal}{\prd} \textbf{\bibinfo{volume}{81}},
  \bibinfo{pages}{123530} (\bibinfo{year}{2010}), \eprint{0905.4720}.

\bibitem[{\citenamefont{Acharya et~al.}(2010)\citenamefont{Acharya, Bobkov, and
  Kumar}}]{Acharya:2010zx}
\bibinfo{author}{\bibfnamefont{B.~S.} \bibnamefont{Acharya}},
  \bibinfo{author}{\bibfnamefont{K.}~\bibnamefont{Bobkov}}, \bibnamefont{and}
  \bibinfo{author}{\bibfnamefont{P.}~\bibnamefont{Kumar}},
  \bibinfo{journal}{JHEP} \textbf{\bibinfo{volume}{11}}, \bibinfo{pages}{105}
  (\bibinfo{year}{2010}), \eprint{1004.5138}.

\bibitem[{\citenamefont{{Graham} et~al.}(2015)}]{2015ARNPS..65..485G}
\bibinfo{author}{\bibfnamefont{P.~W.} \bibnamefont{{Graham}}}
  \bibnamefont{et~al.}, \bibinfo{journal}{Annual Review of Nuclear and Particle
  Science} \textbf{\bibinfo{volume}{65}}, \bibinfo{pages}{485}
  (\bibinfo{year}{2015}), \eprint{1602.00039}.

\bibitem[{\citenamefont{{Ruz} et~al.}(2015)\citenamefont{{Ruz}, {Vogel}, and
  {Pivovaroff}}}]{2015PhPro..61..153R}
\bibinfo{author}{\bibfnamefont{J.}~\bibnamefont{{Ruz}}},
  \bibinfo{author}{\bibfnamefont{J.~K.} \bibnamefont{{Vogel}}},
  \bibnamefont{and} \bibinfo{author}{\bibfnamefont{M.~J.}
  \bibnamefont{{Pivovaroff}}}, \bibinfo{journal}{Physics Procedia}
  \textbf{\bibinfo{volume}{61}}, \bibinfo{pages}{153} (\bibinfo{year}{2015}).

\bibitem[{\citenamefont{{Bastidon} and {for the ALPS
  collaboration}}(2015)}]{Bastidon:2015efa}
\bibinfo{author}{\bibfnamefont{N.}~\bibnamefont{{Bastidon}}} \bibnamefont{and}
  \bibinfo{author}{\bibfnamefont{I.}~\bibnamefont{{for the ALPS
  collaboration}}} (\bibinfo{year}{2015}), \eprint{1509.02070}.

\bibitem[{\citenamefont{{Majorovits} et~al.}(2016)}]{Majorovits:2016yvk}
\bibinfo{author}{\bibfnamefont{B.}~\bibnamefont{{Majorovits}}}
  \bibnamefont{et~al.} (\bibinfo{year}{2016}), \eprint{1611.04549}.

\bibitem[{\citenamefont{{Asztalos} et~al.}(2010)}]{2010PhRvL.104d1301A}
\bibinfo{author}{\bibfnamefont{S.~J.} \bibnamefont{{Asztalos}}}
  \bibnamefont{et~al.}, \bibinfo{journal}{\prl} \textbf{\bibinfo{volume}{104}},
  \bibinfo{eid}{041301} (\bibinfo{year}{2010}), \eprint{0910.5914}.

\bibitem[{\citenamefont{{Budker} et~al.}(2014)}]{Budker:2013hfa}
\bibinfo{author}{\bibfnamefont{D.}~\bibnamefont{{Budker}}}
  \bibnamefont{et~al.}, \bibinfo{journal}{Physical Review X}
  \textbf{\bibinfo{volume}{4}}, \bibinfo{eid}{021030} (\bibinfo{year}{2014}),
  \eprint{1306.6089}.

\bibitem[{\citenamefont{Hu et~al.}(2000)\citenamefont{Hu, Barkana, and
  Gruzinov}}]{Hu:2000ke}
\bibinfo{author}{\bibfnamefont{W.}~\bibnamefont{Hu}},
  \bibinfo{author}{\bibfnamefont{R.}~\bibnamefont{Barkana}}, \bibnamefont{and}
  \bibinfo{author}{\bibfnamefont{A.}~\bibnamefont{Gruzinov}},
  \bibinfo{journal}{\prl} \textbf{\bibinfo{volume}{85}}, \bibinfo{pages}{1158}
  (\bibinfo{year}{2000}), \eprint{astro-ph/0003365}.

\bibitem[{\citenamefont{Park et~al.}(2012)\citenamefont{Park, Hwang, and
  Noh}}]{PhysRevD.86.083535}
\bibinfo{author}{\bibfnamefont{C.-G.} \bibnamefont{Park}},
  \bibinfo{author}{\bibfnamefont{J.-c.} \bibnamefont{Hwang}}, \bibnamefont{and}
  \bibinfo{author}{\bibfnamefont{H.}~\bibnamefont{Noh}},
  \bibinfo{journal}{\prd} \textbf{\bibinfo{volume}{86}},
  \bibinfo{pages}{083535} (\bibinfo{year}{2012}).

\bibitem[{\citenamefont{Hlozek et~al.}(2015)\citenamefont{Hlozek, Grin, Marsh,
  and Ferreira}}]{Hlozek:2014lca}
\bibinfo{author}{\bibfnamefont{R.}~\bibnamefont{Hlozek}},
  \bibinfo{author}{\bibfnamefont{D.}~\bibnamefont{Grin}},
  \bibinfo{author}{\bibfnamefont{D.~J.~E.} \bibnamefont{Marsh}},
  \bibnamefont{and} \bibinfo{author}{\bibfnamefont{P.~G.}
  \bibnamefont{Ferreira}}, \bibinfo{journal}{\prd}
  \textbf{\bibinfo{volume}{91}}, \bibinfo{pages}{103512}
  (\bibinfo{year}{2015}), \eprint{1410.2896}.

\bibitem[{\citenamefont{Hlozek and others}(2017)\citenamefont{Hlozek
  et~al.}}]{Hlozek:2016lzm}
\bibinfo{author}{\bibfnamefont{R.}~\bibnamefont{Hlozek}} \bibnamefont{et~al.},
  \bibinfo{journal}{\prd} \textbf{\bibinfo{volume}{95}},
  \bibinfo{pages}{123511} (\bibinfo{year}{2017}), \eprint{1607.08208}.

\bibitem[{\citenamefont{Hlozek et~al.}(2018)\citenamefont{Hlozek, Marsh, and
  Grin}}]{Hlozek:2017zzf}
\bibinfo{author}{\bibfnamefont{R.}~\bibnamefont{Hlozek}},
  \bibinfo{author}{\bibfnamefont{D.~J.~E.} \bibnamefont{Marsh}},
  \bibnamefont{and} \bibinfo{author}{\bibfnamefont{D.}~\bibnamefont{Grin}},
  \bibinfo{journal}{\mnras} \textbf{\bibinfo{volume}{476}},
  \bibinfo{pages}{3063} (\bibinfo{year}{2018}), \eprint{1708.05681}.

\bibitem[{\citenamefont{Sikivie}(1983)}]{PhysRevLett.51.1415}
\bibinfo{author}{\bibfnamefont{P.}~\bibnamefont{Sikivie}},
  \bibinfo{journal}{\prl} \textbf{\bibinfo{volume}{51}}, \bibinfo{pages}{1415}
  (\bibinfo{year}{1983}).

\bibitem[{\citenamefont{Raffelt and Stodolsky}(1988)}]{PhysRevD.37.1237}
\bibinfo{author}{\bibfnamefont{G.}~\bibnamefont{Raffelt}} \bibnamefont{and}
  \bibinfo{author}{\bibfnamefont{L.}~\bibnamefont{Stodolsky}},
  \bibinfo{journal}{\prd} \textbf{\bibinfo{volume}{37}}, \bibinfo{pages}{1237}
  (\bibinfo{year}{1988}).

\bibitem[{\citenamefont{Anselm}(1988)}]{PhysRevD.37.2001}
\bibinfo{author}{\bibfnamefont{A.~A.} \bibnamefont{Anselm}},
  \bibinfo{journal}{\prd} \textbf{\bibinfo{volume}{37}}, \bibinfo{pages}{2001}
  (\bibinfo{year}{1988}).

\bibitem[{\citenamefont{Cordes and Lazio}(2002)}]{Cordes:2002wz}
\bibinfo{author}{\bibfnamefont{J.~M.} \bibnamefont{Cordes}} \bibnamefont{and}
  \bibinfo{author}{\bibfnamefont{T.~J.~W.} \bibnamefont{Lazio}}
  (\bibinfo{year}{2002}), \eprint{astro-ph/0207156}.

\bibitem[{\citenamefont{{Jansson} and
  {Farrar}}(2012{\natexlab{a}})}]{2012ApJ...757...14J}
\bibinfo{author}{\bibfnamefont{R.}~\bibnamefont{{Jansson}}} \bibnamefont{and}
  \bibinfo{author}{\bibfnamefont{G.~R.} \bibnamefont{{Farrar}}},
  \bibinfo{journal}{\apj} \textbf{\bibinfo{volume}{757}}, \bibinfo{eid}{14}
  (\bibinfo{year}{2012}{\natexlab{a}}), \eprint{1204.3662}.

\bibitem[{\citenamefont{{Jansson} and
  {Farrar}}(2012{\natexlab{b}})}]{2012ApJ...761L..11J}
\bibinfo{author}{\bibfnamefont{R.}~\bibnamefont{{Jansson}}} \bibnamefont{and}
  \bibinfo{author}{\bibfnamefont{G.~R.} \bibnamefont{{Farrar}}},
  \bibinfo{journal}{\apjl} \textbf{\bibinfo{volume}{761}}, \bibinfo{eid}{L11}
  (\bibinfo{year}{2012}{\natexlab{b}}), \eprint{1210.7820}.

\bibitem[{\citenamefont{Mukherjee
  et~al.}(2019{\natexlab{a}})\citenamefont{Mukherjee, Spergel, Khatri, and
  Wandelt}}]{Mukherjee:2019dsu}
\bibinfo{author}{\bibfnamefont{S.}~\bibnamefont{Mukherjee}},
  \bibinfo{author}{\bibfnamefont{D.~N.} \bibnamefont{Spergel}},
  \bibinfo{author}{\bibfnamefont{R.}~\bibnamefont{Khatri}}, \bibnamefont{and}
  \bibinfo{author}{\bibfnamefont{B.~D.} \bibnamefont{Wandelt}}
  (\bibinfo{year}{2019}{\natexlab{a}}), \eprint{1908.07534}.

\bibitem[{\citenamefont{{Delabrouille} et~al.}(2019)}]{Delabrouille2019WP}
\bibinfo{author}{\bibfnamefont{J.}~\bibnamefont{{Delabrouille}}}
  \bibnamefont{et~al.}, \bibinfo{journal}{Voyage 2050 Survey}
  (\bibinfo{year}{2019}).

\bibitem[{\citenamefont{Mukherjee
  et~al.}(2019{\natexlab{b}})\citenamefont{Mukherjee, Khatri, and
  Wandelt}}]{Mukherjee:2018zzg}
\bibinfo{author}{\bibfnamefont{S.}~\bibnamefont{Mukherjee}},
  \bibinfo{author}{\bibfnamefont{R.}~\bibnamefont{Khatri}}, \bibnamefont{and}
  \bibinfo{author}{\bibfnamefont{B.~D.} \bibnamefont{Wandelt}},
  \bibinfo{journal}{\jcap} \textbf{\bibinfo{volume}{1906}},
  \bibinfo{pages}{031} (\bibinfo{year}{2019}{\natexlab{b}}),
  \eprint{1811.11177}.

\bibitem[{\citenamefont{{Rybicki} and {dell'Antonio}}(1994)}]{RybickiDell94}
\bibinfo{author}{\bibfnamefont{G.~B.} \bibnamefont{{Rybicki}}}
  \bibnamefont{and} \bibinfo{author}{\bibfnamefont{I.~P.}
  \bibnamefont{{dell'Antonio}}}, \bibinfo{journal}{\apj}
  \textbf{\bibinfo{volume}{427}}, \bibinfo{pages}{603} (\bibinfo{year}{1994}),
  \eprint{astro-ph/9312006}.

\bibitem[{\citenamefont{{Dubrovich} and {Stolyarov}}(1995)}]{DubroVlad95}
\bibinfo{author}{\bibfnamefont{V.~K.} \bibnamefont{{Dubrovich}}}
  \bibnamefont{and} \bibinfo{author}{\bibfnamefont{V.~A.}
  \bibnamefont{{Stolyarov}}}, \bibinfo{journal}{\aap}
  \textbf{\bibinfo{volume}{302}}, \bibinfo{pages}{635} (\bibinfo{year}{1995}).

\bibitem[{\citenamefont{Kholupenko et~al.}(2005)}]{Kholu2005}
\bibinfo{author}{\bibfnamefont{E.~E.} \bibnamefont{Kholupenko}}
  \bibnamefont{et~al.}, \bibinfo{journal}{Gravitation and Cosmology}
  \textbf{\bibinfo{volume}{11}}, \bibinfo{pages}{161} (\bibinfo{year}{2005}),
  \eprint{astro-ph/0509807}.

\bibitem[{\citenamefont{{Wong} et~al.}(2006)\citenamefont{{Wong}, {Seager}, and
  {Scott}}}]{Wong2006}
\bibinfo{author}{\bibfnamefont{W.~Y.} \bibnamefont{{Wong}}},
  \bibinfo{author}{\bibfnamefont{S.}~\bibnamefont{{Seager}}}, \bibnamefont{and}
  \bibinfo{author}{\bibfnamefont{D.}~\bibnamefont{{Scott}}},
  \bibinfo{journal}{\mnras} \textbf{\bibinfo{volume}{367}},
  \bibinfo{pages}{1666} (\bibinfo{year}{2006}), \eprint{astro-ph/0510634}.

\bibitem[{\citenamefont{{Rubi{\~n}o-Mart{\'{\i}}n}
  et~al.}(2006)\citenamefont{{Rubi{\~n}o-Mart{\'{\i}}n}, {Chluba}, and
  {Sunyaev}}}]{Jose2006}
\bibinfo{author}{\bibfnamefont{J.~A.}
  \bibnamefont{{Rubi{\~n}o-Mart{\'{\i}}n}}},
  \bibinfo{author}{\bibfnamefont{J.}~\bibnamefont{{Chluba}}}, \bibnamefont{and}
  \bibinfo{author}{\bibfnamefont{R.~A.} \bibnamefont{{Sunyaev}}},
  \bibinfo{journal}{\mnras} \textbf{\bibinfo{volume}{371}},
  \bibinfo{pages}{1939} (\bibinfo{year}{2006}).

\bibitem[{\citenamefont{{Chluba} and {Sunyaev}}(2006)}]{Chluba2006}
\bibinfo{author}{\bibfnamefont{J.}~\bibnamefont{{Chluba}}} \bibnamefont{and}
  \bibinfo{author}{\bibfnamefont{R.~A.} \bibnamefont{{Sunyaev}}},
  \bibinfo{journal}{\aap} \textbf{\bibinfo{volume}{446}}, \bibinfo{pages}{39}
  (\bibinfo{year}{2006}), \eprint{astro-ph/0508144}.

\bibitem[{\citenamefont{{Rubi{\~n}o-Mart{\'{\i}}n}
  et~al.}(2008)\citenamefont{{Rubi{\~n}o-Mart{\'{\i}}n}, {Chluba}, and
  {Sunyaev}}}]{Jose2008}
\bibinfo{author}{\bibfnamefont{J.~A.}
  \bibnamefont{{Rubi{\~n}o-Mart{\'{\i}}n}}},
  \bibinfo{author}{\bibfnamefont{J.}~\bibnamefont{{Chluba}}}, \bibnamefont{and}
  \bibinfo{author}{\bibfnamefont{R.~A.} \bibnamefont{{Sunyaev}}},
  \bibinfo{journal}{\aap} \textbf{\bibinfo{volume}{485}}, \bibinfo{pages}{377}
  (\bibinfo{year}{2008}).

\bibitem[{\citenamefont{{Ali-Ha{\"i}moud}}(2013)}]{Yacine2013RecSpec}
\bibinfo{author}{\bibfnamefont{Y.}~\bibnamefont{{Ali-Ha{\"i}moud}}},
  \bibinfo{journal}{\prd} \textbf{\bibinfo{volume}{87}}, \bibinfo{eid}{023526}
  (\bibinfo{year}{2013}), \eprint{1211.4031}.

\bibitem[{\citenamefont{{Chluba} and {Sunyaev}}(2008)}]{Chluba2008T0}
\bibinfo{author}{\bibfnamefont{J.}~\bibnamefont{{Chluba}}} \bibnamefont{and}
  \bibinfo{author}{\bibfnamefont{R.~A.} \bibnamefont{{Sunyaev}}},
  \bibinfo{journal}{\aap} \textbf{\bibinfo{volume}{478}}, \bibinfo{pages}{L27}
  (\bibinfo{year}{2008}), \eprint{0707.0188}.

\bibitem[{\citenamefont{{Zeldovich} et~al.}(1968)\citenamefont{{Zeldovich},
  {Kurt}, and {Syunyaev}}}]{Zeldovich68}
\bibinfo{author}{\bibfnamefont{Y.~B.} \bibnamefont{{Zeldovich}}},
  \bibinfo{author}{\bibfnamefont{V.~G.} \bibnamefont{{Kurt}}},
  \bibnamefont{and} \bibinfo{author}{\bibfnamefont{R.~A.}
  \bibnamefont{{Syunyaev}}}, \bibinfo{journal}{ZhETF}
  \textbf{\bibinfo{volume}{55}}, \bibinfo{pages}{278} (\bibinfo{year}{1968}).

\bibitem[{\citenamefont{{Peebles}}(1968)}]{Peebles68}
\bibinfo{author}{\bibfnamefont{P.~J.~E.} \bibnamefont{{Peebles}}},
  \bibinfo{journal}{\apj} \textbf{\bibinfo{volume}{153}}, \bibinfo{pages}{1}
  (\bibinfo{year}{1968}).

\bibitem[{\citenamefont{{Sunyaev} and
  {Zeldovich}}(1970{\natexlab{d}})}]{Sunyaev1970}
\bibinfo{author}{\bibfnamefont{R.~A.} \bibnamefont{{Sunyaev}}}
  \bibnamefont{and} \bibinfo{author}{\bibfnamefont{Y.~B.}
  \bibnamefont{{Zeldovich}}}, \bibinfo{journal}{\apss}
  \textbf{\bibinfo{volume}{7}}, \bibinfo{pages}{3}
  (\bibinfo{year}{1970}{\natexlab{d}}).

\bibitem[{\citenamefont{{Peebles} and {Yu}}(1970)}]{Peebles1970}
\bibinfo{author}{\bibfnamefont{P.~J.~E.} \bibnamefont{{Peebles}}}
  \bibnamefont{and} \bibinfo{author}{\bibfnamefont{J.~T.} \bibnamefont{{Yu}}},
  \bibinfo{journal}{\apj} \textbf{\bibinfo{volume}{162}}, \bibinfo{pages}{815}
  (\bibinfo{year}{1970}).

\bibitem[{\citenamefont{{Hu} et~al.}(1995)\citenamefont{{Hu}, {Scott},
  {Sugiyama}, and {White}}}]{Hu1995}
\bibinfo{author}{\bibfnamefont{W.}~\bibnamefont{{Hu}}},
  \bibinfo{author}{\bibfnamefont{D.}~\bibnamefont{{Scott}}},
  \bibinfo{author}{\bibfnamefont{N.}~\bibnamefont{{Sugiyama}}},
  \bibnamefont{and} \bibinfo{author}{\bibfnamefont{M.}~\bibnamefont{{White}}},
  \bibinfo{journal}{\prd} \textbf{\bibinfo{volume}{52}}, \bibinfo{pages}{5498}
  (\bibinfo{year}{1995}).

\bibitem[{\citenamefont{{Lewis} et~al.}(2006)\citenamefont{{Lewis}, {Weller},
  and {Battye}}}]{Lewis2006}
\bibinfo{author}{\bibfnamefont{A.}~\bibnamefont{{Lewis}}},
  \bibinfo{author}{\bibfnamefont{J.}~\bibnamefont{{Weller}}}, \bibnamefont{and}
  \bibinfo{author}{\bibfnamefont{R.}~\bibnamefont{{Battye}}},
  \bibinfo{journal}{\mnras} \textbf{\bibinfo{volume}{373}},
  \bibinfo{pages}{561} (\bibinfo{year}{2006}), \eprint{astro-ph/0606552}.

\bibitem[{\citenamefont{{Rubi{\~n}o-Mart{\'{\i}}n}
  et~al.}(2010)\citenamefont{{Rubi{\~n}o-Mart{\'{\i}}n}, {Chluba}, {Fendt}, and
  {Wandelt}}}]{Jose2010}
\bibinfo{author}{\bibfnamefont{J.~A.}
  \bibnamefont{{Rubi{\~n}o-Mart{\'{\i}}n}}},
  \bibinfo{author}{\bibfnamefont{J.}~\bibnamefont{{Chluba}}},
  \bibinfo{author}{\bibfnamefont{W.~A.} \bibnamefont{{Fendt}}},
  \bibnamefont{and} \bibinfo{author}{\bibfnamefont{B.~D.}
  \bibnamefont{{Wandelt}}}, \bibinfo{journal}{\mnras}
  \textbf{\bibinfo{volume}{403}}, \bibinfo{pages}{439} (\bibinfo{year}{2010}).

\bibitem[{\citenamefont{{Shaw} and {Chluba}}(2011)}]{Shaw2011}
\bibinfo{author}{\bibfnamefont{J.~R.} \bibnamefont{{Shaw}}} \bibnamefont{and}
  \bibinfo{author}{\bibfnamefont{J.}~\bibnamefont{{Chluba}}},
  \bibinfo{journal}{\mnras} \textbf{\bibinfo{volume}{415}},
  \bibinfo{pages}{1343} (\bibinfo{year}{2011}), \eprint{1102.3683}.

\bibitem[{\citenamefont{{Chluba} et~al.}(2010)\citenamefont{{Chluba}, {Vasil},
  and {Dursi}}}]{Chluba2010}
\bibinfo{author}{\bibfnamefont{J.}~\bibnamefont{{Chluba}}},
  \bibinfo{author}{\bibfnamefont{G.~M.} \bibnamefont{{Vasil}}},
  \bibnamefont{and} \bibinfo{author}{\bibfnamefont{L.~J.}
  \bibnamefont{{Dursi}}}, \bibinfo{journal}{\mnras}
  \textbf{\bibinfo{volume}{407}}, \bibinfo{pages}{599} (\bibinfo{year}{2010}),
  \eprint{1003.4928}.

\bibitem[{\citenamefont{{Desjacques} et~al.}(2015)\citenamefont{{Desjacques},
  {Chluba}, {Silk}, {de Bernardis}, and {Dor{\'e}}}}]{Vince2015}
\bibinfo{author}{\bibfnamefont{V.}~\bibnamefont{{Desjacques}}},
  \bibinfo{author}{\bibfnamefont{J.}~\bibnamefont{{Chluba}}},
  \bibinfo{author}{\bibfnamefont{J.}~\bibnamefont{{Silk}}},
  \bibinfo{author}{\bibfnamefont{F.}~\bibnamefont{{de Bernardis}}},
  \bibnamefont{and}
  \bibinfo{author}{\bibfnamefont{O.}~\bibnamefont{{Dor{\'e}}}},
  \bibinfo{journal}{\mnras} \textbf{\bibinfo{volume}{451}},
  \bibinfo{pages}{4460} (\bibinfo{year}{2015}).

\bibitem[{\citenamefont{{Sathyanarayana Rao}
  et~al.}(2015)\citenamefont{{Sathyanarayana Rao}, {Subrahmanyan}, {Udaya
  Shankar}, and {Chluba}}}]{Mayuri2015}
\bibinfo{author}{\bibfnamefont{M.}~\bibnamefont{{Sathyanarayana Rao}}},
  \bibinfo{author}{\bibfnamefont{R.}~\bibnamefont{{Subrahmanyan}}},
  \bibinfo{author}{\bibfnamefont{N.}~\bibnamefont{{Udaya Shankar}}},
  \bibnamefont{and} \bibinfo{author}{\bibfnamefont{J.}~\bibnamefont{{Chluba}}},
  \bibinfo{journal}{\apj} \textbf{\bibinfo{volume}{810}}, \bibinfo{eid}{3}
  (\bibinfo{year}{2015}).

\bibitem[{\citenamefont{{Sunyaev} and
  {Zeldovich}}(1972{\natexlab{b}})}]{Sunyaev1972b}
\bibinfo{author}{\bibfnamefont{R.~A.} \bibnamefont{{Sunyaev}}}
  \bibnamefont{and} \bibinfo{author}{\bibfnamefont{Y.~B.}
  \bibnamefont{{Zeldovich}}}, \bibinfo{journal}{\aap}
  \textbf{\bibinfo{volume}{20}}, \bibinfo{pages}{189}
  (\bibinfo{year}{1972}{\natexlab{b}}).

\bibitem[{\citenamefont{{Hu} et~al.}(1994{\natexlab{b}})\citenamefont{{Hu},
  {Scott}, and {Silk}}}]{Hu1994pert}
\bibinfo{author}{\bibfnamefont{W.}~\bibnamefont{{Hu}}},
  \bibinfo{author}{\bibfnamefont{D.}~\bibnamefont{{Scott}}}, \bibnamefont{and}
  \bibinfo{author}{\bibfnamefont{J.}~\bibnamefont{{Silk}}},
  \bibinfo{journal}{\prd} \textbf{\bibinfo{volume}{49}}, \bibinfo{pages}{648}
  (\bibinfo{year}{1994}{\natexlab{b}}), \eprint{arXiv:astro-ph/9305038}.

\bibitem[{\citenamefont{{Cen} and {Ostriker}}(1999)}]{Cen1999}
\bibinfo{author}{\bibfnamefont{R.}~\bibnamefont{{Cen}}} \bibnamefont{and}
  \bibinfo{author}{\bibfnamefont{J.~P.} \bibnamefont{{Ostriker}}},
  \bibinfo{journal}{\apj} \textbf{\bibinfo{volume}{514}}, \bibinfo{pages}{1}
  (\bibinfo{year}{1999}), \eprint{arXiv:astro-ph/9806281}.

\bibitem[{\citenamefont{{Miniati} et~al.}(2000)\citenamefont{{Miniati}, {Ryu},
  {Kang}, {Jones}, {Cen}, and {Ostriker}}}]{Miniati2000}
\bibinfo{author}{\bibfnamefont{F.}~\bibnamefont{{Miniati}}},
  \bibinfo{author}{\bibfnamefont{D.}~\bibnamefont{{Ryu}}},
  \bibinfo{author}{\bibfnamefont{H.}~\bibnamefont{{Kang}}},
  \bibinfo{author}{\bibfnamefont{T.~W.} \bibnamefont{{Jones}}},
  \bibinfo{author}{\bibfnamefont{R.}~\bibnamefont{{Cen}}}, \bibnamefont{and}
  \bibinfo{author}{\bibfnamefont{J.~P.} \bibnamefont{{Ostriker}}},
  \bibinfo{journal}{\apj} \textbf{\bibinfo{volume}{542}}, \bibinfo{pages}{608}
  (\bibinfo{year}{2000}).

\bibitem[{\citenamefont{{Oh} et~al.}(2003)\citenamefont{{Oh}, {Cooray}, and
  {Kamionkowski}}}]{Oh2003}
\bibinfo{author}{\bibfnamefont{S.~P.} \bibnamefont{{Oh}}},
  \bibinfo{author}{\bibfnamefont{A.}~\bibnamefont{{Cooray}}}, \bibnamefont{and}
  \bibinfo{author}{\bibfnamefont{M.}~\bibnamefont{{Kamionkowski}}},
  \bibinfo{journal}{\mnras} \textbf{\bibinfo{volume}{342}},
  \bibinfo{pages}{L20} (\bibinfo{year}{2003}).

\bibitem[{\citenamefont{{Khatri} and
  {Sunyaev}}(2015{\natexlab{b}})}]{Khatri2015y}
\bibinfo{author}{\bibfnamefont{R.}~\bibnamefont{{Khatri}}} \bibnamefont{and}
  \bibinfo{author}{\bibfnamefont{R.}~\bibnamefont{{Sunyaev}}},
  \bibinfo{journal}{\jcap} \textbf{\bibinfo{volume}{8}}, \bibinfo{eid}{013}
  (\bibinfo{year}{2015}{\natexlab{b}}), \eprint{1505.00781}.

\bibitem[{\citenamefont{{Hill} et~al.}(2015)}]{Hill2015}
\bibinfo{author}{\bibfnamefont{J.~C.} \bibnamefont{{Hill}}}
  \bibnamefont{et~al.}, \bibinfo{journal}{\prl} \textbf{\bibinfo{volume}{115}},
  \bibinfo{eid}{261301} (\bibinfo{year}{2015}), \eprint{1507.01583}.

\bibitem[{\citenamefont{{Battaglia} et~al.}(2010)\citenamefont{{Battaglia},
  {Bond}, {Pfrommer}, {Sievers}, and {Sijacki}}}]{Battaglia2010}
\bibinfo{author}{\bibfnamefont{N.}~\bibnamefont{{Battaglia}}},
  \bibinfo{author}{\bibfnamefont{J.~R.} \bibnamefont{{Bond}}},
  \bibinfo{author}{\bibfnamefont{C.}~\bibnamefont{{Pfrommer}}},
  \bibinfo{author}{\bibfnamefont{J.~L.} \bibnamefont{{Sievers}}},
  \bibnamefont{and}
  \bibinfo{author}{\bibfnamefont{D.}~\bibnamefont{{Sijacki}}},
  \bibinfo{journal}{\apj} \textbf{\bibinfo{volume}{725}}, \bibinfo{pages}{91}
  (\bibinfo{year}{2010}), \eprint{1003.4256}.

\bibitem[{\citenamefont{{Le Brun} et~al.}(2014)\citenamefont{{Le Brun},
  {McCarthy}, {Schaye}, and {Ponman}}}]{LeBrun2014}
\bibinfo{author}{\bibfnamefont{A.~M.~C.} \bibnamefont{{Le Brun}}},
  \bibinfo{author}{\bibfnamefont{I.~G.} \bibnamefont{{McCarthy}}},
  \bibinfo{author}{\bibfnamefont{J.}~\bibnamefont{{Schaye}}}, \bibnamefont{and}
  \bibinfo{author}{\bibfnamefont{T.~J.} \bibnamefont{{Ponman}}},
  \bibinfo{journal}{\mnras} \textbf{\bibinfo{volume}{441}},
  \bibinfo{pages}{1270} (\bibinfo{year}{2014}), \eprint{1312.5462}.

\bibitem[{\citenamefont{{Wright}}(1979)}]{Wright1979}
\bibinfo{author}{\bibfnamefont{E.~L.} \bibnamefont{{Wright}}},
  \bibinfo{journal}{\apj} \textbf{\bibinfo{volume}{232}}, \bibinfo{pages}{348}
  (\bibinfo{year}{1979}).

\bibitem[{\citenamefont{{Rephaeli}}(1995)}]{Rephaeli1995}
\bibinfo{author}{\bibfnamefont{Y.}~\bibnamefont{{Rephaeli}}},
  \bibinfo{journal}{\apj} \textbf{\bibinfo{volume}{445}}, \bibinfo{pages}{33}
  (\bibinfo{year}{1995}).

\bibitem[{\citenamefont{{Sazonov} and {Sunyaev}}(1998)}]{Sazonov1998}
\bibinfo{author}{\bibfnamefont{S.~Y.} \bibnamefont{{Sazonov}}}
  \bibnamefont{and} \bibinfo{author}{\bibfnamefont{R.~A.}
  \bibnamefont{{Sunyaev}}}, \bibinfo{journal}{\apj}
  \textbf{\bibinfo{volume}{508}}, \bibinfo{pages}{1} (\bibinfo{year}{1998}).

\bibitem[{\citenamefont{{Itoh} et~al.}(1998)\citenamefont{{Itoh}, {Kohyama},
  and {Nozawa}}}]{Itoh98}
\bibinfo{author}{\bibfnamefont{N.}~\bibnamefont{{Itoh}}},
  \bibinfo{author}{\bibfnamefont{Y.}~\bibnamefont{{Kohyama}}},
  \bibnamefont{and} \bibinfo{author}{\bibfnamefont{S.}~\bibnamefont{{Nozawa}}},
  \bibinfo{journal}{\apj} \textbf{\bibinfo{volume}{502}}, \bibinfo{pages}{7}
  (\bibinfo{year}{1998}), \eprint{arXiv:astro-ph/9712289}.

\bibitem[{\citenamefont{{Challinor} and {Lasenby}}(1998)}]{Challinor1998}
\bibinfo{author}{\bibfnamefont{A.}~\bibnamefont{{Challinor}}} \bibnamefont{and}
  \bibinfo{author}{\bibfnamefont{A.}~\bibnamefont{{Lasenby}}},
  \bibinfo{journal}{\apj} \textbf{\bibinfo{volume}{499}}, \bibinfo{pages}{1}
  (\bibinfo{year}{1998}), \eprint{arXiv:astro-ph/9711161}.

\bibitem[{\citenamefont{{Chluba}
  et~al.}(2012{\natexlab{c}})\citenamefont{{Chluba}, {Nagai}, {Sazonov}, and
  {Nelson}}}]{ChlubaSZpack}
\bibinfo{author}{\bibfnamefont{J.}~\bibnamefont{{Chluba}}},
  \bibinfo{author}{\bibfnamefont{D.}~\bibnamefont{{Nagai}}},
  \bibinfo{author}{\bibfnamefont{S.}~\bibnamefont{{Sazonov}}},
  \bibnamefont{and} \bibinfo{author}{\bibfnamefont{K.}~\bibnamefont{{Nelson}}},
  \bibinfo{journal}{\mnras} \textbf{\bibinfo{volume}{426}},
  \bibinfo{pages}{510} (\bibinfo{year}{2012}{\natexlab{c}}),
  \eprint{1205.5778}.

\bibitem[{\citenamefont{{Kay} et~al.}(2008)\citenamefont{{Kay}, {Powell},
  {Liddle}, and {Thomas}}}]{Kay2008}
\bibinfo{author}{\bibfnamefont{S.~T.} \bibnamefont{{Kay}}},
  \bibinfo{author}{\bibfnamefont{L.~C.} \bibnamefont{{Powell}}},
  \bibinfo{author}{\bibfnamefont{A.~R.} \bibnamefont{{Liddle}}},
  \bibnamefont{and} \bibinfo{author}{\bibfnamefont{P.~A.}
  \bibnamefont{{Thomas}}}, \bibinfo{journal}{\mnras}
  \textbf{\bibinfo{volume}{386}}, \bibinfo{pages}{2110} (\bibinfo{year}{2008}),
  \eprint{0706.3668}.

\bibitem[{\citenamefont{{Lee} et~al.}(2019)\citenamefont{{Lee}, {Chluba}, and
  {Kay}}}]{Lee2019}
\bibinfo{author}{\bibfnamefont{E.}~\bibnamefont{{Lee}}},
  \bibinfo{author}{\bibfnamefont{J.}~\bibnamefont{{Chluba}}}, \bibnamefont{and}
  \bibinfo{author}{\bibfnamefont{S.}~\bibnamefont{{Kay}}}, \bibinfo{journal}{in
  prep}  (\bibinfo{year}{2019}).

\bibitem[{\citenamefont{{Zeldovich} et~al.}(1972)\citenamefont{{Zeldovich},
  {Illarionov}, and {Sunyaev}}}]{Zeldovich1972}
\bibinfo{author}{\bibfnamefont{Y.~B.} \bibnamefont{{Zeldovich}}},
  \bibinfo{author}{\bibfnamefont{A.~F.} \bibnamefont{{Illarionov}}},
  \bibnamefont{and} \bibinfo{author}{\bibfnamefont{R.~A.}
  \bibnamefont{{Sunyaev}}}, \bibinfo{journal}{SJETP}
  \textbf{\bibinfo{volume}{35}}, \bibinfo{pages}{643} (\bibinfo{year}{1972}).

\bibitem[{\citenamefont{{Shimon} and {Rephaeli}}(2002)}]{Shimon2002}
\bibinfo{author}{\bibfnamefont{M.}~\bibnamefont{{Shimon}}} \bibnamefont{and}
  \bibinfo{author}{\bibfnamefont{Y.}~\bibnamefont{{Rephaeli}}},
  \bibinfo{journal}{\apj} \textbf{\bibinfo{volume}{575}}, \bibinfo{pages}{12}
  (\bibinfo{year}{2002}), \eprint{astro-ph/0204355}.

\bibitem[{\citenamefont{{Colafrancesco}
  et~al.}(2003)\citenamefont{{Colafrancesco}, {Marchegiani}, and
  {Palladino}}}]{Colafrancesco2003}
\bibinfo{author}{\bibfnamefont{S.}~\bibnamefont{{Colafrancesco}}},
  \bibinfo{author}{\bibfnamefont{P.}~\bibnamefont{{Marchegiani}}},
  \bibnamefont{and}
  \bibinfo{author}{\bibfnamefont{E.}~\bibnamefont{{Palladino}}},
  \bibinfo{journal}{\aap} \textbf{\bibinfo{volume}{397}}, \bibinfo{pages}{27}
  (\bibinfo{year}{2003}), \eprint{arXiv:astro-ph/0211649}.

\bibitem[{\citenamefont{{Colafrancesco}
  et~al.}(2006)\citenamefont{{Colafrancesco}, {Profumo}, and
  {Ullio}}}]{Colafrancesco06}
\bibinfo{author}{\bibfnamefont{S.}~\bibnamefont{{Colafrancesco}}},
  \bibinfo{author}{\bibfnamefont{S.}~\bibnamefont{{Profumo}}},
  \bibnamefont{and} \bibinfo{author}{\bibfnamefont{P.}~\bibnamefont{{Ullio}}},
  \bibinfo{journal}{\aap} \textbf{\bibinfo{volume}{455}}, \bibinfo{pages}{21}
  (\bibinfo{year}{2006}), \eprint{astro-ph/0507575}.

\bibitem[{\citenamefont{{Fixsen} et~al.}(1998)}]{1998ApJ...508..123F}
\bibinfo{author}{\bibfnamefont{D.~J.} \bibnamefont{{Fixsen}}}
  \bibnamefont{et~al.}, \bibinfo{journal}{\apj} \textbf{\bibinfo{volume}{508}},
  \bibinfo{pages}{123} (\bibinfo{year}{1998}), \eprint{astro-ph/9803021}.

\bibitem[{\citenamefont{{Serra} et~al.}(2014)\citenamefont{{Serra}, {Lagache},
  {Dor{\'e}}, {Pullen}, and {White}}}]{2014A&A...570A..98S}
\bibinfo{author}{\bibfnamefont{P.}~\bibnamefont{{Serra}}},
  \bibinfo{author}{\bibfnamefont{G.}~\bibnamefont{{Lagache}}},
  \bibinfo{author}{\bibfnamefont{O.}~\bibnamefont{{Dor{\'e}}}},
  \bibinfo{author}{\bibfnamefont{A.}~\bibnamefont{{Pullen}}}, \bibnamefont{and}
  \bibinfo{author}{\bibfnamefont{M.}~\bibnamefont{{White}}},
  \bibinfo{journal}{\aap} \textbf{\bibinfo{volume}{570}}, \bibinfo{eid}{A98}
  (\bibinfo{year}{2014}), \eprint{1404.1933}.

\bibitem[{\citenamefont{{Kovetz} et~al.}(2017)}]{2017arXiv170909066K}
\bibinfo{author}{\bibfnamefont{E.~D.} \bibnamefont{{Kovetz}}}
  \bibnamefont{et~al.}, \bibinfo{journal}{arXiv e-prints}
  (\bibinfo{year}{2017}), \eprint{1709.09066}.

\bibitem[{\citenamefont{{Kovetz} et~al.}(2019)}]{2019BAAS...51c.101K}
\bibinfo{author}{\bibfnamefont{E.}~\bibnamefont{{Kovetz}}}
  \bibnamefont{et~al.}, \bibinfo{journal}{\baas} \textbf{\bibinfo{volume}{51}},
  \bibinfo{eid}{101} (\bibinfo{year}{2019}), \eprint{1903.04496}.

\bibitem[{\citenamefont{{Carilli} and {Walter}}(2013)}]{2013ARA&A..51..105C}
\bibinfo{author}{\bibfnamefont{C.~L.} \bibnamefont{{Carilli}}}
  \bibnamefont{and} \bibinfo{author}{\bibfnamefont{F.}~\bibnamefont{{Walter}}},
  \bibinfo{journal}{\araa} \textbf{\bibinfo{volume}{51}}, \bibinfo{pages}{105}
  (\bibinfo{year}{2013}), \eprint{1301.0371}.

\bibitem[{\citenamefont{{Pullen} et~al.}(2018)\citenamefont{{Pullen}, {Serra},
  {Chang}, {Dor{\'e}}, and {Ho}}}]{2018MNRAS.478.1911P}
\bibinfo{author}{\bibfnamefont{A.~R.} \bibnamefont{{Pullen}}},
  \bibinfo{author}{\bibfnamefont{P.}~\bibnamefont{{Serra}}},
  \bibinfo{author}{\bibfnamefont{T.-C.} \bibnamefont{{Chang}}},
  \bibinfo{author}{\bibfnamefont{O.}~\bibnamefont{{Dor{\'e}}}},
  \bibnamefont{and} \bibinfo{author}{\bibfnamefont{S.}~\bibnamefont{{Ho}}},
  \bibinfo{journal}{\mnras} \textbf{\bibinfo{volume}{478}},
  \bibinfo{pages}{1911} (\bibinfo{year}{2018}), \eprint{1707.06172}.

\bibitem[{\citenamefont{{Yang} et~al.}(2019)\citenamefont{{Yang}, {Pullen}, and
  {Switzer}}}]{2019arXiv190401180Y}
\bibinfo{author}{\bibfnamefont{S.}~\bibnamefont{{Yang}}},
  \bibinfo{author}{\bibfnamefont{A.~R.} \bibnamefont{{Pullen}}},
  \bibnamefont{and} \bibinfo{author}{\bibfnamefont{E.~R.}
  \bibnamefont{{Switzer}}}, \bibinfo{journal}{arXiv e-prints}
  (\bibinfo{year}{2019}), \eprint{1904.01180}.

\bibitem[{\citenamefont{{Padmanabhan}}(2019)}]{2019MNRAS.tmp.1857P}
\bibinfo{author}{\bibfnamefont{H.}~\bibnamefont{{Padmanabhan}}},
  \bibinfo{journal}{\mnras} p. \bibinfo{pages}{1857} (\bibinfo{year}{2019}),
  \eprint{1811.01968}.

\bibitem[{\citenamefont{{Mashian} et~al.}(2016)\citenamefont{{Mashian}, {Loeb},
  and {Sternberg}}}]{2016MNRAS.458L..99M}
\bibinfo{author}{\bibfnamefont{N.}~\bibnamefont{{Mashian}}},
  \bibinfo{author}{\bibfnamefont{A.}~\bibnamefont{{Loeb}}}, \bibnamefont{and}
  \bibinfo{author}{\bibfnamefont{A.}~\bibnamefont{{Sternberg}}},
  \bibinfo{journal}{\mnras} \textbf{\bibinfo{volume}{458}},
  \bibinfo{pages}{L99} (\bibinfo{year}{2016}), \eprint{1601.02618}.

\bibitem[{\citenamefont{{Serra} et~al.}(2016)\citenamefont{{Serra}, {Dor{\'e}},
  and {Lagache}}}]{2016ApJ...833..153S}
\bibinfo{author}{\bibfnamefont{P.}~\bibnamefont{{Serra}}},
  \bibinfo{author}{\bibfnamefont{O.}~\bibnamefont{{Dor{\'e}}}},
  \bibnamefont{and}
  \bibinfo{author}{\bibfnamefont{G.}~\bibnamefont{{Lagache}}},
  \bibinfo{journal}{\apj} \textbf{\bibinfo{volume}{833}}, \bibinfo{eid}{153}
  (\bibinfo{year}{2016}), \eprint{1608.00585}.

\bibitem[{\citenamefont{{Switzer}}(2017)}]{2017ApJ...838...82S}
\bibinfo{author}{\bibfnamefont{E.~R.} \bibnamefont{{Switzer}}},
  \bibinfo{journal}{\apj} \textbf{\bibinfo{volume}{838}}, \bibinfo{eid}{82}
  (\bibinfo{year}{2017}), \eprint{1703.07832}.

\bibitem[{\citenamefont{{Switzer} et~al.}(2019)\citenamefont{{Switzer},
  {Anderson}, {Pullen}, and {Yang}}}]{2019ApJ...872...82S}
\bibinfo{author}{\bibfnamefont{E.~R.} \bibnamefont{{Switzer}}},
  \bibinfo{author}{\bibfnamefont{C.~J.} \bibnamefont{{Anderson}}},
  \bibinfo{author}{\bibfnamefont{A.~R.} \bibnamefont{{Pullen}}},
  \bibnamefont{and} \bibinfo{author}{\bibfnamefont{S.}~\bibnamefont{{Yang}}},
  \bibinfo{journal}{\apj} \textbf{\bibinfo{volume}{872}}, \bibinfo{eid}{82}
  (\bibinfo{year}{2019}), \eprint{1812.06223}.

\bibitem[{\citenamefont{{Moradinezhad Dizgah} and
  {Keating}}(2019)}]{2019ApJ...872..126M}
\bibinfo{author}{\bibfnamefont{A.}~\bibnamefont{{Moradinezhad Dizgah}}}
  \bibnamefont{and} \bibinfo{author}{\bibfnamefont{G.~K.}
  \bibnamefont{{Keating}}}, \bibinfo{journal}{\apj}
  \textbf{\bibinfo{volume}{872}}, \bibinfo{eid}{126} (\bibinfo{year}{2019}),
  \eprint{1810.02850}.

\bibitem[{\citenamefont{{Decarli} et~al.}(2019)}]{2019arXiv190309164D}
\bibinfo{author}{\bibfnamefont{R.}~\bibnamefont{{Decarli}}}
  \bibnamefont{et~al.}, \bibinfo{journal}{arXiv e-prints}
  (\bibinfo{year}{2019}), \eprint{1903.09164}.

\bibitem[{\citenamefont{{Loeb}}(2001)}]{Loeb2001}
\bibinfo{author}{\bibfnamefont{A.}~\bibnamefont{{Loeb}}},
  \bibinfo{journal}{\apjl} \textbf{\bibinfo{volume}{555}}, \bibinfo{pages}{L1}
  (\bibinfo{year}{2001}), \eprint{astro-ph/0103505}.

\bibitem[{\citenamefont{{Zaldarriaga} and {Loeb}}(2002)}]{Zaldarriaga2002}
\bibinfo{author}{\bibfnamefont{M.}~\bibnamefont{{Zaldarriaga}}}
  \bibnamefont{and} \bibinfo{author}{\bibfnamefont{A.}~\bibnamefont{{Loeb}}},
  \bibinfo{journal}{\apj} \textbf{\bibinfo{volume}{564}}, \bibinfo{pages}{52}
  (\bibinfo{year}{2002}), \eprint{astro-ph/0105345}.

\bibitem[{\citenamefont{{Basu} et~al.}(2004)\citenamefont{{Basu},
  {Hern{\'a}ndez-Monteagudo}, and {Sunyaev}}}]{Kaustuv2004}
\bibinfo{author}{\bibfnamefont{K.}~\bibnamefont{{Basu}}},
  \bibinfo{author}{\bibfnamefont{C.}~\bibnamefont{{Hern{\'a}ndez-Monteagudo}}},
  \bibnamefont{and} \bibinfo{author}{\bibfnamefont{R.~A.}
  \bibnamefont{{Sunyaev}}}, \bibinfo{journal}{\aap}
  \textbf{\bibinfo{volume}{416}}, \bibinfo{pages}{447} (\bibinfo{year}{2004}).

\bibitem[{\citenamefont{{Rubi{\~n}o-Mart{\'{\i}}n}
  et~al.}(2005)\citenamefont{{Rubi{\~n}o-Mart{\'{\i}}n},
  {Hern{\'a}ndez-Monteagudo}, and {Sunyaev}}}]{Jose2005}
\bibinfo{author}{\bibfnamefont{J.~A.}
  \bibnamefont{{Rubi{\~n}o-Mart{\'{\i}}n}}},
  \bibinfo{author}{\bibfnamefont{C.}~\bibnamefont{{Hern{\'a}ndez-Monteagudo}}},
  \bibnamefont{and} \bibinfo{author}{\bibfnamefont{R.~A.}
  \bibnamefont{{Sunyaev}}}, \bibinfo{journal}{\aap}
  \textbf{\bibinfo{volume}{438}}, \bibinfo{pages}{461} (\bibinfo{year}{2005}).

\bibitem[{\citenamefont{{Hern{\'a}ndez-Monteagudo} and
  {Sunyaev}}(2005)}]{Carlos2005}
\bibinfo{author}{\bibfnamefont{C.}~\bibnamefont{{Hern{\'a}ndez-Monteagudo}}}
  \bibnamefont{and} \bibinfo{author}{\bibfnamefont{R.~A.}
  \bibnamefont{{Sunyaev}}}, \bibinfo{journal}{\mnras}
  \textbf{\bibinfo{volume}{359}}, \bibinfo{pages}{597} (\bibinfo{year}{2005}),
  \eprint{astro-ph/0405487}.

\bibitem[{\citenamefont{{Hern{\'a}ndez-Monteagudo}
  et~al.}(2006)\citenamefont{{Hern{\'a}ndez-Monteagudo}, {Verde}, and
  {Jimenez}}}]{Carlos2006}
\bibinfo{author}{\bibfnamefont{C.}~\bibnamefont{{Hern{\'a}ndez-Monteagudo}}},
  \bibinfo{author}{\bibfnamefont{L.}~\bibnamefont{{Verde}}}, \bibnamefont{and}
  \bibinfo{author}{\bibfnamefont{R.}~\bibnamefont{{Jimenez}}},
  \bibinfo{journal}{\apj} \textbf{\bibinfo{volume}{653}}, \bibinfo{pages}{1}
  (\bibinfo{year}{2006}), \eprint{astro-ph/0604324}.

\bibitem[{\citenamefont{{Hern{\'a}ndez-Monteagudo}
  et~al.}(2007{\natexlab{a}})\citenamefont{{Hern{\'a}ndez-Monteagudo},
  {Rubi{\~n}o-Mart{\'{\i}}n}, and {Sunyaev}}}]{Carlos2007}
\bibinfo{author}{\bibfnamefont{C.}~\bibnamefont{{Hern{\'a}ndez-Monteagudo}}},
  \bibinfo{author}{\bibfnamefont{J.~A.}
  \bibnamefont{{Rubi{\~n}o-Mart{\'{\i}}n}}}, \bibnamefont{and}
  \bibinfo{author}{\bibfnamefont{R.~A.} \bibnamefont{{Sunyaev}}},
  \bibinfo{journal}{\mnras} \textbf{\bibinfo{volume}{380}},
  \bibinfo{pages}{1656} (\bibinfo{year}{2007}{\natexlab{a}}).

\bibitem[{\citenamefont{{Hern{\'a}ndez-Monteagudo}
  et~al.}(2017)\citenamefont{{Hern{\'a}ndez-Monteagudo}, {Maio}, {Ciardi}, and
  {Sunyaev}}}]{Carlos2017}
\bibinfo{author}{\bibfnamefont{C.}~\bibnamefont{{Hern{\'a}ndez-Monteagudo}}},
  \bibinfo{author}{\bibfnamefont{U.}~\bibnamefont{{Maio}}},
  \bibinfo{author}{\bibfnamefont{B.}~\bibnamefont{{Ciardi}}}, \bibnamefont{and}
  \bibinfo{author}{\bibfnamefont{R.~A.} \bibnamefont{{Sunyaev}}},
  \bibinfo{journal}{arXiv e-prints}  (\bibinfo{year}{2017}),
  \eprint{1707.01910}.

\bibitem[{\citenamefont{{Yu} et~al.}(2001)\citenamefont{{Yu}, {Spergel}, and
  {Ostriker}}}]{Yu2001}
\bibinfo{author}{\bibfnamefont{Q.}~\bibnamefont{{Yu}}},
  \bibinfo{author}{\bibfnamefont{D.~N.} \bibnamefont{{Spergel}}},
  \bibnamefont{and} \bibinfo{author}{\bibfnamefont{J.~P.}
  \bibnamefont{{Ostriker}}}, \bibinfo{journal}{\apj}
  \textbf{\bibinfo{volume}{558}}, \bibinfo{pages}{23} (\bibinfo{year}{2001}),
  \eprint{astro-ph/0103149}.

\bibitem[{\citenamefont{{Lewis}}(2013)}]{Lewis2013}
\bibinfo{author}{\bibfnamefont{A.}~\bibnamefont{{Lewis}}},
  \bibinfo{journal}{\jcap} \textbf{\bibinfo{volume}{8}}, \bibinfo{eid}{053}
  (\bibinfo{year}{2013}), \eprint{1307.8148}.

\bibitem[{\citenamefont{{Righi}
  et~al.}(2008{\natexlab{a}})\citenamefont{{Righi}, {Hern{\'a}ndez-Monteagudo},
  and {Sunyaev}}}]{Righi2008b}
\bibinfo{author}{\bibfnamefont{M.}~\bibnamefont{{Righi}}},
  \bibinfo{author}{\bibfnamefont{C.}~\bibnamefont{{Hern{\'a}ndez-Monteagudo}}},
  \bibnamefont{and} \bibinfo{author}{\bibfnamefont{R.~A.}
  \bibnamefont{{Sunyaev}}}, \bibinfo{journal}{\aap}
  \textbf{\bibinfo{volume}{489}}, \bibinfo{pages}{489}
  (\bibinfo{year}{2008}{\natexlab{a}}), \eprint{0805.2174}.

\bibitem[{\citenamefont{{Hern{\'a}ndez-Monteagudo}
  et~al.}(2007{\natexlab{b}})\citenamefont{{Hern{\'a}ndez-Monteagudo},
  {Haiman}, {Jimenez}, and {Verde}}}]{Hernandezetal2007b}
\bibinfo{author}{\bibfnamefont{C.}~\bibnamefont{{Hern{\'a}ndez-Monteagudo}}},
  \bibinfo{author}{\bibfnamefont{Z.}~\bibnamefont{{Haiman}}},
  \bibinfo{author}{\bibfnamefont{R.}~\bibnamefont{{Jimenez}}},
  \bibnamefont{and} \bibinfo{author}{\bibfnamefont{L.}~\bibnamefont{{Verde}}},
  \bibinfo{journal}{\apjl} \textbf{\bibinfo{volume}{660}}, \bibinfo{pages}{L85}
  (\bibinfo{year}{2007}{\natexlab{b}}).

\bibitem[{\citenamefont{{Gong} et~al.}(2012)}]{Gongetal2012}
\bibinfo{author}{\bibfnamefont{Y.}~\bibnamefont{{Gong}}} \bibnamefont{et~al.},
  \bibinfo{journal}{\apj} \textbf{\bibinfo{volume}{745}}, \bibinfo{eid}{49}
  (\bibinfo{year}{2012}), \eprint{1107.3553}.

\bibitem[{\citenamefont{{Aghanim} et~al.}(2019)}]{Nabila2018}
\bibinfo{author}{\bibfnamefont{N.}~\bibnamefont{{Aghanim}}}
  \bibnamefont{et~al.}, \bibinfo{journal}{ESA F-class mission proposal}
  (\bibinfo{year}{2019}).

\bibitem[{\citenamefont{{Abitbol} et~al.}(2017)\citenamefont{{Abitbol},
  {Chluba}, {Hill}, and {Johnson}}}]{abitbol_pixie}
\bibinfo{author}{\bibfnamefont{M.~H.} \bibnamefont{{Abitbol}}},
  \bibinfo{author}{\bibfnamefont{J.}~\bibnamefont{{Chluba}}},
  \bibinfo{author}{\bibfnamefont{J.~C.} \bibnamefont{{Hill}}},
  \bibnamefont{and} \bibinfo{author}{\bibfnamefont{B.~R.}
  \bibnamefont{{Johnson}}}, \bibinfo{journal}{\mnras}
  \textbf{\bibinfo{volume}{471}}, \bibinfo{pages}{1126} (\bibinfo{year}{2017}).

\bibitem[{\citenamefont{{Masi} et~al.}(2003)}]{Masi2003}
\bibinfo{author}{\bibfnamefont{S.}~\bibnamefont{{Masi}}} \bibnamefont{et~al.},
  \bibinfo{journal}{Memorie della Societ\'a Astronomica Italiana}
  \textbf{\bibinfo{volume}{74}}, \bibinfo{pages}{96} (\bibinfo{year}{2003}).

\bibitem[{\citenamefont{{Schillaci} et~al.}(2014)}]{Schillaci2014}
\bibinfo{author}{\bibfnamefont{A.}~\bibnamefont{{Schillaci}}}
  \bibnamefont{et~al.}, \bibinfo{journal}{\aap} \textbf{\bibinfo{volume}{565}},
  \bibinfo{eid}{A125} (\bibinfo{year}{2014}).

\bibitem[{\citenamefont{{Kogut} et~al.}(2006)}]{Kogut2006ARCADE}
\bibinfo{author}{\bibfnamefont{A.}~\bibnamefont{{Kogut}}} \bibnamefont{et~al.},
  \bibinfo{journal}{New Astronomy Reviews} \textbf{\bibinfo{volume}{50}},
  \bibinfo{pages}{925} (\bibinfo{year}{2006}), \eprint{arXiv:astro-ph/0609373}.

\bibitem[{\citenamefont{{Seiffert} et~al.}(2011)}]{arcade2}
\bibinfo{author}{\bibfnamefont{M.}~\bibnamefont{{Seiffert}}}
  \bibnamefont{et~al.}, \bibinfo{journal}{\apj} \textbf{\bibinfo{volume}{734}},
  \bibinfo{eid}{6} (\bibinfo{year}{2011}).

\bibitem[{\citenamefont{{Maffei} et~al.}(2019)}]{Bruno2019}
\bibinfo{author}{\bibfnamefont{B.}~\bibnamefont{{Maffei}}}
  \bibnamefont{et~al.}, \bibinfo{journal}{Balloon proposal to CNES}
  (\bibinfo{year}{2019}).

\bibitem[{\citenamefont{{Chluba}
  et~al.}(2017{\natexlab{b}})\citenamefont{{Chluba}, {Hill}, and
  {Abitbol}}}]{Chluba2017Moments}
\bibinfo{author}{\bibfnamefont{J.}~\bibnamefont{{Chluba}}},
  \bibinfo{author}{\bibfnamefont{J.~C.} \bibnamefont{{Hill}}},
  \bibnamefont{and} \bibinfo{author}{\bibfnamefont{M.~H.}
  \bibnamefont{{Abitbol}}}, \bibinfo{journal}{\mnras}
  \textbf{\bibinfo{volume}{472}}, \bibinfo{pages}{1195}
  (\bibinfo{year}{2017}{\natexlab{b}}), \eprint{1701.00274}.

\bibitem[{\citenamefont{{Remazeilles} and
  {Chluba}}(2018{\natexlab{b}})}]{Remazeilles2018mu}
\bibinfo{author}{\bibfnamefont{M.}~\bibnamefont{{Remazeilles}}}
  \bibnamefont{and} \bibinfo{author}{\bibfnamefont{J.}~\bibnamefont{{Chluba}}},
  \bibinfo{journal}{\mnras} \textbf{\bibinfo{volume}{478}},
  \bibinfo{pages}{807} (\bibinfo{year}{2018}{\natexlab{b}}),
  \eprint{1802.10101}.

\bibitem[{\citenamefont{{Remazeilles} et~al.}(2018)}]{Remazeilles2018}
\bibinfo{author}{\bibfnamefont{M.}~\bibnamefont{{Remazeilles}}}
  \bibnamefont{et~al.}, \bibinfo{journal}{\jcap} \textbf{\bibinfo{volume}{4}},
  \bibinfo{eid}{023} (\bibinfo{year}{2018}), \eprint{1704.04501}.

\bibitem[{\citenamefont{{Righi}
  et~al.}(2008{\natexlab{b}})\citenamefont{{Righi}, {Hern{\'a}ndez-Monteagudo},
  and {Sunyaev}}}]{Righi2008}
\bibinfo{author}{\bibfnamefont{M.}~\bibnamefont{{Righi}}},
  \bibinfo{author}{\bibfnamefont{C.}~\bibnamefont{{Hern{\'a}ndez-Monteagudo}}},
  \bibnamefont{and} \bibinfo{author}{\bibfnamefont{R.~A.}
  \bibnamefont{{Sunyaev}}}, \bibinfo{journal}{\aap}
  \textbf{\bibinfo{volume}{478}}, \bibinfo{pages}{685}
  (\bibinfo{year}{2008}{\natexlab{b}}), \eprint{0707.0288}.

\bibitem[{\citenamefont{{Planck Collaboration}}(2019)}]{Planck2019STISO}
\bibinfo{author}{\bibnamefont{{Planck Collaboration}}}, \bibinfo{journal}{arXiv
  e-prints}  (\bibinfo{year}{2019}), \eprint{1906.02552}.

\bibitem[{\citenamefont{{Inomata} and {Nakama}}(2019)}]{Inomata2018}
\bibinfo{author}{\bibfnamefont{K.}~\bibnamefont{{Inomata}}} \bibnamefont{and}
  \bibinfo{author}{\bibfnamefont{T.}~\bibnamefont{{Nakama}}},
  \bibinfo{journal}{\prd} \textbf{\bibinfo{volume}{99}}, \bibinfo{eid}{043511}
  (\bibinfo{year}{2019}), \eprint{1812.00674}.

\end{thebibliography}

\end{document}